%% file: Ayan_abun_draft.tex
\documentclass[fleqn,usenatbib]{mnras}

\usepackage{mathptmx}

\usepackage[T1]{fontenc}
\usepackage{ae,aecompl}

\usepackage{graphicx}	
\usepackage{amsmath}	
\usepackage{amssymb}	
\usepackage{caption}
\usepackage{subcaption}
\usepackage{hyperref}
\usepackage{multicol}        
\usepackage{bm}		
\usepackage{pdflscape}	
\usepackage{booktabs} 
\usepackage{ifpdf} 

\newcommand{\galaxy}{RCS0327}
\newcommand{\knotE}{RCS0327-E}
\newcommand{\etal}{et al.}
\newcommand{\pc}{pc}

\newcommand{\Myr}{Myr}

\newcommand{\kmps}{km\,s$^{-1}$}
\newcommand{\Msun}{M$_{\odot}$\,}
\newcommand{\Msunpyr}{\,M$_{\odot}$yr$^{-1}$}
\newcommand{\logq}{$\log{(q)}$}
\newcommand{\lpok}{$\log{(P/k)}$}
\newcommand{\Ne}{n$_{e}$}
\newcommand{\Te}{T$_{e}$}
\newcommand{\logOH}{12+log(O/H)}
\newcommand{\z}{$z$}
\newcommand{\JWST}{\textit{JWST}}
\newcommand{\HST}{\textit{HST}}
\newcommand{\htr}{\ion{H}{ii} region}
\newcommand{\htrs}{\ion{H}{ii} regions}

\usepackage[usenames, dvipsnames]{color}
\usepackage[normalem]{ulem}

\defcitealias{Rigby:2011aa}{R11}
\defcitealias{Kewley:2002fk}{KD02}
\defcitealias{Whitaker:2014aa}{W14}
\defcitealias{Blanc:2015aa}{B15}
\defcitealias{Pettini:2004qe}{PP04}
\defcitealias{Shi:2007aa}{S07}
\defcitealias{Dopita:2016aa}{D16}
\defcitealias{Kobulnicky:2004aa}{KK04}
\defcitealias{Izotov:2006ab}{I06}
\defcitealias{Garnett:1995aa}{G95a}
\defcitealias{Garnett:1995ab}{G95b}
\defcitealias{Levesque:2014aa}{LR14}
\defcitealias{Kewley:2008aa}{KE08}
\defcitealias{Bian:2018aa}{BKD18}
\defcitealias{Sanders:2016aa}{S16}
\defcitealias{Strom:2018aa}{S18}
\defcitealias{Kewley:2019aa}{K19a}
\defcitealias{Kewley:2019ab}{K19b}

\title[ISM properties from UV-optical diagnostics]{Rest-frame UV and optical emission line diagnostics of ionised gas properties: a test case in a star-forming knot of a lensed galaxy at  $z \sim 1.7$}

\author[A. Acharyya {\etal}]
{Ayan Acharyya$^{1,2}$\thanks{E-mail: ayan.acharyya@anu.edu.au},
Lisa J.~Kewley$^{1,2}$,
Jane R. Rigby$^{3}$,
Matthew Bayliss$^{4}$,
\newauthor
Fuyan Bian$^{1,5}$,
David Nicholls$^{1}$,
Christoph Federrath$^{1,2}$,
Melanie Kaasinen$^{6}$,
\newauthor
Michael Florian$^{3}$,
and Guillermo A. Blanc$^{7,8}$
\\
$^1$Research School of Astronomy and Astrophysics, Australian National University, Canberra, ACT~2611, Australia\\
$^2$ARC Centre of Excellence for All Sky Astrophysics in 3 Dimensions (ASTRO 3D)\\
$^3$Astrophysics Science Division, Goddard Space Flight Center, 8800 Greenbelt Rd., Greenbelt, MD 20771, USA\\
$^4$MIT Kavli Institute for Astrophysics and Space Research, 77 Massachusetts Ave., Cambridge, MA 02139, USA\\
$^5$European Southern Observatory, Alonso de C\'{o}rdova 3107, Casilla 19001, Vitacura, Santiago 19, Chile \\
$^6$Max Planck Institute f\"ur Astronomie, K\"onigstuhl 17, 69117, Heidelberg, Germanyy\\
$^7$Observatories of the Carnegie Institution for Science, 813 Santa Barbara Street, Pasadena, CA 91101, USA \\
$^8$Departamento de Astronomía, Universidad de Chile, Camino del Observatorio 1515, Las Condes, Santiago, Chile
}

\date{Accepted XXX. Received YYY; in original form ZZZ}

\pubyear{2017}

\begin{document}
\label{firstpage}
\pagerange{\pageref{firstpage}--\pageref{lastpage}}
\maketitle

\begin{abstract}
We examine the diagnostic power of rest-frame ultraviolet (UV) nebular emission lines, and compare them to more commonly used rest-frame optical emission lines, using the test case of a single star-forming knot of the bright lensed galaxy RCSGA 032727-132609 at redshift $z \sim 1.7$. This galaxy has complete coverage of all the major rest-frame UV and optical emission lines from Magellan/MagE and Keck/NIRSPEC. Using the full suite of diagnostic lines, we infer the physical properties: nebular electron temperature ({\Te}), electron density ({\Ne}), oxygen abundance ($\log{(\mathrm{O/H})}$), ionisation parameter ({\logq}) and interstellar medium (ISM) pressure ({\lpok}). We examine the effectiveness of the different UV, optical and joint UV-optical spectra in constraining the physical conditions. Using UV lines alone we can reliably estimate {\logq}, but the same is difficult for $\log{(\mathrm{O/H})}$. UV lines yield a higher ($\sim$1.5 dex) {\lpok} than the optical lines, as the former probes a further inner nebular region than the latter. For this comparison, we extend the existing Bayesian inference code IZI, adding to it the capability to infer ISM pressure simultaneously with metallicity and ionisation parameter. This work anticipates future rest-frame UV spectral datasets from the James Webb Space Telescope ({\JWST}) at high redshift and from the Extremely Large Telescope (ELT) at moderate redshift.
\end{abstract}




\section{Introduction}
Understanding the evolution of inter-stellar medium (ISM) properties requires measuring galaxies at both lower redshifts ($z \leqslant 1$) where the star formation has been quenched due to feedback from several physical processes \citep[e.g.][]{Cicone:2014aa, Bluck:2014aa, Schwamb:2016aa, Leslie:2016aa} including turbulence \citep[e.g.][]{Federrath:2012aa, Federrath:2017aa, Zhou:2017aa}, and at higher redshifts ($z \sim 2$) where the star-formation rate (SFR) is at its peak \citep[e.g.][]{Uzgil:2016aa, Kewley:2016aa, Kulas:2010aa}. 

The star formation history \citep[e.g.][]{Schulte-Ladbeck:2003aa}, ISM conditions \citep[e.g.][]{Tremonti:2004aa, Lamareille:2004aa, Gallazzi:2005aa, Contursi:2017aa} and kinematics \citep[e.g.][]{Cicone:2016aa} of low-redshift galaxies have been studied extensively thanks to large, targeted surveys \citep[e.g.][]{Ho:2016ab, Medling:2018aa, Ellison:2018aa, Belfiore:2017aa}. Not only are low-{\z} galaxies brighter then the distant ones, but in many cases we have spatially resolved properties. In contrast, the physical conditions of high-{\z} ({\z}>1) galaxies are still poorly understood, because these galaxies are fainter and harder to observe. Spatially resolved studies of high-{\z} galaxies are difficult with current telescopes and the resolution is poorer than for the local samples \citep[e.g.][]{Forster-Schreiber:2009aa, Wisnioski:2015aa}. The advent of the next generation of large ground-based (Giant Magellan Telescope, GMT \citep{Johns:2004aa}; Extremely Large Telescope, ELT \citep{Sanders:2013aa}; Thirty Metre Tescope, TMT \citep{Hook:2009aa}) and space-based telescopes (\textit{James Webb Space Telescope}, {\JWST} \citep{Gardner:2006aa}) will facilitate spectroscopic observations of galaxies with unprecedented spatial resolution ($\sim 0.1''$), out to very high redshifts ($z \sim 6 - 10$) and detection of the first generation of galaxies ($z \sim 15$). 

{\JWST} will be able to observe the rest-frame optical emission lines in a multiplexed way with its NIRSpec instrument for galaxies at $z \leqslant 6.4$. Moreover, for $z \geqslant 3$ {\JWST} can additionally capture diagnostic  rest-frame UV lines. At moderate redshifts ($z \geqslant 3$) {\JWST} /NIRSpec will be able to capture both rest-frame optical and rest-frame UV diagnostics. At higher redshifts ($z \geqslant 6.4$), the rest-frame optical lines redshift out of the NIRSpec bandpass and are no longer accessible for multiplexed spectroscopy, though they can be captured singly by the MIRI (Mid-Infrared Instrument Instrument). For the highest redshift (7 < {\z} < 10) galaxies, all {\JWST} may spectroscopically detect are the rest-frame UV lines. This motivates the development of a suite of UV emission line diagnostics, as well as a full UV$+$optical suite of diagnostics to be used at moderate redshifts.

Even prior to the onset of the next generation telescopes, we can preview the measurements that will be possible with such facilities by studying galaxies that are strongly magnified through lensing. Gravitational lensing has been used by astronomers to quantify the ISM conditions in galaxies at $z \geqslant$ 1.5 \citep[e.g.][]{Yuan:2009aa, Bian:2010, Jones:2013aa, Yuan:2013ab, Canameras:2015aa, Leethochawalit:2016aa, James:2018aa, Rigby:2009aa, Rigby:2011aa, Bayliss:2014aa, Jones:2015ab, Wang:2017aa}. In addition to magnifying background galaxies, gravitational lensing stretches the images of the galaxies into extended arcs, making it possible to spatially resolve individual star-forming knots \citep[Figure 1 of][]{Bayliss:2014aa, Canameras:2015aa, Swinbank:2015aa, Sharda:2018aa}.

Diagnostics using rest-frame optical nebular lines have been extensively applied to {\htrs} and galaxies \citep[e.g.][henceforth \citetalias{Kewley:2002fk}, \citetalias{Kobulnicky:2004aa}, \citetalias{Pettini:2004qe}]{Kewley:2002fk, Pagel:1979aa, McGaugh:1991aa, Zaritsky:1994aa, Kewley:2001aa, Kobulnicky:2004aa, Pettini:2004qe, Sanders:2016aa}. UV diagnostics appear promising \citep[e.g.][henceforth \citetalias{Izotov:2006ab}, \citetalias{Garnett:1995aa} and \citetalias{Garnett:1995ab} respectively]{Jaskot:2016aa, Feltre:2016aa, Stark:2014aa, Byler:2018aa, Izotov:2006aa, Garnett:1995aa, Garnett:1995ab} but have not yet been extensively tested. Moreover, the atomic data and stellar atmospheric libraries used for the calibrations of the earlier works have now been updated. These updates affects the suite of UV and optical diagnostics developed to date.

Recent studies have presented improvements to the rest-frame UV and optical diagnostics. \citet[][hereafter \citetalias{Kewley:2019aa}]{Kewley:2019aa} present a set of rest-frame optical and UV line diagnostics for electron density and ISM pressure using the updated version of the MAPPINGS photoionzation models \citep{Sutherland:2013aa} (explained in section \ref{optdiag}). Using these models, \citet[][hereafter \citetalias{Kewley:2019ab}]{Kewley:2019ab} further present a set of optical and UV diagnostics for ionisation parameter and metallicity. 
Nicholls {\etal} (in prep; N18 hereafter) propose new methods for determining the nebular electron temperature using rest-frame UV oxygen emission lines combined with improved theoretical models. \citet[hereafter \citetalias{Dopita:2016aa}]{Dopita:2016aa} present a robust technique to isolate the dependence of oxygen abundance on a set of nebular optical lines (independent of ionisation parameter and ISM pressure) applicable for high-{\z} ($z \sim 2$) galaxies. These new  diagnostics, together with the existing diagnostics, constitute a comprehensive suite that can be employed to determine ISM properties of high-{\z} galaxies. However, the new diagnostics have not yet been tested using observations of galaxies.

In addition to strong emission line diagnostics (SEL), Bayesian techniques are also becoming increasingly important in inferring ionised gas properties due to their ability to probe asymmetry and non-trivial topography in the probability distributions of the properties. Recent Bayesian estimation tools like IZI \citep{Blanc:2015aa}, BOND \citep{Vale:2016aa}, HII-CHI-mistry \citep{Perez-Montero:2015aa} and NebulaBayes \citep{Thomas:2018aa} have proven useful in inferring nebular gas properties. However, in light of the recent development of SEL diagnostics, particularly the rest-frame UV diagnostics, it is necessary to test the agreement between the Bayesian and SEL techniques. Hence, in this work we compare the SEL diagnostics with a new, extended version of IZI, as described later in Section~\ref{izidiag}.

The purpose of this paper is to test these new diagnostics by applying them to a star-forming knot of a {\z}$\sim$2 galaxy with full coverage of these diagnostic lines. Herein we exploit the advantage that spatially resolved spectroscopy has over other high-z samples. Choosing a single star-forming knot of the lensed galaxy {\galaxy} for the testing allows us to probe a small ($\sim$100 pc) spatial region, which can be expected to have fairly homogeneous ISM properties with high fraction of {\htr} relative to diffuse gas.

This paper is organized as follows. In Section~\ref{sampsel} we justify the selection of a single star-forming knot of {\galaxy} as the test case and describe the selection of the local galaxies used for comparison. Section~\ref{obs} describes the observation and data reduction for both the Keck/NIRSPEC and Magellan/MagE data. We explain the tools used for our analysis, including line fitting algorithms, various diagnostics used and the results obtained from them in Section~\ref{analysis}. The comparison among various results are discussed in Section~\ref{disc} followed by a summary of our work in Section~\ref{sum}.

We use a solar oxygen abundance {\logOH}\ = 8.72 \citep{Asplund:2009dp} throughout the paper. We assume a standard flat $\Lambda$ cold dark matter cosmology with $H_0$ = 70 km s$^{-1}$ Mpc$^{-1}$ and matter density \mbox{$\Omega_M$ = 0.27}. For the emission lines, we adopt the sign convention of negative equivalent width and use the wavelengths from the NIST database, the \citet{Leitherer:2011aa} atlas, and MAPPINGS v5.

\section{Sample selection}\label{sampsel}
The primary limitation in the development of rest-frame UV spectral diagnostics has been the lack of high-quality spectra. Project MEGaSaURA \citep{Rigby:2018aa} has obtained high signal to noise, moderate spectral resolution (R $\sim$ 3000) spectra for 15 bright gravitationally lensed galaxies. From that sample, we select the spectrum of knot-E of {\galaxy} (henceforth referred to as {\knotE}) for our pilot study, for the following reasons.
\begin{itemize}
\item RCSGA 032727-132609 (henceforth {\galaxy}) is a very bright lensed galaxy with an r-band magnitude of 19.1 \citep{Wuyts:2010aa, Sharon:2012aa}.
\item The fact that the lensed galaxy appears as a very extended (38") arc makes it possible to resolve and target individual star-forming knots. Spectra of four knots have been published; we select knot-E \citep[see][Figure 1]{Sharon:2012aa} for our analysis in this paper because it has the highest signal-to-noise ratio (SNR).
\item The spectrum of {\galaxy} has a SNR per resolution element $\sim$ 20 at $\lambda_{\mathrm{obs}}$ = 5000 \AA, sufficient to clearly detect the rest-frame UV emission lines.
\item Rest-frame optical spectra for this object have been obtained from Keck/NIRSPEC \citep[][ henceforth \citetalias{Rigby:2011aa}]{Rigby:2011aa}, Keck/OSIRIS \citep{Wuyts:2014ab} and {\HST}/WFC3 G141 grism \citep[hereafter \citetalias{Whitaker:2014aa}]{Whitaker:2014aa}. \citet{Wuyts:2014ab} also report optical fluxes for knots B, U from Magellan/FIRE observations and that for knot U from re-extracted Keck/NIRSPEC observations. \citetalias{Whitaker:2014aa} report grism fluxes from several other star-forming knots too. However, the largest number of optical$+$UV emission lines are detected in Knot E, which allow us to compare the extensively used optical diagnostics to the new UV diagnostics.
\end{itemize}

{\galaxy} was first discovered \citep{Wuyts:2010aa} in a dedicated search for highly magnified giant arcs \citep{Bayliss:2012aa} in the Red Sequence Cluster Survey 2 \citep[RCS2;][]{Gilbank:2011aa}. \citet{Sharon:2012aa} performed a source plane reconstruction of {\galaxy}, based on {\HST}/WFC3 imaging data, down to a scale of $\sim 100$ {\pc}. Due to its apparent brightness, this galaxy has been subjected to extensive subsequent spectroscopic analyses. \citetalias{Rigby:2011aa} constrained the nebular physical properties of {\galaxy} using the spatially summed Keck/NIRSPEC (rest-frame optical) spectra. \citet{Wuyts:2014aa} studied the stellar populations and concluded that {\galaxy} is a starburst galaxy \citep[SFR $30-50$ {\Msunpyr};][]{Wuyts:2012aa} with a young \citep[$5 - 100$ {\Myr};][]{Wuyts:2014aa} stellar population that has just started to build up its stellar mass \citep[$10^{7} -  5\times10^{8}$ {\Msun};][]{Wuyts:2014aa}. \citetalias{Whitaker:2014aa} analyzed spatial variations in {\HST}/WFC3 grism spectra of {\galaxy} and reported no appreciable knot-to-knot variation in reddening, and an enhanced star formation rate ($\sim 2$ dex above the star-formation main sequence) due to an ongoing interaction. They also found spectroscopic evidence of the presence of O stars in most knots (except knots E and F which have lower \ion{He}{i}/H$\beta$ ratios), which is consistent with the young stellar population scenario. \citet{Bordoloi:2016aa} examined the galactic outflows for this galaxy as traced by the \ion{Mg}{ii} and \ion{Fe}{ii} emission and its spatial variation, finding large outflow velocities ($\sim$ 170-250 {\kmps}) and mass outflow rates ($\gtrsim$ 30-50 {\Msunpyr}). However, \citet{Rigby:2014} report a lack of correlation between the \ion{Mg}{ii} and Ly-$\alpha$ emissions, which implies that the source of \ion{Mg}{ii} emission is not nebular, but may instead be resonantly scattered continuum. Overall, the picture that has emerged is that {\galaxy} hosts a young stellar population that is driving a large-scale outflow.

In this work, we analyze the spectra of one particular knot of star-formation (knot E) within {\galaxy}. A unique feature of this analysis is that the spectra cover a $\sim$100 pc \citep{Sharon:2012aa} star-forming region (classified using the Baldwin, Phillips \& Terlevich (BPT) diagram and \citetalias{Kewley:2002fk} line). Our work has the distinct advantage in that the physical conditions can be expected to be fairly homogeneous within the small ($\sim$100 pc) spatial region that the spectra probe, rather than being averaged across several kiloparsecs, which ensures fair comparison to photoionisation models. Moreover, the relative contribution from {\htrs} with respect to diffuse ionised gas is expected to be very high because we integrate over a small star-forming knot, rather than the whole galaxy. 

We compare the rest-frame UV and rest-frame optical emission line fluxes to newly-developed diagnostics (\citetalias{Kewley:2019aa},b, N18), to constrain the ionisation parameter ({\logq}), electron density ({\Ne}), ISM pressure ({\lpok}), electron temperature ({\Te}) and oxygen abundance ({\logOH}). We also consider how the results would differ if we had only the UV dataset, or only the optical dataset, using a Bayesian approach (detailed in Section~\ref{izidiag}). Moreover, given that the [\ion{O}{ii}]$\lambda\lambda$3727,9 doublet will be within the wavelength coverage of {\JWST} up to $z\sim$ 12, we also investigate the effect of including the [\ion{O}{ii}]$\lambda\lambda$3727,9 lines with the set of rest-frame UV emission lines.

\section{Observations}\label{obs}
We use the rest-frame optical spectra from the NIRSPEC instrument on Keck and rest-frame ultraviolet spectra from the MagE instrument on Magellan.

\subsection{Rest-frame optical spectroscopy from NIRSPEC on Keck}\label{opt_data}
Near-infrared spectra of {\knotE}, 
covering the rest-frame optical, were obtained on UT 2010 Feb.\ 4 with the
NIRSPEC spectrograph \citep{McLean:1998aa} on the Keck II telescope. 
The spectra were originally published in \citetalias{Rigby:2011aa}, along
with a detailed description of the observation and data reduction procedures.
Since only a basic lensing model was available at that time, the spectrum
in the long-slit was summed across the spatial direction.
Subsequently, high-resolution images with the \textit{Hubble Space Telescope}
({\HST}) revealed that the NIRSPEC observations had captured multiple 
physically distinct knots of star formation \citep[see][Figure 6]{Sharon:2012aa}, 
the brightest of which they labeled Knot E and Knot U.  
Guided by the {\HST}--enabled lensing model, 
\citet{Wuyts:2014ab} re-extracted the spectra, producing spectra for these two physical regions. 
Subsequently, \citetalias{Whitaker:2014aa} improved the measurement of the reddening for {\galaxy} .
Therefore, we take the NIRSPEC line fluxes reported for Knot E by  \citet{Wuyts:2014ab}, and apply the  
reddening of E(B-V)$_{\rm gas} = 0.40 \pm 0.07$ measured by \citetalias{Whitaker:2014aa}, 
to compute updated cross-filter scaling factors, as we describe in more detail in Section~\ref{nirspecfitting}.

\subsection{Rest-frame UV spectroscopy from MagE on Magellan}\label{mage_data}
Optical spectra of {\knotE} \citep[see][Figure 1]{Sharon:2012aa}, covering the rest-frame ultraviolet, were 
obtained on multiple nights in the range UT 2008-07-31 to UT 2010-12-10 
with the MagE instrument \citep{Marshall:2008} on the Magellan Clay telescope.  
The observations and data reduction are described in \citet{Rigby:2018aa}. The data were reduced using the MagE pipeline, which 
is part of the Carnegie Python Distribution\footnote{http://code.obs.carnegiescience.edu}.
The full spectra from each observation were obtained by combining the weighted average of different spectral orders. Observations from different nights were then combined via a weighted average to obtain the resulting rest-frame UV spectra of {\knotE} used in this paper.
The spectra were flux-calibrated by comparing to spectrophotometric standard stars, 
as described in \citet{Rigby:2018aa}. The spectra were corrected for Milky Way reddening.
The effective spectral resolution of the final combined spectrum, 
measured from the widths of night sky lines, is $R= 3650 \pm 120$ (median and absolute median deviation).

\section{Flux measurements}\label{linefit}

\begin{table}
	\centering
	\input{Tables/lineflux_restopt_detected}
	\label{tab:optflux}
\end{table}

\subsection{Keck/NIRSPEC spectral line fits}\label{nirspecfitting}
For each NIRSPEC filter setting (N1, N3, and N6, roughly corresponding to J-band, H-band, and Ks-band), all emission lines were fit simultaneously with Gaussian profiles by \citet[][see their Sections 2.3 and 2.4]{Wuyts:2014ab}.  Relative flux offsets are expected in the NIRSPEC spectra across the three grating settings, due to slit losses associated with variable seeing.  As a result, the fluxes of the NIRSPEC-1 and NIRSPEC-6 spectra had to be adjusted  with respect to the flux calibration in NIRSPEC-3. Section 3.2 of \citetalias{Rigby:2011aa} describes how the Balmer lines were used to perform this process, using the at-the-time best measurement of reddening, $E(B-V)_{gas}=0.23 \pm 0.23$. This procedure produced offset factors of 1.15 for N1 and 0.61 for N6, both with respect to N3.


\begin{table*}
	\centering
	\input{Tables/lineflux_restUV_detected}
	\label{tab:uvflux}
\end{table*}

\begin{table}
	\centering
	\input{Tables/lineflux_restUV_detected_interv}
	\label{tab:interv}
\end{table}

\begin{figure*}
	\centering
	\includegraphics[scale=0.65, trim=0.1cm 0.1cm 0.0cm 0.0cm,clip=true, angle=0]{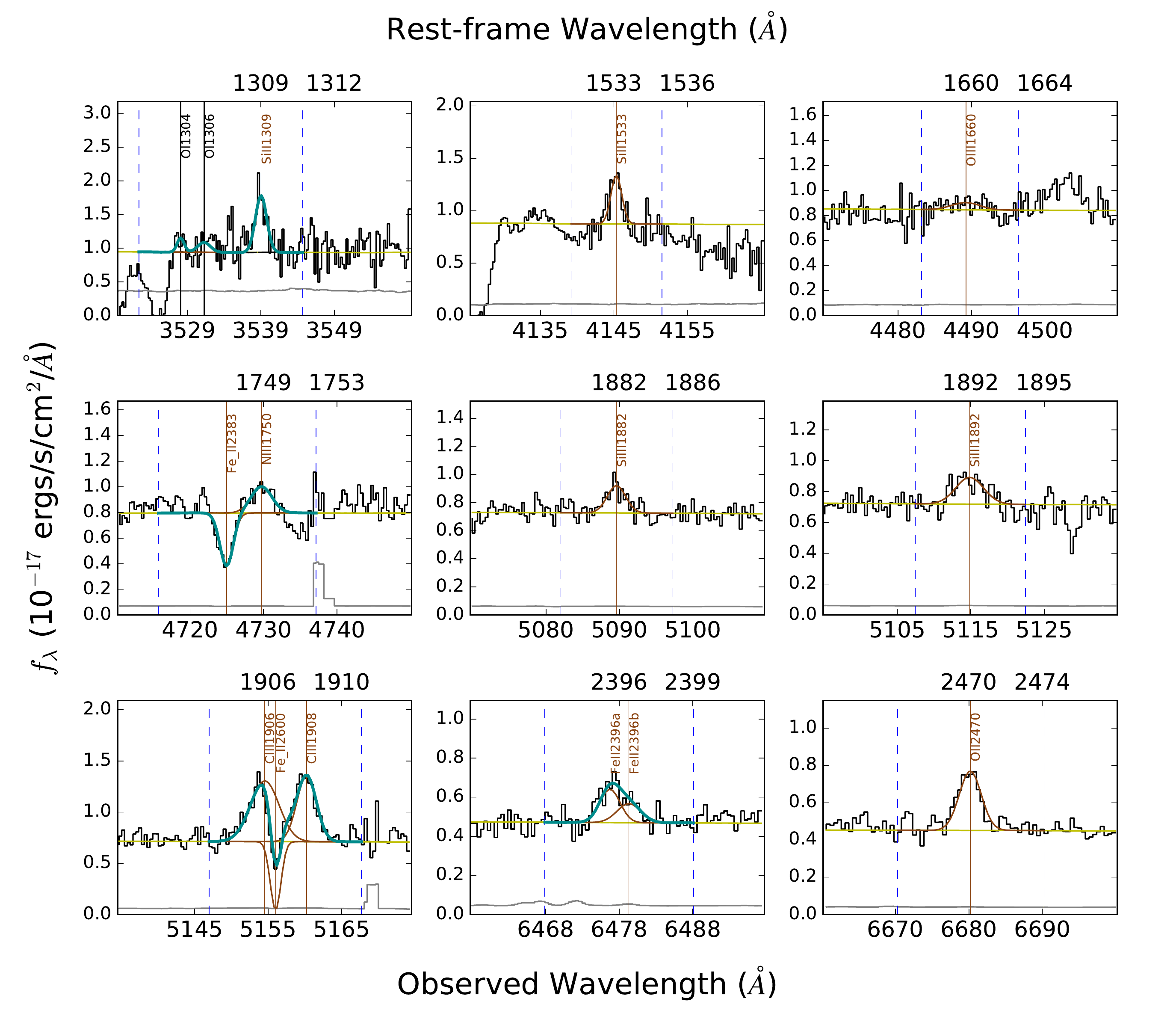}
	\caption{ 
		Gaussian fits to the MagE spectra for individual lines. The black line is the observed spectrum, whereas grey and yellow denote the 1 $\sigma$ uncertainty spectrum and the continuum respectively. Brown solid lines represent the Gaussian fit of individual lines. Thick dark cyan lines are the sum of the Gaussian fits, wherever there are multiple lines. The fitting routine works on a portion of the spectrum 
		as shown bounded by blue dashed lines. The fitted value of the central wavelength is denoted by a vertical brown solid line, if the line is detected (i.e. > 3 $\sigma$) and by a black line if not (i.e. < 3 $\sigma$). A single  Gaussian is fit to a single line or multiple Gaussians are simultaneously fit to a group of lines, depending on the separation between neighboring lines. This is visible, for example, in the \ion{C}{iii}] 1906-1908 doublet fitting.
	}
	\label{fig: indiv-fit}
\end{figure*}

\begin{figure*}
	\centering
	\includegraphics[scale=0.7, trim=0.1cm 0.1cm 0.0cm 0.0cm,clip=true, angle=0]{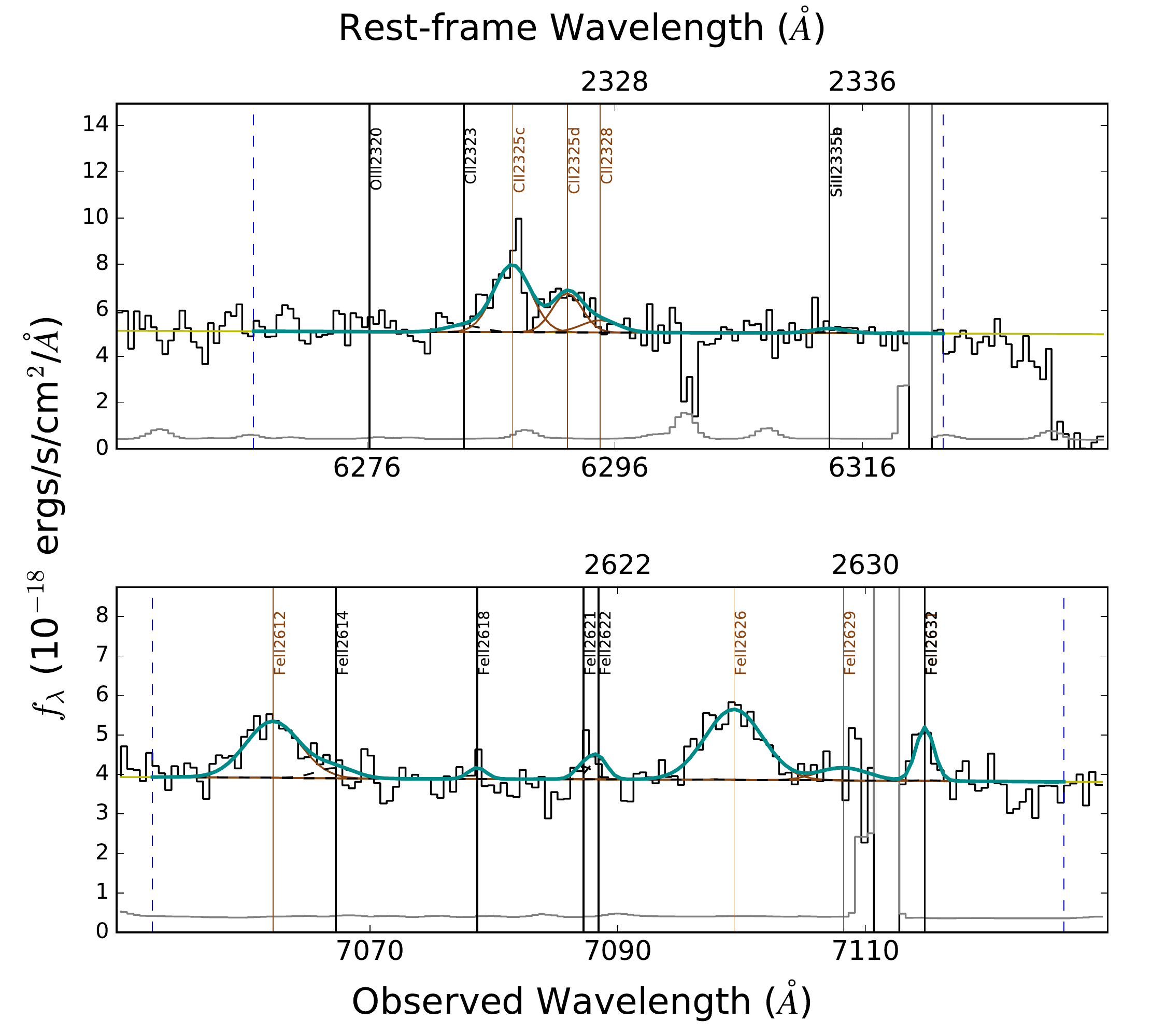}
	\caption{ 
		Our fits of the \ion{C}{ii}] 2323 (top) and FeII 2600 (bottom) complex. The colour coding is as in Figure \ref{fig: indiv-fit}. There are >3 $\sigma$ detections for only some of the lines (brown vertical lines) in this complex. For the rest (black vertical lines) we quote upper limits in Table~\ref{tab:uvflux}. Blue solid lines indicate the initial guess/es of the line center/s provided to the fitting code. For each line/group, only the portion bounded by blue dashed lines is fit.
	}
	\label{fig: Fe-comp}
\end{figure*}

\subsection{Magellan/MagE spectral line fits}\label{mage fitting}
We fit the continuum with an automatic routine that masks the positions of all expected spectral features, including from known intervening absorption systems, and then apply boxcar smoothing. The result is almost identical to the hand-fit spline continuum described in \citet{Rigby:2018aa}. We subsequently fit the emission lines in the combined, continuum normalized spectrum using a Python-based, automated line fitting code, explained below.

We simultaneously fit all neighboring spectral lines with one Gaussian profile per line employing the python tool \texttt{scipy.optimize.curve\_fit} that implements the non-linear least-squares method. A neighbor is defined as follows. Each line centroid is assumed to have a window of $\pm 5$ spectral resolution elements, both blueward and redward. If the windows of any two adjacent lines overlap they are neighbors. A neighbor of a neighbor is considered in the same group of lines, which are fit simultaneously. As an example, if there are 4 lines with separation between each adjacent pair of line centroids less than $\pm 10$ resolution elements (5 resolution elements from the window of each line), the lines are considered as a single group. This group is fit with a quadruple Gaussian function with $4\times 3 = 12$ parameters, where the parameters are height, width and centroid of the Gaussians for each line in the group. We set the continuum value to unity\footnote{The code is capable of fitting the continuum value, for each group, as an additional parameter if desired.} because, in this case, we use a continuum normalized spectrum. Initially, the nebular redshift is measured by fitting the strongest emission lines (for {\knotE} it is the $[$\ion{C}{iii}$]\lambda\lambda$1906,8 doublet) with sufficiently large allowance for the fitted redshift. Subsequently, an initial guess for each line centroid $\lambda_{in}$ is provided to \texttt{curve\_fit()} by redshifting the rest-frame vacuum wavelengths by the nebular redshift. The centroids are forced to be fit within a window of $\lambda_{in} \pm 3\sigma$ wavelength interval corresponding to the uncertainties in the nebular redshift. The line width is fit within upper and lower bounds of 300 {\kmps} and one spectral resolution element respectively. The amplitude of the Gaussians were allowed to vary freely. We measure rest-frame equivalent widths ($W_r$) from both the fitted Gaussian parameters and by direct summation. In this paper, we quote the $W_r$ values derived using the former method.

In order to determine the significance of the $W_r$ measurements ($W_{\mathrm{r,signi}}$), we define a quantity $W_{\mathrm{r,Schneider}}$ as the $W_r$ derived by interpolating a rolling average of the error weighted $W_r$ at every point throughout the spectrum, following \citet[Section 6.2]{Schneider:1993}. The $W_{\mathrm{r,signi}}$ is then defined as the ratio of the measured $W_r$ of a line to the $W_{\mathrm{r,Schneider}}$ computed at the line center. The rolling average technique gives us a quantifiable estimate of the spectral noise. We consider all lines that meet the criteria (a) $W_{\mathrm{r,signi}} >$3, and (b) SNR > 1, to be detected\footnote{There are cases where only criteria (a) is satisfied, especially when the code fits a broad emission in the noisy part of the spectrum.}. Features not satisfying these criteria are considered to be non-detections, for which we quote $3\times W_{\mathrm{r,Schneider}}$ as the $3\sigma$ upper limits of $W_r$s. We repeat the same operation on measured line flux values ($f$) to derive $f_{\mathrm{Schneider}}$, $f_{\mathrm{signi}}$ and $f_{\mathrm{uplim}}$. In the absence of flux ($f$), we translate $f_{\mathrm{uplim}}$ to lower/upper limits on line ratios and consequently to limits on the ISM properties.

We present the emission line fluxes and upper limits of nebular lines in the MagE spectrum in Table~\ref{tab:uvflux} and show the Gaussian fits to the MagE data in Figures~\ref{fig: indiv-fit} and \ref{fig: Fe-comp}. Some of the emission lines of interest in our spectrum are affected by intervening absorption lines. We fit the intervening lines simultaneously with the emission lines, to properly account for the missing (absorbed) emission line fluxes (e.g. see bottom-left panel of Figure~\ref{fig: indiv-fit}). The intervening absorption features, along with their strengths and redshifts are presented in Table~\ref{tab:interv}. These intervening lines have been studied by \citet{Lopez:2018aa}.

\begin{table*}
	\centering
	\caption{Complete list of all UV and optical diagnostics and the line ratios involved.}
	\label{tab:masterlist}
	\begin{tabular}{llll}
		\hline
		Name of diagnostic & Inferred quantity &  Line ratio used & Reference\\
		\hline
		\hline
		Te\_O3a\_I06 & {\Te} & [\ion{O}{iii}]$\lambda \lambda$4959,5007/[\ion{O}{iii}]$\lambda$4363 & \cite{Izotov:2006aa} \\
		Te\_O3a\_N18 & " & [\ion{O}{iii}]$\lambda \lambda$5007/[\ion{O}{iii}]$\lambda$4363 & Nicholls {\etal} \textit{(in prep)} \\
		Te\_O3b & " & [\ion{O}{iii}]$\lambda $5007/\ion{O}{iii}]$\lambda\lambda$ 1660,6 & " \\
		\hline
		Te\_O3a\_O32 & {\logOH} & Te\_O3a\_I06 ratios (as above) and  [\ion{O}{iii}]$\lambda\lambda$4959,5007/H$\beta$ \& & \citet{Izotov:2006aa}  \\
		(Direct method)& & [\ion{O}{ii}]$\lambda\lambda$3727,9/H$\beta$ &  \\
		Te\_O3b\_O32 & " & Te\_O3b ratios (as above) and  [\ion{O}{iii}]$\lambda\lambda$4959,5007/H$\beta$ \& & " \\
		(Direct method)& & [\ion{O}{ii}]$\lambda\lambda$3727,9/H$\beta$ &  \\
		KD02N2O2 & " & [\ion{N}{ii}]$\lambda$6584/[\ion{O}{ii}]$\lambda\lambda$3727,3729 & \cite{Kewley:2002fk} \\
		PP03N2 & " & ([\ion{O}{iii}]$\lambda$5007/H$\beta$)/([\ion{N}{ii}]$\lambda$6584/H$\alpha$) & \cite{Pettini:2004qe} \\
		PPN2 & " & [\ion{N}{ii}]$\lambda$6584/H$\alpha$ & " \\
		S07 & " & [\ion{Ne}{iii}]$\lambda$ 3869/[\ion{O}{ii}]$\lambda\lambda$3727,3729 & \cite{Shi:2007aa} \\
		BKD18 & " & [\ion{Ne}{iii}]$\lambda$ 3869/[\ion{O}{ii}]$\lambda\lambda$3727,3729 & \cite{Bian:2018aa} \\		D16 & " & [\ion{N}{ii}]$\lambda$6584/[\ion{S}{ii}]$\lambda\lambda$6717,31 \&& \cite{Dopita:2016aa} \\
		& & [\ion{N}{ii}]/H$\alpha$ &  \\
		KK04 & 
		{\logOH}, {\logq} & ([\ion{O}{ii}]$\lambda$3727+[\ion{O}{iii}]$\lambda\lambda$4959,5007)/H$\beta$ \& & \cite{Kobulnicky:2004aa} \\
		\hline
		LR14Ne3O2 & {\logq} & [\ion{Ne}{iii}]$\lambda$ 3869/[\ion{O}{ii}]$\lambda\lambda$3727,3729 & \cite{Levesque:2014aa} \\
		LR14O3O2 & " & [\ion{O}{iii}]$\lambda\lambda$4959,5007/[\ion{O}{ii}]$\lambda\lambda$3727,9 & " \\
		KD02O32 & " & [\ion{O}{iii}]$\lambda$5007/[\ion{O}{ii}]$\lambda\lambda$3727,9 & \cite{Kewley:2002fk} \\
		S18O32 & " & [\ion{O}{iii}]]$\lambda\lambda$4959,5007/[\ion{O}{ii}]$\lambda\lambda$3727,9 & \cite{Strom:2018aa} \\
		S18Ne3O2 & " & [\ion{Ne}{iii}]$\lambda$ 3869/[\ion{O}{ii}]$\lambda\lambda$3727,9 & " \\
		S18O3Hb & " & [\ion{O}{iii}]$\lambda$5007/H$\beta$ & " \\
		\citetalias{Kewley:2019ab}Ne3O2 & " & [\ion{Ne}{iii}]$\lambda$ 3869/[\ion{O}{ii}]$\lambda\lambda$3727,9 & \citet{Kewley:2019ab} \\
		\citetalias{Kewley:2019ab}O3Hb & " & [\ion{O}{iii}]$\lambda$5007/H$\beta$ & " \\
		\citetalias{Kewley:2019ab}O32a & " & [\ion{O}{iii}]$\lambda$5007/[\ion{O}{ii}]$\lambda$3727 & " \\
		\citetalias{Kewley:2019ab}O32b & " & [\ion{O}{iii}]$\lambda\lambda$1660,6/[\ion{O}{ii}]$\lambda$2470a,b & " \\
		\citetalias{Kewley:2019ab}C32a & " & [\ion{C}{iii}]$\lambda\lambda$1906,8/[\ion{C}{ii}]$\lambda$1335 & " \\
		\citetalias{Kewley:2019ab}C32b & " & [\ion{C}{iii}]$\lambda\lambda$1906,8,8/[\ion{C}{ii}]$\lambda$2323-8 & "  \\
		\hline
		Ost\_O2 & {\Ne} & [\ion{O}{ii}]$\lambda$3729/[\ion{O}{ii}]$\lambda$3727 & \cite{Osterbrock:1989} \\
		S16O2 & " & [\ion{O}{ii}]$\lambda$3729/[\ion{O}{ii}]$\lambda$3727 & \cite{Sanders:2016aa} \\
		S16S2 & " & [\ion{S}{ii}]$\lambda$6731/[\ion{S}{ii}]$\lambda$6717 & "\\
		\citetalias{Kewley:2019aa}O2 & {\Ne}, {\lpok} & [\ion{O}{ii}]$\lambda$3729/[\ion{O}{ii}]$\lambda$3727 & \citet{Kewley:2019aa} \\
		\citetalias{Kewley:2019aa}S2 & " & [\ion{S}{ii}]$\lambda$6731/[\ion{S}{ii}]$\lambda$6717 & " \\
		\citetalias{Kewley:2019aa}C3 & " & [\ion{C}{iii}]$\lambda$1908/[\ion{C}{iii}]$\lambda$1906 & " \\
		\citetalias{Kewley:2019aa}Si3 & " & [\ion{Si}{iii}]$\lambda$1892/[\ion{C}{iii}]$\lambda$1882 & " \\
		\hline
	\end{tabular}
\end{table*}

\begin{figure*}
\centering
\caption{
	Comparison of individual diagnostics for the physical parameters -- electron temperature, ionisation parameter, ISM pressure, electron density and gas phase oxygen abundance. For each panel the curves indicate the normalized, scaled (to unity) probability density function (PDF) of the measured physical quantity generated by performing every diagnostic 10$^4$ times. For each realization we randomly draw the line fluxes from a Gaussian distribution with mean and width equal to the measured flux and corresponding uncertainty, respectively (see Section~\ref{diag_uncert}). Different colours denote different diagnostics. The median of each PDF is shown with a filled circle of the corresponding colour. PDFs with dashed lines denote rest-frame UV diagnostics whereas solid lines denote optical diagnostics. The line ratios used for each diagnostic can be looked up in Table~\ref{tab:masterlist}. For diagnostics involving lines for which we only have upper limits, we do not plot the PDF. Instead, we show the median by thick dashed vertical lines. Whether these are the upper or lower limits of the physical parameter, are denoted by arrows (right arrow: lower limit, left arrow: upper limit). We demonstrate that the UV and optical diagnostics for {\Te}, {\logq} and $\log{(\rm O/H)}$ broadly agree, with some exceptions where the diagnostics either could not be transformed to the common reference frame (for $\log{(\rm O/H)}$) or used the Ne3O2 index (for {\logq}) (detailed in Section~\ref{analysis}). For {\lpok} and {\Ne}, however, we find that the UV diagnostics probe different (denser, higher pressure) physical nebular regions than their optical counterparts (detailed in Section~\ref{disc_press}).
}
\includegraphics[scale=1., trim=0.0cm 0.0cm 0.0cm 0.0cm,clip=true,angle=0]{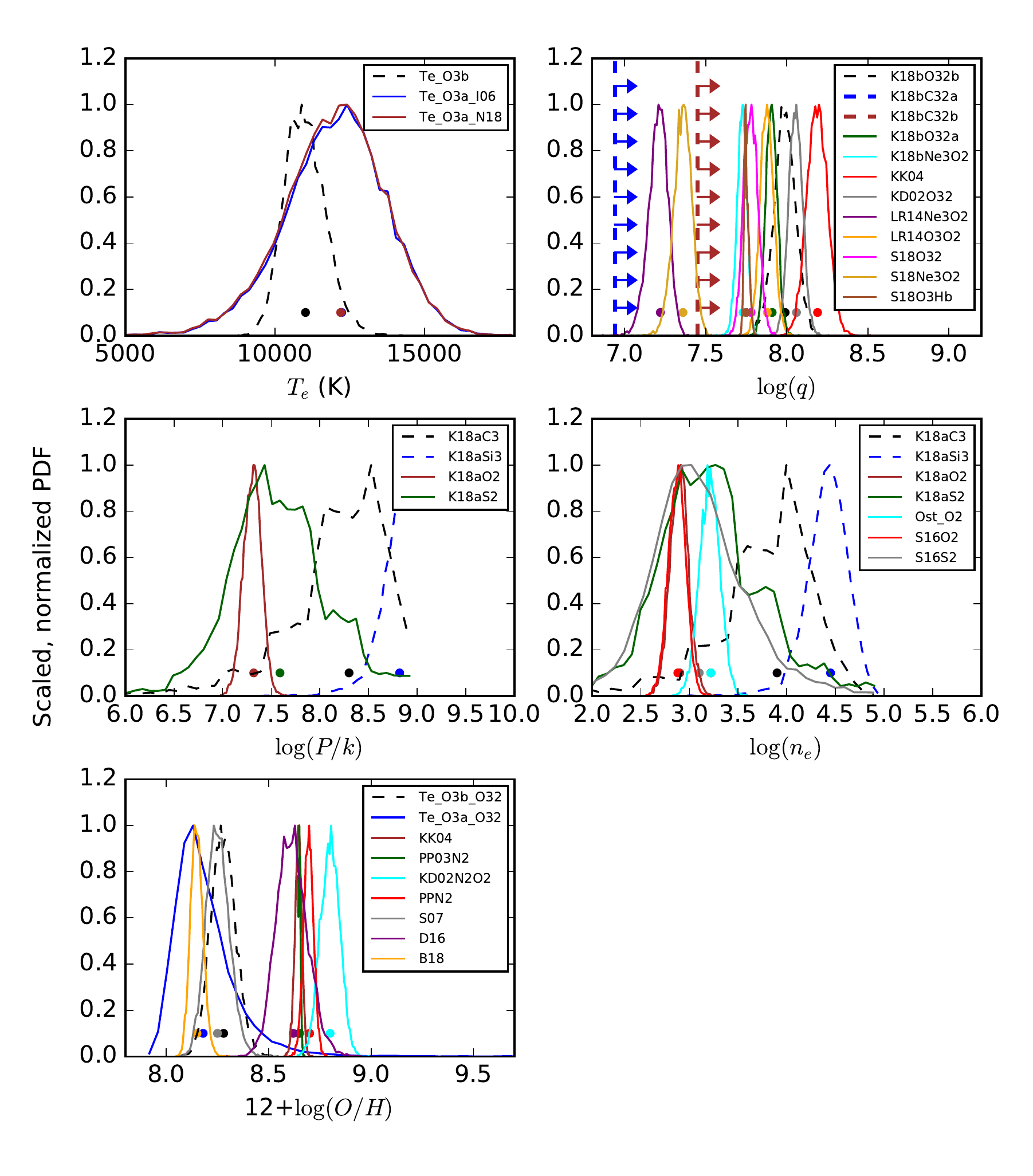}
\label{fig: indiv_diag}
\end{figure*}

\begin{landscape}
\begin{table}
\centering
\caption{Inferred physical parameters. The diagnostics (see Table~\ref{tab:masterlist} for a list) are discussed in Section~\ref{abundiag_opt} \& \ref{abundiag_uv}. For all further analysis, we consider log(O/H)+12 values in column 3 as the final values, if  available. Otherwise we use the values from column 2. We use the {\logOH}$=$8.23 and {\lpok}$=$8.0 branch of the \citetalias{Kewley:2019ab} calibrations, for all the \citetalias{Kewley:2019ab} diagnostics. K18C32a,b methods have zero uncertainty because they are lower limits on {\logq}. The values quoted for the IZIP methods correspond to Figure~\ref{fig: pizi_prior}, where a [\ion{N}{ii}]$\lambda$6584/H$_{\alpha}$ based prior on $Z$ has been used.}
\label{tab:phys_list}
\begin{tabular}{lcc|lc|lc|lc|lc}
\hline
\multicolumn{3}{c}{Oxygen abundance} & \multicolumn{2}{c}{Ionization parameter} & \multicolumn{2}{c}{ISM pressure} & \multicolumn{2}{c}{Elecron density} & \multicolumn{2}{c}{Electron temperature} \\
\hline
Name of  &  {\logOH}\ & {\logOH}\ & Name of & $\log{(q (cm/s))}$ & Name of & $\log{(P/k (K/cm^3))}$ & Name of &   $\log{(n_e (cm^{-3}))}$ & Name of & {\Te} ($\times$10$^4$ K) \\
diagnostic &  & in KK04 frame & diagnostic & & diagnostic &  & diagnostic &  & diagnostic & \\
\hline
\multicolumn{11}{c}{Optical diagnostics} \\
\hline
Direct (Te\_O3a\_O32) & 8.19 $^{+0.16}_{-0.10}$ & - & KK04 & 8.19 $^{+0.05}_{-0.05}$ & \citetalias{Kewley:2019aa}O2 & 7.31 $^{+0.09}_{-0.09}$ & Osterbrock O2 & 3.21 $^{+0.10}_{-0.10}$ & Te\_O3a\_I06 & 1.22 $^{+0.16}_{-0.17}$ \\
KK04 & 8.65 $^{+0.02}_{-0.02}$ & 8.65 $^{+0.02}_{-0.02}$ & KD02O32 & 8.06 $^{+0.04}_{-0.04}$ & \citetalias{Kewley:2019aa}S2 & 7.59 $^{+0.51}_{-0.44}$ & \citetalias{Kewley:2019aa}O2 & 2.90 $^{+0.09}_{-0.09}$ & Te\_O3a\_N18 &1.22 $^{+0.16}_{-0.17}$ \\
PPN2 & 8.32 $^{+0.02}_{-0.02}$ & 8.70 $^{+0.02}_{-0.02}$ & LR14Ne3O2 & 7.21 $^{+0.06}_{-0.06}$ & - & - & \citetalias{Kewley:2019aa}S2 & 3.22 $^{+0.55}_{-0.46}$ & - & - \\
PPO3N2 &8.21 $^{+0.01}_{-0.01}$ & 8.65 $^{+0.01}_{-0.01}$ & LR14O32 & 7.88 $^{+0.04}_{-0.03}$ & - & - & S16O2 & 2.88 $^{+0.09}_{-0.09}$ & - & - \\
KD02N2O2 & 8.71 $^{+0.05}_{-0.06}$ & 8.80 $^{+0.05}_{-0.05}$ & \citetalias{Kewley:2019ab}O32a & 7.78 $^{+0.03}_{-0.03}$ & - & - & S16S2 & 3.1 $^{+0.5}_{-0.41}$ & - & -\\
S07 & 8.26 $^{+0.25}_{-0.26}$ & - & \citetalias{Kewley:2019ab}Ne3O2 & 7.36 $^{+0.05}_{-0.06}$ & - & - & - & - & - & -\\
D16 & 8.62 $^{+0.13}_{-0.14}$ & - & \citetalias{Kewley:2019ab}O3Hb & 7.75 $^{+0.02}_{-0.01}$ & - & - & - &  & - & -\\
BKD18 & 8.15 $^{+0.03}_{-0.03}$ & - & S18O32 & 7.91 $^{+0.04}_{-0.03}$ & - & - & - &  & - & -\\
- & - & - & S18Ne3O2 & 7.73 $^{+0.03}_{-0.03}$ & - & - & - &  & - & -\\
- & - & - & S18O3Hb & 7.85 $^{+0.01}_{-0.01}$ & - & - & - &  & - & -\\
\hline
\multicolumn{11}{c}{UV diagnostic} \\
\hline
Direct(Te\_O3b\_O32) & 8.29 $^{+0.07}_{-0.06}$ & - & K18C32a & $\geq$ 6.94 & \citetalias{Kewley:2019aa}C3 & 8.31 $^{+0.37}_{-0.56}$ & \citetalias{Kewley:2019aa}C3 & 3.91 $^{+0.36}_{-0.56}$ & Te\_O3b & 1.10 $^{+0.07}_{-0.06}$ \\
- & - & - & K18C32b & $\geq$ 7.45 & \citetalias{Kewley:2019aa}Si3 & 8.82 $^{+0.12}_{-0.19}$ & \citetalias{Kewley:2019aa}Si3 & 4.45 $^{+0.19}_{-0.21}$ & - & - \\
- & - & - & K18O32b & 7.99 $^{+0.07}_{-0.07}$ & - & - & - & - & - & - \\
\hline
\multicolumn{11}{c}{IZIP diagnostic} \\
\hline
All lines &  8.56 $^{+0.07}_{-0.03}$ & - & - & 8.21 $^{+0.12}_{-0.12}$ & - & 6.55 $^{+0.31}_{-0.31}$ & - & - & - & - \\
Only optical lines & 8.53 $^{+0.03}_{-0.03}$ & - & - & 8.01 $^{+0.12}_{-0.12}$ & - & 7.06 $^{+0.31}_{-0.51}$ & - & - & - & - \\
Only UV lines & 8.04 $^{+0.13}_{-0.26}$ & - & - & 8.01 $^{+0.29}_{-0.29}$ & - & unconstrained & - & - & - & -  \\
UV+[\ion{O}{ii}]3727,9 & 8.11 $^{+0.13}_{-0.26}$ & - & - & unconstrained & - & 7.88 $^{+0.31}_{-0.20}$ & - & - & - & - \\
\hline
\hline
\end{tabular}
\end{table}
\end{landscape}

\section{Strong Emission Line Diagnostics}\label{analysis}
\citetalias{Whitaker:2014aa} measured a luminosity-weighted average reddening value of E(B-V) = $0.4 \pm 0.07$. We use the \citetalias{Whitaker:2014aa} E(B-V) value and \citet{Cardelli:1989} reddening law to correct the NIRSPEC and MagE fluxes for extinction. The quoted uncertainities in the measured fluxes are inferred directly from the Gaussian fits to the emission lines. To obtain the dereddened flux uncertainty, we scale the measured flux uncertainty with the same dereddening factor as applied to the flux measurements. We take into account the uncertainties in both the dereddened flux and the E(B-V) via a Monte Carlo approach, as described in Section~\ref{diag_uncert}.

 We list the emission line ratios used and corresponding labels for all the diagnostics used in Table~\ref{tab:masterlist}. The following sections describe the diagnostics used. Figure~\ref{fig: indiv_diag} shows the results and Table~\ref{tab:phys_list} quotes the corresponding values.
 
\subsection{Deriving uncertainties}\label{diag_uncert}
For each emission line involved in a particular diagnostic, we randomly draw a flux value from a Gaussian distribution which has a mean equal to the measured (non-dereddened) flux and a width equal to the 1$\sigma$ uncertainty in the measurement. We also randomly draw a value of E(B-V) from the measured range of 0.4$\pm$0.07. Then we deredden the fluxes relative to the reddest line involved in that particular diagnostic, using the randomly drawn E(B-V) for that particular iteration. For instance, if the diagnostic involves the [\ion{O}{iii}]$\lambda \lambda$ 4959,5007/[\ion{O}{iii}]$\lambda$ 4363 ratio, we de-redden the [\ion{O}{iii}]$\lambda$ 4363 and [\ion{O}{iii}]$\lambda$ 4959 fluxes to bring them to the reference frame of the [\ion{O}{iii}]$\lambda$ 5007 line. Because only the line ratios are relevant for our purposes, and not the overall shape of the spectrum, the relative de-reddening is performed to ensure that we are not over-estimating the de-reddening uncertainties. We calculate all the diagnostics, explained in the preceding sections, with this set of de-reddened line fluxes and repeat the process 10$^4$ times. This leaves us with 10$^4$ different values of each parameter we are trying to estimate, which we convert to a probability distribution function (PDF). We quote the median values of each distribution, along with the  16$^{th}$ and 84$^{th}$ percentiles as the 1$\sigma$ lower and upper limits of that parameter. To calculate the mean value and uncertainties of a quantity from multiple SEL diagnostics, we simply add the PDFs from the concerned diagnostics and then quote the 50$^{th}$, 16$^{th}$ and 84$^{th}$ percentiles of the summed PDF. We adopt this approach to account for asymmetry (non-Gaussianity) in the uncertainties of physical parameters, which indeed is the case for some of the properties.

\subsection{Rest-frame optical diagnostics}\label{optdiag}
We use the nebular [\ion{O}{ii}]$\lambda \lambda$ 3727,3729, [\ion{O}{iii}]$\lambda \lambda$ 4959,5007, [\ion{O}{iii}]$\lambda$ 4363, [\ion{N}{ii}]$\lambda$ 6583 and [\ion{S}{ii}]$\lambda \lambda$ 6716, 6731 lines from the rest-frame optical NIRSPEC spectrum (Table~\ref{tab:optflux}) to determine the physical quantities in {\knotE}.

\subsubsection{Electron temperature}
We measure the electron temperature ({\Te}), in the [\ion{O}{iii}] nebular zone, using the [\ion{O}{iii}]$\lambda \lambda$4959,5007/[\ion{O}{iii}]$\lambda$4363 ratio, following equations 1 and 2 of \citetalias{Izotov:2006ab}, iteratively. This method has almost no dependence on the {\Ne}, determined using [\ion{S}{ii}] line fluxes, and can therefore constrain {\Te} even in absence of the [\ion{S}{ii}] lines. The [\ion{O}{iii}]$\lambda \lambda$ 4959,5007 and [\ion{O}{iii}]$\lambda$4363 emission lines originate from the $^{1}S \rightarrow\ ^{1}D$ and  $^{1}D \rightarrow\ ^{3}P$ transitions, which can be completely constrained by atomic physics from a given {\Te}. Therefore, the [\ion{O}{iii}]$\lambda \lambda$ 4959,5007/[\ion{O}{iii}]$\lambda$4363 ratio is an excellent {\Te} diagnostic in the low-density ({\Ne} < 10$^5$ cm$^{-3}$) regime. Using the \citetalias{Izotov:2006ab} diagnostic we derive {\Te} = 1.22 $^{+0.16}_{-0.17}$ $\times$ 10$^4$ K.

We also use the [\ion{O}{iii}]$\lambda$4363/[\ion{O}{iii}]$\lambda$5007 ratio calibration from N18 and which also yields {\Te} = 1.22 $^{+0.16}_{-0.17}$ $\times$ 10$^4$ K. The N18 calibrations are based on the latest version of MAPPINGS v5.1 \citep[see][]{Sutherland:1993aa, Dopita:2013aa} photoionisation code. The MAPPINGS photoionisation code self-consistently computes the ionisation structure of the nebulae, accounting for dust absorption, radiation pressure, grain charging, and photoelectric heating of small grains \citep{Groves:2004aa}. 

N18 is based on the latest atomic data for an ensemble of atoms at constant temperature and density. The \citetalias{Izotov:2006ab}  diagnostic \citep[from][]{Aller:1984aa} is based on the same model, but uses older atomic data. As a result, both are emission-weighted average temperatures, despite the existence of temperature gradients in real nebulae. The remarkable agreement between {\Te} derived from \citetalias{Izotov:2006ab} and N18 methods is expected because the oxygen atomic data for the relevant lines has not changed substantially.

\subsubsection{Ionization parameter}\label{qdiag:opt}
The ionisation parameter $q$ is defined as the ratio of incident ionising photon flux to the hydrogen density at the inner boundary of the ionised shell. It is a measure of the hardness of the ionising radiation and bolometric luminosity of the ionising source. We quote $q$ in units of cm s$^{-1}$ throughout this paper. The ionization parameter can also be represented as a dimensionless quantity $U$, by dividing $q$ by the speed of light. 

We measure {\logq} using ten different diagnostics (Table~\ref{tab:masterlist}): four employ the O$_{32}$ ([\ion{O}{iii}]$\lambda\lambda$4959,5007/[\ion{O}{ii}]$\lambda\lambda$3727,9) index from the calibrations of \citetalias{Kobulnicky:2004aa}, \citetalias{Kewley:2002fk}, \citet[][henceforth \citetalias{Levesque:2014aa}]{Levesque:2014aa}, \citet[][henceforth \citetalias{Strom:2018aa}]{Strom:2018aa} and new photoionisation models of \citetalias{Kewley:2019ab}, two are based on the Ne3O2 ([\ion{Ne}{iii}]$\lambda$3869/[\ion{O}{ii}]$\lambda$3727) calibrations by \citetalias{Levesque:2014aa}, \citetalias{Strom:2018aa} and \citetalias{Kewley:2019ab}, and another using the [\ion{O}{iii}]$\lambda$ 5007/H-$\beta$ ratio diagnostic from \citetalias{Strom:2018aa} and \citetalias{Kewley:2019ab}. Averaging over all the optical SEL diagnostics, we derive a weighted mean {\logq} = 7.8$^{+0.2}_{-0.5}$. We compare the different diagnostics in the top-right panel of Figure~\ref{fig: indiv_diag} and quote the values in Table~\ref{tab:phys_list}.

Reddening is a concern for the methods involving the O$_{32}$ ratio, because the [\ion{O}{ii}] and [\ion{O}{iii}] wavelengths are widely separated, and hence the dereddened flux obtained for these lines greatly depends on the E(B-V) value and extinction law assumed. As a possible solution, \citetalias{Levesque:2014aa} proposed a new {\logq} diagnostic using the Ne3O2 index, which uses lines with smaller wavelength separation and makes for a more powerful diagnostic in spite of reddening concerns. \citetalias{Kewley:2019ab} adds to the \citetalias{Levesque:2014aa} calibrations by using updated photoionisation models. We use both methods and compare the results.

The O$_{32}$ and Ne3O2 line ratios are sensitive to metallicity. We use the mean metallicity obtained from all the abundance diagnostics to specify the metallicity branch of the {\logq} calibrations in the \citetalias{Levesque:2014aa} and \citetalias{Kewley:2019ab} methods. \citetalias{Kewley:2019ab} point out that the Ne3O2 index should only be used when reliable estimates of ISM pressure and metallicity are available. We use the mean ISM pressure derived using our pressure diagnostics, to define the ISM pressure for use in the K18 methods.

The Ne3O2 diagnostic consistently yields lower ($\approx$0.7 dex) {\logq} values than the other diagnostics, for both \citetalias{Kewley:2019ab} and \citetalias{Levesque:2014aa} calibrations. This is because the Ne3O2 ratio is extremely sensitive to ISM pressure and metallicity. {\knotE} yields {\lpok} $\approx$ 7.5 and 8.5 for the low and high ionisation zones, respectively. The \citetalias{Kewley:2019ab} calibrations quoted in this paper correspond to a mean pressure branch of {\lpok} = 8.0. However, the {\lpok} = 7.5 branch of the \citetalias{Kewley:2019ab} calibrations yields a $\approx$0.2 dex higher {\logq} and the  {\lpok} = 8.5 branch yields $\approx$0.2 dex lower {\logq}, compared to the {\lpok} = 8.0 branch. Thus, using {\lpok} = 7.5, brings the Ne3O2 ratio into agreement with the mean {\logq} value derived from other diagnostics. The \citetalias{Levesque:2014aa} diagnostic assumes fixed {\Ne} = 100 cm$^{-3}$ ({\lpok} $\approx$6), which is not a good approximation for {\knotE}. \citetalias{Sanders:2016aa}, on the other hand, include all galaxies, irrespective of their metallicty and pressure, while fitting {\logq} as a function of Ne3O2. This leads to a $\sim 0.2$ dex intrinsic scatter in the Ne3O2 calibration of \citetalias{Sanders:2016aa}. Consequently, the {\logq} thus derived, agrees well (within $<$0.1 dex) with the mean {\logq}, but has a large intrinsic scatter which is not reflected in the quoted uncertainty. The example of {\knotE} demonstrates the drawbacks of the assumptions in individual emission line diagnostics. It is in such cases that Bayesian inference methods (Section~\ref{izidiag}) can be more useful, provided the relevant emission lines are available.

\subsubsection{Electron density}\label{nediag_opt}
We compute the electron density {\Ne} \,in {\knotE} using the O2 ([\ion{O}{ii}]$\lambda$3729/[\ion{O}{ii}]$\lambda$3727) and S2 ([\ion{S}{ii}]$\lambda$6731/[\ion{S}{ii}]$\lambda$6717) calibrations from \citet[][hereafter \citetalias{Sanders:2016aa}]{Sanders:2016aa} and from the constant density models of \citetalias{Kewley:2019aa}. The \citetalias{Kewley:2019aa} models cover a range of $\log{(n_e/\mathrm{cm^{-3}})}$ = 0 to 5, in increments of 0.5 dex. The \citetalias{Sanders:2016aa} diagnostics are based on a 5-level atom approximation of the \ion{O}{ii} and \ion{S}{ii} ions which yield $\log{(n_e)}$ = 2.9$^{+0.1}_{-0.1}$ and 3.1$^{+0.5}_{-0.4}$ from the O2 and S2 ratios respectively, where {\Ne} is in units of cm$^{-3}$.

The $\log{(n_e)}$ we obtain using the \citetalias{Kewley:2019aa} O2 and S2 ratios, are in excellent agreement with the corresponding \citetalias{Sanders:2016aa} values. \citetalias{Kewley:2019aa} provide a 3D (metallicity, ionisation parameter and pressure/density) grid of models with predicted emission line fluxes for every combination of the three parameters. As such, obtaining electron density (or pressure) given a line ratio, requires the metallicity ($Z$) and ionisation parameter ($q$) as inputs. We use the mean of metallicity and {\logq} values, measured using the different SEL diagnostics, to constrain the $Z$ and {\logq}. Thereafter, we interpolate the line ratios as a function of the electron density (or pressure) to obtain our desired physical quantity.

For comparison, we also use the theoretical O2 vs {\Ne} curves from \citet{Osterbrock:1989} that are based on single atom models. We infer a weighted mean $\log{(n_e)}$ = 3.0$^{+0.4}_{-0.2}$ using rest-optical diagnostics. The middle-right panel in Figure~\ref{fig: indiv_diag} shows our {\Ne} measurements and Table~\ref{tab:phys_list} quotes the corresponding values.

The electron density {\Ne} is fundamentally related to the ISM pressure \citep[e.g. see][for discussion]{Dopita:2006ab, Kewley:2019aa}. Assuming a constant electron temperature {\Te}, the ISM pressure $P$ is related to the total density $n = P/T_ek$ and the total density $n$ is related to the electron density through $n = n_e(1 + (4X + Y)/(2X + 2))$ where $X$ and $Y$ are the mass fractions of Hydrogen and Helium respectively. Thus, for a fixed {\Te} the ISM pressure is directly proportional to {\Ne}. In reality, however, neither the electron temperature nor the density is constant. The ISM is often clumpy and has fluctuations and/or gradients in temperature and density. The constant density models of \citetalias{Kewley:2019aa}  allow for the temperature structure within the {\htr} but they do not allow the electron density to vary. Hence, \citetalias{Kewley:2019aa} point out that the constant density models are likely to be less realistic than the constant pressure models because typical {\htrs} have shorter sound crossing timescale than cooling/heating timescale, allowing the pressure to equalise throughout the nebula. 

\subsubsection{ISM pressure}\label{pdiag_opt}
We use the O2 ([\ion{O}{ii}]$\lambda$3729/[\ion{O}{ii}]$\lambda$3727) and S2 ([\ion{S}{ii}]$\lambda$6731/[\ion{S}{ii}]$\lambda$6717) calibrations  of \citetalias{Kewley:2019aa} to measure the ISM pressure, given in terms of {\lpok} where $k$ is the Boltzmann constant. We derive a weighted average {\lpok} = 7.4$^{+0.6}_{-0.2}$ for {\knotE} from the rest-frame optical diagnostics, where $P/k$ is in units of $\mathrm{K\ cm^{-3}}$. The middle-left panel in Figure~\ref{fig: indiv_diag} shows the PDFs for all diagnostics.

Reddening is not a concern for either of these sets of closely spaced lines. Nevertheless, we performed reddening corrections (as described in Section~\ref{analysis}) for consistency. \citetalias{Kewley:2019aa} point out that the S2 ratio changes by $\sim 1$ dex within a range of 5.5 $\leq \log{(P/k)} \leq$ 9.0 and the O2 index drops by $\sim 1$ dex within 5.5 $\leq \log{(P/k)} \leq$ 8.0, demonstrating that these diagnostics are extremely sensitive to the pressure for this range.

\citetalias{Kewley:2019aa} used plane-parallel MAPPINGS v5.1 {\htr} models at constant pressure, with {\lpok} ranging from 4.0 to 9.0 in increments of 0.5. MAPPINGS calculates detailed electron temperature and density structure within the {\htr} for each of these models at a fixed ISM pressure. \citetalias{Kewley:2019aa} point out that the constant pressure models are more realistic than the constant density models used for the electron density calibrations (Section~\ref{nediag_opt}) and recommend using the former. We refer the reader to \citetalias{Kewley:2019aa} for a detailed description of the models.

\subsubsection{Gas phase oxygen abundance}\label{abundiag_opt}
We measure the gas phase oxygen abundance from the available set of optical lines in eight different ways (Table~\ref{tab:masterlist}). We use the combined method of \citetalias{Kewley:2002fk} (Section 6 \citetalias{Kewley:2002fk}), which, for our abundance regime, uses the [\ion{N}{ii}]$\lambda$6584/[\ion{O}{ii}]$\lambda\lambda$3727,3729 ratio.  We also employ the iterative method of \citetalias{Kobulnicky:2004aa} which uses the R$_{23}$ (([\ion{O}{ii}]$\lambda$3727+[\ion{O}{iii}]$\lambda\lambda$4959,5007)/H$\beta$) and O$_{32}$ indices to solve for both {\logOH} and {\logq}. Both these works stem from MAPPINGS photoionisation models of \ion{H}{ii} regions. We additionally use the N2  ([\ion{N}{ii}]$\lambda$6584/H$\alpha$) and O3N2  (([\ion{O}{iii}]$\lambda$5007/H$\beta$)/([\ion{N}{ii}]$\lambda$6584/H$\alpha$)) calibrations from \citetalias{Pettini:2004qe} to compare the abundance values derived using different calibration methods. We also use the empirical Ne3O2 (\ion{Ne}{iii}]$\lambda$ 3869/[\ion{O}{ii}]$\lambda\lambda$3727,9) calibrations from \citet[hereafter \citetalias{Shi:2007aa}]{Shi:2007aa} and \citet[hereafter \citetalias{Bian:2018aa}]{Bian:2018aa}, and the theoretical [\ion{N}{ii}]/[\ion{S}{ii}] and [\ion{N}{ii}]/H$\alpha$ ratios, following \citetalias{Dopita:2016aa}. Both the \citetalias{Bian:2018aa} and \citetalias{Dopita:2016aa} methods are suitable for high redshift ($z\gtrsim$2) galaxies. In addition to these SEL methods, which depend on photoionisation models (with the exception of \citetalias{Shi:2007aa} and BKD18), we also use the direct estimation of the abundance from the electron temperature {\Te}, following the  \citetalias{Izotov:2006ab} procedure.

Each method has its own drawbacks. The $T_e$ method, \citetalias{Kobulnicky:2004aa} and \citetalias{Kewley:2002fk} methods are sensitive to reddening corrections because they involve lines with widely spaced wavelengths. The \citetalias{Kobulnicky:2004aa} R$_{23}$ diagnostic is double valued and requires an initial guess of abundance, which we provide by using the [\ion{N}{ii}]$\lambda$6584/H$\alpha$ ratio. The R$_{23}$ index is also sensitive to ionisation parameter {\logq}. We use R$_{23}$ in conjunction with the O$_{32}$ index to iteratively solve for both log(O/H)+12 and {\logq}. The \citetalias{Shi:2007aa}, BKD18, \citetalias{Dopita:2016aa} and both the \citetalias{Pettini:2004qe} diagnostics, do not suffer from reddening issues because they use lines that are closely spaced in wavelength. 

In addition to the above shortcomings, all the methods have systematics offsets, relative to one another, on their zero-points \citep{Kewley:2008aa, Bian:2017aa}. This is because of the different photoionisation models and samples of {\htrs} used to derive the diagnostics. The discrepancy between the strong line diagnostics and the {\Te} method are well known and are mainly attributed to the assumption of a constant temperature in the {\Te} methods. \citep[e.g.][]{Stasinska:2002aa, Lopez:2012aa}. Additionally, the existence of a temperature gradient within the {\htrs} may lead the {\Te} method to systematically underestimate the oxygen abundance because of the assumption of a one or two-zone temperature model \citep{Stasinska:2005aa}. Therefore, it is sensible to compare among these methods, only after we have corrected for the relative offsets. 

We correct for this offset following \citetalias{Kewley:2008aa} prescription which was developed using local SDSS galaxies. We convert the {\logOH} values from the empirical and theoretical calibrations to the reference frame of the \citetalias{Kobulnicky:2004aa} method. The choice for this common reference frame was motivated by the fact that \citetalias{Kobulnicky:2004aa} take into account the dependence of the metallicity sensitive lines on the ionisation parameter. However, \citet[][hereafter \citetalias{Kewley:2008aa}]{Kewley:2008aa} do not prescribe a conversion scheme from the direct {\Te} method to the other SEL diagnostics. Therefore, we quote the {\Te} metallicity as it is, without any conversion. We also transform the \citetalias{Rigby:2011aa} abundance value to the \citetalias{Kobulnicky:2004aa} frame, in order to facilitate comparison between our work and \citetalias{Rigby:2011aa}. Note that we cannot account for any potential relative offsets in the \citetalias{Shi:2007aa}, BKD18 and \citetalias{Dopita:2016aa} methods following the \citetalias{Kewley:2008aa} prescription because \citetalias{Kewley:2008aa} predates both.

Table~\ref{tab:phys_list} lists the values of {\logOH} computed using various diagnostics and the bottom-left panel in Figure~\ref{fig: indiv_diag} shows the corresponding PDFs. \citet{Wuyts:2014ab} inferred a {\logOH} = 8.28 $\pm$ 0.02 for {\knotE}, which transforms to {\logOH} = 8.65 $\pm$ 0.03 in the \citetalias{Kobulnicky:2004aa} frame. We plot the \citet{Wuyts:2014ab} value as a dotted line in Figure~\ref{fig: indiv_diag} for comparison. Averaging over all diagnostics, we derive a weighted average {\logOH} = 8.6$^{+0.1}_{-0.4}$ for {\knotE} in the \citetalias{Kobulnicky:2004aa} frame.

The optical SEL diagnostics yield a weighted mean {\logOH} = 8.6$^{+0.1}_{-0.4}$. Both the {\Te} methods and the Ne3O2 methods (\citetalias{Shi:2007aa} and BKD18) result in lower ($\sim$0.4 dex) abundance. With the exception of BKD18, the other three measurements have large uncertainties associated with them (broad PDFs in Figure~\ref{fig: indiv_diag}). As discussed in Section~\ref{abundiag_opt}, the discrepancy between the {\Te} and strong line methods is well known and can be attributed to the temperature fluctuations and gradients in the {\htr}, which the {\Te} methods do not take into account. \citetalias{Shi:2007aa} and BKD18 use the {\Te} method to calibrate their diagnostic and hence, suffer from the same discrepancy. The higher uncertainties of the {\Te} and \citetalias{Shi:2007aa} methods is because the [\ion{O}{iii}]$\lambda$4363 line is often weak or undetected. Moreover, the strong line methods use additional information in terms of {\logq} or a specific branch of metallicity, leading to lower uncertainties. 

\subsection{Rest-frame UV diagnostics}\label{uvdiag}
The rest-frame UV diagnostics use the \ion{O}{iii}]$\lambda\lambda$1660,6, \ion{C}{iii}]$\lambda$1907, [\ion{C}{iii}]$\lambda$ 1909 and \ion{Si}{ii}]$\lambda\lambda$1883,92 line fluxes measured from the Magellan/MagE spectra (Table~\ref{tab:uvflux}). 
The majority of these diagnostics (refer Table~\ref{tab:masterlist}) are from recent works of \citetalias{Kewley:2019aa},b and N18, which use the latest, improved version 5.1 of MAPPINGS. The new MAPPINGS v5.1 uses the latest available atomic data from the CHIANTI8 atomic database \citep{DelZanna:2015aa}, which is a prime factor that governs the nebular emission line strengths. As described in detail in Section~\ref{optdiag}, the plane parallel isobaric {\htr} models have been used for both N18 and K18 (except for density diagnostics, where constant density models have been used). Moreover, reddening corrections are not important for the diagnostics involving only rest-frame UV lines, because these pairs of lines have closely spaced wavelengths.
 
\subsubsection{Electron temperature}
We derive electron temperature ({\Te}) using the [\ion{O}{iii}]$\lambda $ 5007/\ion{O}{iii}]$\lambda\lambda$ 1660,6 ratio from the theoretical calibrations of N18. This method relies on the fact that the  $^{5}S \rightarrow\ ^{3}P$ ($\lambda $ 5007) and  $^{1}D \rightarrow\ ^{3}P$ ($\lambda\lambda$ 1660,6) transition rate ratio depends only on one physical parameter, the {\Te}. The other dependencies of the ratio are constants that can be derived from atomic physics. The UV N18 calibrations yield a {\Te} = 1.10 $^{+0.07}_{-0.06}$ $\times$ 10$^4$ K which is $\sim 1000$\,K lower than that derived from the optical {\Te} diagnostics. The optical and UV {\Te} measurements agree to within $1 \sigma$. 

The large wavelength baseline renders the  [\ion{O}{iii}]$\lambda $ 5007/\ion{O}{iii}]$\lambda\lambda$ 1660,6 ratio highly susceptible to uncertainties in reddening corrections. An uncertainty of $\pm$0.07 in E(B-V) leads to $\sim$+33\%, -25\% uncertainty in this ratio. Moreover, discrepancies in the relative flux calibration from the rest-frame UV to the rest-frame optical spectra can contribute to uncertainties in {\Te}.

Although we refer to the  [\ion{O}{iii}]$\lambda$5007/\ion{O}{iii}]$\lambda\lambda$1660,6 method as a UV diagnostic, it still requires [\ion{O}{iii}]$\lambda$5007. Therefore, no {\Te} diagnostic used in this paper is completely independent of rest-optical spectra. However, the method involving only the optical lines requires the [\ion{O}{iii}]$\lambda$4363 line, which is barely detected (SNR = 2.75) in {\knotE}. Consequently, the [\ion{O}{iii}]$\lambda$5007/\ion{O}{iii}]$\lambda\lambda$1660,6 ratio provides a better constraint on {\Te}. Taking a weighted average of all UV-optical methods, we find a mean {\Te} = 1.2$^{+0.2}_{-0.1}$ $\times 10^4$ K for {\knotE}.

\subsubsection{Ionization parameter}\label{qdiag_uv}
\citetalias{Kewley:2019ab} outline many rest-frame UV emission line ratios that can potentially be used as ionisation parameter ({\logq}) diagnostics. We choose to use three line ratios: [\ion{C}{iii}]]$\lambda\lambda$1907,9/[\ion{C}{ii}]$\lambda$1335 (blend of 1334.58 \AA, 1335.66 \AA\ and 1335.71 \AA),  [\ion{C}{iii}]]$\lambda\lambda$1907,9/[\ion{C}{ii}]$\lambda$2323-8 (2323.50 \AA, 2324.69 \AA, 2325.40 \AA, 2326.93 \AA, and 2328.12 \AA) and  [\ion{O}{iii}]]$\lambda\lambda$1660,6/[\ion{O}{ii}]$\lambda$2470. We derive {\logq} entirely from UV lines, for the first time, by using these three diagnostics. The top-right panel of Figure~\ref{fig: indiv_diag} shows the PDFs of our {\logq} measurements. The [\ion{C}{II}]$\lambda$1335 and [\ion{C}{II}]$\lambda$2323 group of lines are not detected in {\knotE}. We therefore use the 3$\sigma$ upper limits for the [\ion{C}{ii}] line fluxes to estimate a lower limit for {\logq}, wherever applicable. Lower limits are represented in Figure~\ref{fig: indiv_diag} as dashed vertical lines with horizontal arrows.

The [\ion{C}{iii}]/[\ion{C}{ii}] ratios are very effective measures of the ionisation parameter, especially in the low metallicity regime ({\logOH}\ $\leq$ 8.5), because they have negligible sensitivity to ISM pressure. Moreover, the [\ion{C}{iii}]$\lambda\lambda$1907,9/[\ion{C}{ii}]$\lambda$1335 ratio does not vary with metallicity for low metallicities, making it an ideal {\logq} diagnostic. The [\ion{O}{iii}]$\lambda\lambda$1660,6/[\ion{O}{ii}]$\lambda$2470 ratio is analogous to the O$_{32}$ index in optical. \citetalias{Kewley:2019ab} advise against the use of the [\ion{O}{iii}]/[\ion{O}{ii}] diagnostic in the high pressure ({\lpok} $\geq$7) and low metallicity ({\logOH} $\leq$ 8.23) regime owing to the high sensitivity of these lines (varies $>$ 0.5 dex) to ISM pressure. We derive the ISM pressure of {\knotE}, and use it to interpolate between the pressure grid, thereby minimising the sensitivity issue.

The {\logq} estimated from all rest-frame UV and optical diagnostics (except the Ne3O2 ratio) broadly agree within $\sim$0.7 dex. \citetalias{Kewley:2019ab}O32b is the only UV diagnostic which is not a lower limit, and it yields {\logq} = 8.0$^{+0.1}_{-0.1}$ which agrees with the mean optical result within 1$\sigma$ uncertainty, as do all the individual \citetalias{Kewley:2019ab} diagnostics. The \citetalias{Kewley:2002fk} and \citetalias{Kobulnicky:2004aa} diagnostics were based on an earlier version of MAPPINGS (v3) and as such, yield slightly ($\sim$0.2 dex) higher {\logq} than the \citetalias{Kewley:2019ab} methods, while agreeing to within 1$\sigma$ with each other. The lower limits on {\logq} obtained from the undetected [\ion{C}{ii}] emission lines are consistent with the other diagnostics. Although the scatter in ionization parameter is large ($\sim$0.7 dex), the UV estimates are \textit{not} systematically offset from the optical estimates within the uncertainties, which is encouraging. This clearly demonstrates that it is indeed possible to determine {\logq} using only rest-frame UV spectra, provided at least one of \ion{C}{iii}]$\lambda\lambda$1906,8/[\ion{C}{ii}]$\lambda$1335, \ion{C}{iii}]$\lambda\lambda$1906,8/[\ion{C}{ii}]$\lambda$2325, or [\ion{O}{iii}]$\lambda\lambda$1660,6/[\ion{O}{ii}]$\lambda$2470 ratios is available.

\subsubsection{ISM Pressure} \label{pdiag_uv}
We determine the ISM pressure ({\lpok}) using the \citetalias{Kewley:2019aa} calibrations of rest-frame UV line ratios \ion{C}{iii}]$\lambda$1907/[\ion{C}{iii}]$\lambda$1909 (C3) and \ion{Si}{iii}]$\lambda$1883/\ion{Si}{iii}]$\lambda$1892 (Si3). These calibrations are derived using the plane parallel, isobaric model grid from MAPPINGS, as described Section~\ref{optdiag}. The middle-left panel in Figure~\ref{fig: indiv_diag} shows the PDFs of our ISM pressure measurements and the corresponding values are quoted in Table~\ref{tab:phys_list}. We derive a weighted average {\lpok} = 8.8$^{+0.2}_{-0.6}$ from the UV diagnostics.

In general, the UV diagnostics (dashed lines) yield considerably higher ($\sim$1.4 dex) ISM pressures than the optical diagnostics (solid lines), though there is significant ovelap between the UV and optical PDFs. The discrepancy is the result of the \citetalias{Kewley:2019aa} C3 and Si3 diagnostics probing the high pressure ({\lpok} > 7.5) regime. We further discuss this difference in Section~\ref{disc_press}.

The \ion{Si}{iii} and \ion{C}{iii} ratios approach the high pressure limit at $\sim 1.4$ and cease to be sensitive to pressure. Both the pressure diagnostics are sensitive to {\logq} and metallicity. The \ion{Si}{iii} ratio is more sensitive to the ISM pressure, varying by over an order of magnitude in the range 7.5 < {\lpok} < 9.0, as compared to the  \ion{C}{iii} ratio (which varies by almost an order of magnitude in the same range of {\lpok}). 

\subsubsection{Electron density}\label{nediag_uv}
Due to the interdependency between the electron density ({\Ne}) and the ISM pressure, the line ratios sensitive to one property are sensitive to the other as well. We derive {\Ne} using the same C3 and Si3 ratios, as in the case of pressure, from the \citetalias{Kewley:2019aa} calibrations. However, in this case, the {\htr} models used assume constant density throughout the nebula, which may not be a valid assumption. We infer a weighted mean $\log{(n_e)}$ = 4.3$^{+0.3}_{-0.6}$ using rest-UV diagnostics.

Note that the \citetalias{Kewley:2019aa} C3 and Si3 diagnostics are sensitive only in the high density ({\Ne} > 1000 cm$^{-3}$) regime, which holds for {\knotE}. Both the \ion{Si}{iii} and \ion{C}{iii} ratios have almost no dependence on {\logq} (Figures 1 and 2 in \citetalias{Kewley:2019aa}) but are slightly sensitive to the metallicity, which has to be provided in order to obtain {\Ne}.

Similar to the ISM pressure case, we derive consistently higher {\Ne} ($\sim 1.5$ dex) with the UV lines than the optical lines, which is reflected as a bimodality in the PDFs in Figure~\ref{fig: indiv_diag}. Again, this discrepancy is the result of different emission line species tracing different physical regions of the nebula, as described in Section~\ref{disc_press}. Thus, it is possible to infer {\Ne} using only the rest-frame UV lines, although the values inferred do not represent the same physical region as the optical diagnostics.

\subsubsection{Oxygen abundance}\label{abundiag_uv}
We measure {\logOH} by the direct method following equations 2 and 3 of \citetalias{Izotov:2006ab}. Here we use the {\Te} obtained from the  [\ion{O}{iii}]$\lambda$5007/\ion{O}{iii}]$\lambda\lambda$1660,6 ratio. Because this abundance estimate makes use of the UV lines \ion{O}{iii}]$\lambda\lambda$ 1660,6 we classify this as a rest-frame UV abundance diagnostic. The UV {\Te} diagnostic yields {\logOH} = 8.3 $\pm$ 0.1, which agrees with the optical {\Te} abundance within $1 \sigma$ uncertainty. The bottom-left panel in Figure~\ref{fig: indiv_diag} compares the abundance diagnostics.

\section{Joint Bayesian Diagnostics}\label{izidiag}
\textit{IZI} is an IDL-based software developed by \citet[hereafter \citetalias{Blanc:2015aa}]{Blanc:2015aa} that uses Bayesian inference to simultaneously infer metallicity ($Z$) and ionisation parameter ({\logq}) of the ionised nebular gas. \textit{IZI} requires a set of emission line fluxes observed from the nebulae and a 2D ($Z$ and {\logq}) grid of models, as its inputs.

We extend the publicly available version of \textit{IZI} to 3D -- to enable the metallicity, ionisation parameter and ISM pressure to be inferred simultaneously. This 3D Bayesian method avoids the need for assumptions about {\lpok} and hence constrains the physical properties in a self-consistent way. The 3D \textit{IZI}, referred to as IZIP (Inferring metallicities (Z), Ionization, and Pressure) henceforth, requires a 3D grid of models ($Z$, {\logq} \& {\lpok}) as an input to interpolate. We emphasize that IZIP is simply an extension of \textit{IZI} to include an extra dimension, but otherwise preserves the functionality of the original \textit{IZI} algorithm. We use MAPPINGS-V photoionisation models to produce a grid of $Z$, {\logq}, {\lpok} and emission line fluxes as inputs to IZIP. In the hope that it would benefit the community, we make IZIP publicly available at [URL to be added at proofs stage] and ask future users to consider this paper as the appropriate reference for IZIP.

The input to the MAPPINGS-V models, which in turn is input to IZIP, accounts for the primary and secondary nucleosynthetic components of the N/O and C/O ratios. The application assumes that the nitrogen and carbon in galaxies have a primary and a secondary origin and that there is no N/O or C/O excess. We are assuming that {\knotE} lies along the local relation for the above nucleosynthetic origin. Thus, in effect, IZIP accounts for variations in N/O and C/O as a function of metallicity, when used with the MAPPINGS models as the input.

Most individual emission line diagnostics suffer from a major drawback: they are simultaneously sensitive to metallicity, ionisation parameter and pressure. IZIP simultaneously computes the likelihood of each of {\logOH}, {\logq} and {\lpok} without any assumption about the others. Moreover, IZIP makes use of all the available emission line information simultaneously as opposed to using a specific pair of lines to derive $Z$ and {\logq}. The Bayesian approach allows us to calculate joint and marginalised posterior probability distribution functions (PDFs). PDFs allow for multiple peaks and/or asymmetry, which reflect degeneracies in the relation between line fluxes and nebular properties. These degeneracies are harder to deal with while using a specific emission line pair. IZIP also takes into account upper limits for lines that are not formally detected and translates them to a limit on the derived $Z$, \,$\log{(q)}$, and {\lpok}. This is extremely useful for high-redshift spectroscopic studies for which, often, only upper limits on emission lines are available.

A potential disadvantage of IZIP is that it assumes equal weights on all the available emission lines. In other words, lines that are potentially not sensitive to the concerned nebular property or are also sensitive to other nebular properties are weighted the same as the lines that are only sensitive to the concerned property. Consider, for example, the metallicity. Some metallicity sensitive emission lines are also sensitive to ionisation parameter and/or ISM pressure. Individual nebular diagnostics involving these lines would be considered less robust as they may not be ideal probes of metallicity if {\logq} and {\lpok} are not accurately known. IZIP, however, is unable to make such informed decisions and would treat these lines with the same weights as other lines that are sensitive to metallicity only (and thus are ideal metallicity indicators). This might lead to poor constraints in the derived metallicity because IZIP includes emission lines that are dominated by other properties (.e.g. {\logq}) which may be dependent on, but not necessarily positively correlated with, metallicity. One way to remedy this would be to provide only those lines to IZIP that are sensitive only to the particular physical parameter. We carry out such tests for different physical quantities and present the results in Appendix~\ref{app:izip}. The issue of using less sensitive emission lines is especially relevant for high-{\z} galaxies because high-{\z} spectra often contain only a few emission lines above a desired SNR threshold, all of which may not necessarily be exclusively sensitive to $Z$, {\logq} or {\lpok}. Additionally, not all emission lines have a nebular origin. For instance, \citet{Prochaska:2011aa} suggest that the \ion{Mg}{ii} and \ion{Fe}{ii} lines arise due to scattering in the galactic wind. Hence, we remove all the \ion{Mg}{ii} and \ion{Fe}{ii} lines, before providing the set of emission lines to IZIP.

Another way to force IZIP to give different weights to different lines is by providing user defined priors to IZIP. For instance, to break the degeneracy of the double valued metallicity branches (e.g. R$_{23}$), we use a top-hat prior on $Z$ such that {\logOH} > 8.52 if $[$\ion{N}{ii}$]\lambda$6584/H$\alpha$ > 0.0776 and vice versa, wherever $[$\ion{N}{ii}$]$/H$\alpha$ ratio is avaiable. The choice of this prior is motivated by the fact that $[$\ion{N}{ii}$]$/H$\alpha$ varies monotonically with $Z$. We chose $[$\ion{N}{ii}$]$/H$\alpha$ over the $[$\ion{N}{ii}$]$/$[$\ion{O}{ii}$]$ ratio because the latter could be affected by relative flux calibration issues as the $[$\ion{N}{ii}$]$ and $[$\ion{O}{ii}$]$ lines are captured by different NIRSPEC filters. We also investigate the impact of using a flat prior (i.e. no user-defined prior) on the inferred physical quantities in Section~\ref{sec:disc_pizi}.

\begin{table}
\centering
\caption{List of emission lines provided to IZIP for Bayesian analysis. We provide different combinations of the rest-frame optical and UV emission line fluxes for different cases, as described in Section~\ref{izidiag}. The third column denotes whether the emission line is undetected and hence upper limits are used by IZIP, or if it is a blended doublet.}
\begin{tabular}{lrc}
	\toprule
	Line ID & $\lambda_{\mathrm{rest}}$ (\AA) & Comments \\
	\midrule
	\hline
	\multicolumn{3}{c}{Rest-frame UV} \\ 
	\hline 
      	C II 1335a &                  1334.5770 &	Upper limit	\\
		C II 1335b &                  1335.6630 &	Upper limit	\\
		C II 1335c &                  1335.7080 &	Upper limit	\\
		Si II 1533 &                  1533.4312 &	-	\\
		He II 1640 &                  1640.4170 &	-	\\
		O III] 1660 &                  1660.8090 &	-	\\
		O III] 1666 &                  1666.1500 &	-	\\
		N III] 1750 &                  1749.7000 &	-	\\
		{[}Si III] 1882 &                  1882.7070 &	-	\\
		Si III] 1892 &                  1892.0290 &	-	\\
		{[}C III] 1906 &                  1906.6800 &	-	\\
		C III] 1908 &                  1908.7300 &	-	\\
		N II] 2140 &                  2139.6800 &	-	\\
		{[}O III] 2320 &                  2321.6640 &	-	\\
		C II] 2323 &                  2324.2140 &	Upper limit	\\
		C II] 2325c &                  2326.1130 &	-	\\
		C II] 2325d &                  2327.6450 &	-	\\
		C II] 2328 &                  2328.8380 &	-	\\
		Si II] 2335a &                  2335.1230 &	Upper limit	\\
		Si II] 2335b &                  2335.3210 &	Upper limit	\\
		{[}O II] 2470 &                  2471.0270 &	Unresolved doublet	\\
		He I 2945 &                  2945.1030 &	Upper limit	\\
	\hline
	\multicolumn{3}{c}{Rest-frame optical} \\ 
	\hline 
	{[}O II] 3727,9 &  3727.092, 3729.875 & Unresolved doublet \\
	{[}Ne III] 3869 &                        3869.860 &	-	\\
	H$\zeta$ &                        3890.166 &	-	\\
	H$\delta$ &                        3971.198 &	-	\\
	H$\epsilon$ &                        4102.892 &	-	\\
	H$\gamma$ &                        4341.692 &	-	\\
	O III 4363 &                        4364.435 &	-	\\
	H$\beta$ &                        4862.691 &	-	\\
	{[}O III] 4959 &                        4959.895 &	-	\\
	{[}O III] 5007 &                        5008.239 &	-	\\
	H$\alpha$ &                        6564.632 &	-	\\
	{[}N II] 6584 &                        6585.273 &	-	\\
	{[}S II] 6717 &                        6718.294 &	-	\\
	S II 6731 &                        6732.674 &	-	\\
	{[}Ar III] 7136 &                        7137.770 &	-	\\
\bottomrule
\end{tabular}
\label{tab:pizi_lines}
\end{table}

To investigate how the ISM properties depend on the amount of spectral information available, we supply IZIP with four different sets of emission lines tailored to mimic observations with different wavelength coverage, as follows.
\begin{enumerate}
\item All the UV and optical emission lines (in Table~\ref{tab:pizi_lines}) as input: Using all the nebular spectral information available would help us understand by how much the constraints on the physical properties of {\knotE} are improved on inclusion of the UV information in addition to the existing optical spectra.
\item Only the rest-frame optical emission lines of Table~\ref{tab:pizi_lines}: This combination emulates the scenario when the rest-frame UV is not within the observed wavelength coverage and all the information we have is from the rest-frame optical spectra.
\item Only the rest-frame UV emission lines of Table~\ref{tab:pizi_lines}: We use this combination to study how well the nebular properties can be constrained in a scenario where only the rest-frame UV spectra are available, the redder part of the spectra having been redshifted out of the observed wavelength range.
\item The rest-frame UV emission lines along with the [\ion{O}{ii}]$\lambda\lambda$3727,9 doublet: The [\ion{O}{ii}] doublet would be within wavelength coverage until $z\sim 12$ with {\JWST}. Hence, combined information from both the [\ion{O}{ii}] doublet and the UV spectra could be used to infer the physical properties for high-{\z} galaxies.
\item All the UV and optical emission lines except [\ion{Si}{iii}]$\lambda\lambda$1882,92: Excluding the [\ion{Si}{iii}] doublet allows us to investigate its effect on the inferred nebular properties.
\item Only the UV emission lines except [\ion{Si}{iii}]$\lambda\lambda$1882,92. This combination of emission lines is one of the many combinations we tested, including or excluding certain emission lines each time. Excluding [\ion{Si}{iii}] produced a considerable impact, as we discuss in Section~\ref{sec:izip_res}. We discuss all the other additional tests in Appendix~\ref{app:izip}.
\end{enumerate}

IZIP does not account for extinction by dust and requires the user to provide extinction-corrected flux values. We propagate the uncertainties in the reddening value via a Monte Carlo (MC) technique. The MC approach is better than the analytic error propagation because the latter is only applicable up to first order expansion in Taylor series whereas, given enough iterations, the MC approach better samples the parameter space. We randomly draw from a normal distribution of the measured value of E(B-V) = 0.4 $\pm$ 0.07 \citepalias{Whitaker:2014aa}. After correcting for the reddening using this randomly drawn E(B-V) and \citet{Cardelli:1989} reddening law, we supply the dereddened fluxes to IZIP. We repeat this process 100 times, adding and normalizing the resulting PDFs for each iteration, to give the final marginalised and joint PDFs presented in Section~\ref{sec:izip_res}. The results converge well before the 100 iterations used here.

\noindent
\begin{figure*}
	\centering
	
	\begin{subfigure}[t]{0.47\textwidth}
		\centering
		\includegraphics[scale=0.42, trim=0.cm 0.0cm 0.0cm 0.0cm,clip=true,angle=0]{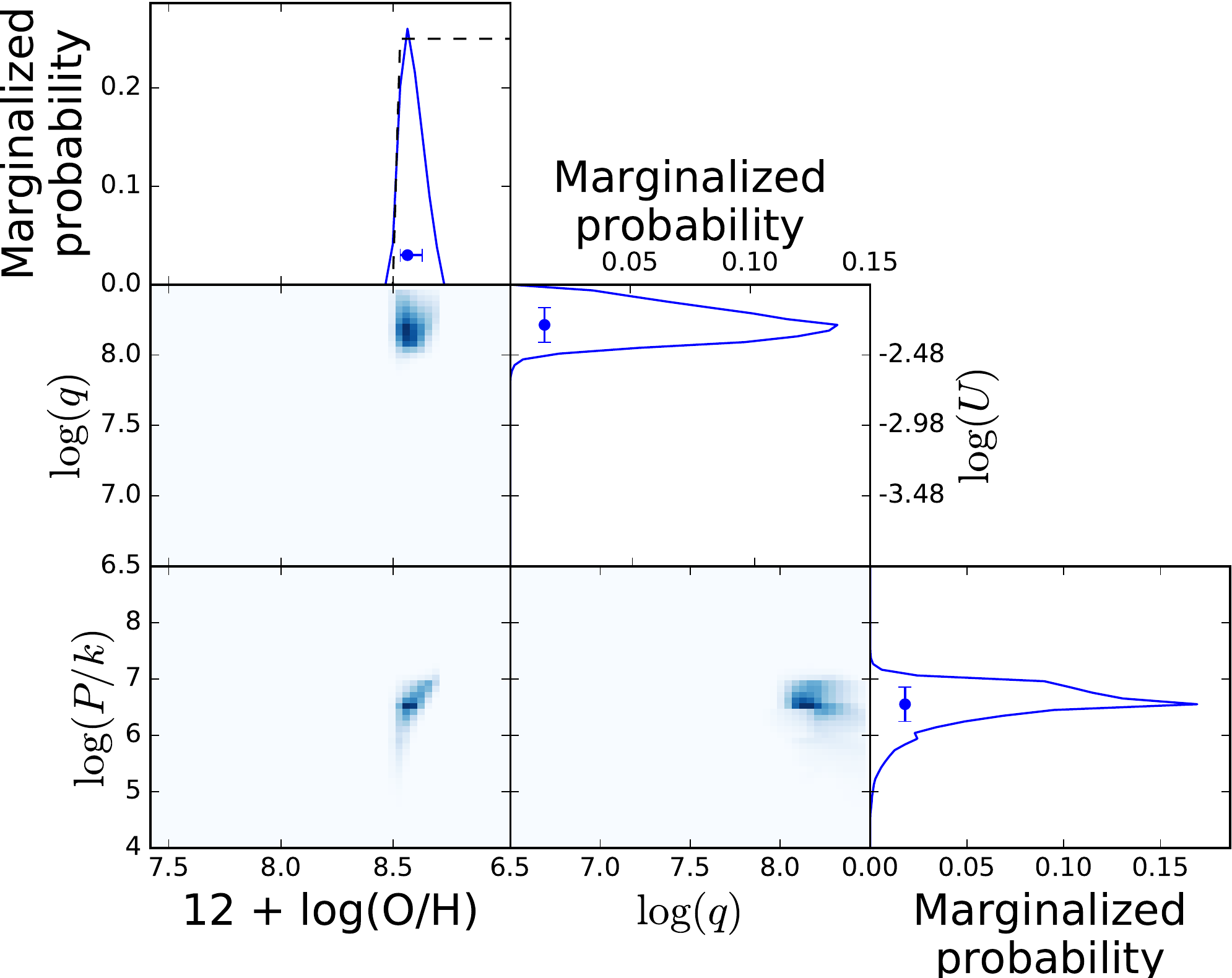}
		\caption{All emission lines used}
		\label{fig: pizi_prior_all}
	\end{subfigure}
	\hspace{0.5cm}
	\begin{subfigure}[t]{0.47\textwidth}
		\centering
		\includegraphics[scale=0.42, trim=0.cm 0.0cm 0.0cm 0.0cm,clip=true,angle=0]{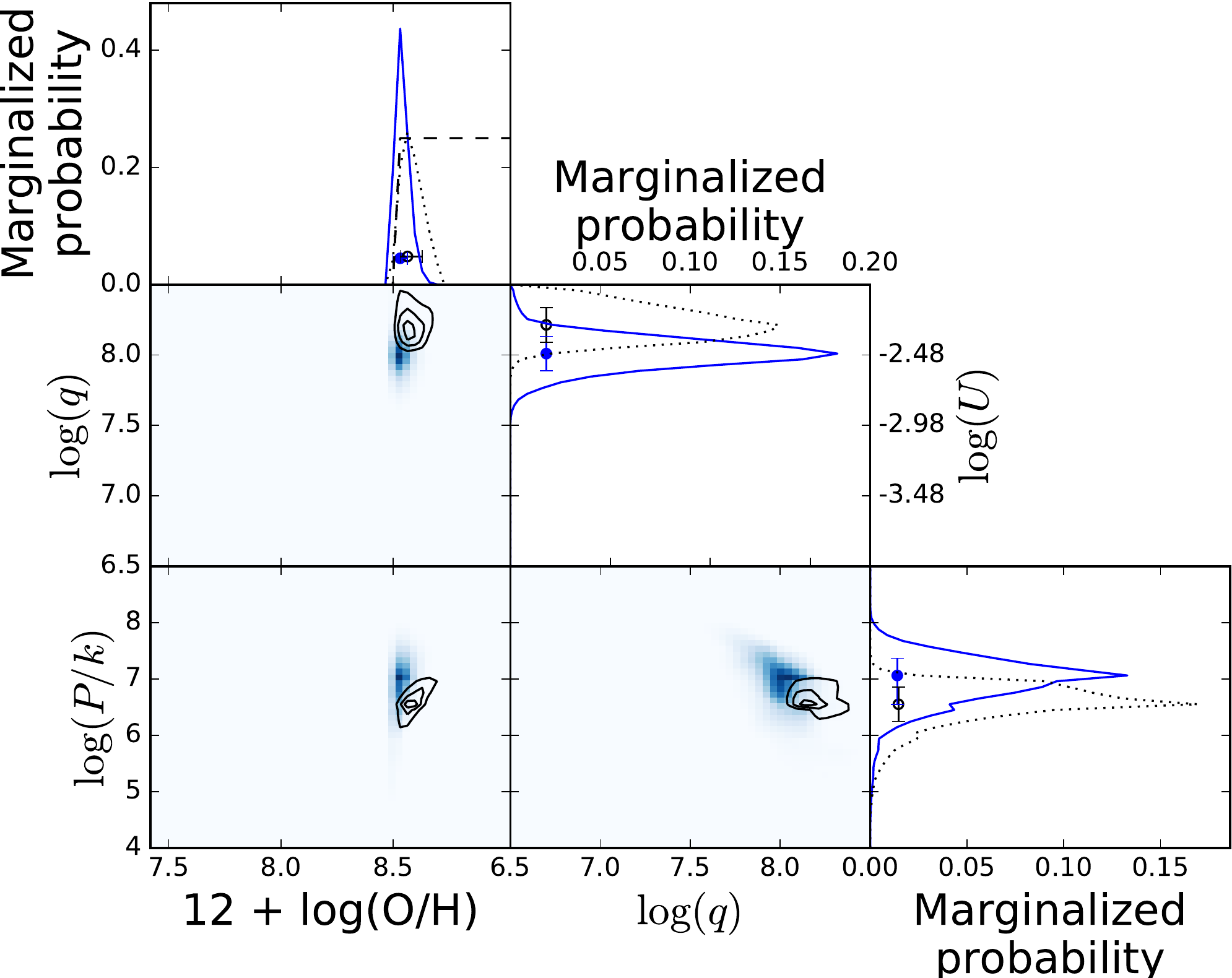}
		\caption{Only optical emission lines used}
		\label{fig: pizi_prior_onlyopt}
	\end{subfigure}	
	\vspace{0.5cm}	
	\begin{subfigure}[t]{0.47\textwidth}
		\centering
		\includegraphics[scale=0.42, trim=0.cm 0.0cm 0.0cm 0.0cm,clip=true,angle=0]{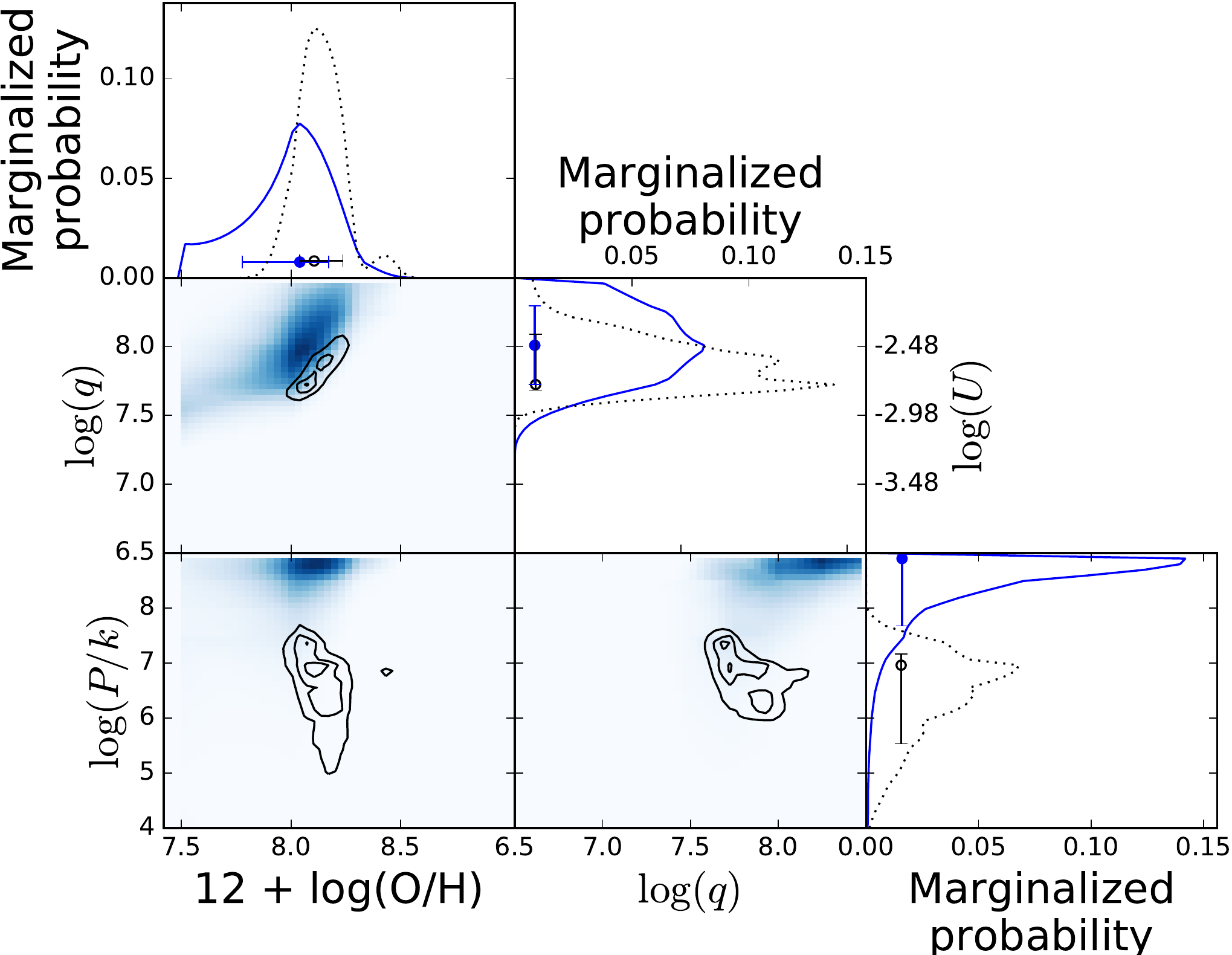}
		\caption{Only UV emission lines used}
		\label{fig: pizi_noprior_onlyUV}
	\end{subfigure}
	\hspace{0.5cm}
	\begin{subfigure}[t]{0.47\textwidth}
		\centering
		\includegraphics[scale=0.42, trim=0.cm 0.0cm 0.0cm 0.0cm,clip=true,angle=0]{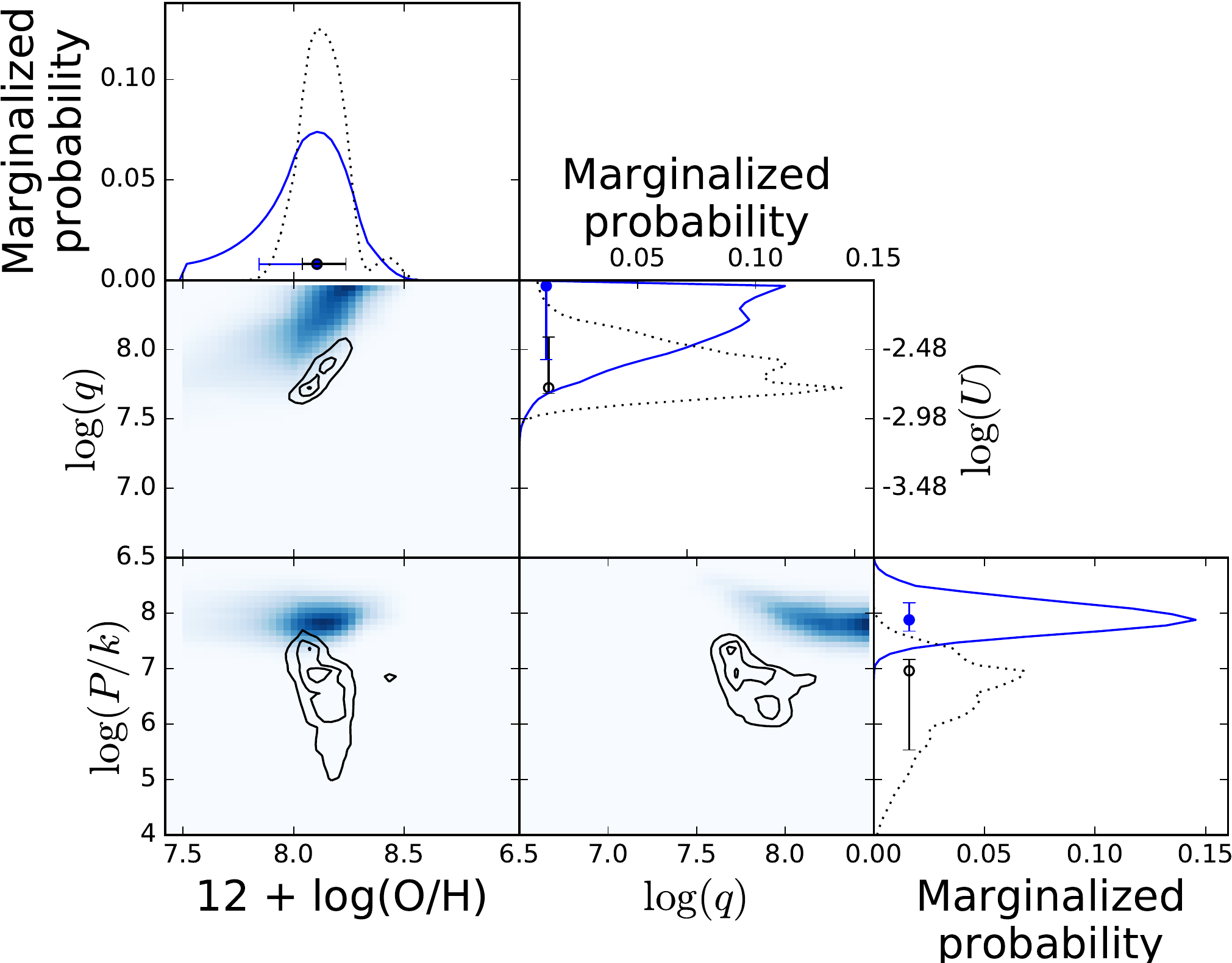}
		\caption{Only UV and [\ion{O}{ii}]$\lambda\lambda$3727,9 emission lines used}
		\label{fig: pizi_noprior_onlyUV+O2}
	\end{subfigure}

	\caption{
	Results from 100 iterations of IZIP (Section~\ref{izidiag}) with uniform priors on {\logq} and {\lpok}. Wherever both [\ion{N}{ii}]$\lambda$6584 and H$\alpha$ are available, we use a top-hat prior on metallicity (black dashed line) such that {\logOH} > 8.52 if [\ion{N}{ii}]$\lambda$6584/H$\alpha$ > 0.0776 and vice versa. For cases where at least one of the lines is unavailable we use a uniform prior on $Z$. In each group of plots: the bottom-left plot shows the 2D joint PDF for ISM pressure on the y-axis and metallicity on the x-axis, the middle-left plot shows the 2D joint PDF for ionisation parameter (y-axis) and metallicity (x-axis) and bottom-middle plot denotes that for ISM pressure (y-axis) and ionisation parameter (x-axis). The remaining plots show the 1D marginalized posterior PDFs for metallicity (top), ionisation parameter (middle) and ISM pressure (bottom). The blue circle represents the peak of the marginalised distribution, with error bars being the 16$^{th}$ and 84$^{th}$ percentiles. The top-left group of plots denote results when all rest-frame UV and optical emission lines are used by IZIP. The top-right and bottom-left groups of plots show the results when IZIP works on only the rest-frame optical and only rest-frame UV line measurements, respectively. The bottom-right group of plots corresponds to the scenario where only rest-frame UV and [\ion{O}{ii}]$\lambda\lambda$3727,9 doublet were provided to IZIP. Fe or Mg emission lines have not been included in the IZIP analysis, as discussed in Section~\ref{izidiag}. The black contours and dotted histograms denote the PDFs (2D and 1D respectively) of the corresponding fiducial cases i.e. when all the emission lines are used. By definition, the fiducial case for Figure~\ref{fig: pizi_prior_onlyopt} is Figure~\ref{fig: pizi_prior_all}. For Figures~\ref{fig: pizi_noprior_onlyUV} and \ref{fig: pizi_noprior_onlyUV+O2}, we consider the configuration using all the lines but not the [\ion{N}{ii}]/H$\alpha$ prior, as the fiducial case (Figure~\ref{fig: pizi_noprior_all}). We demonstrate that the ISM properties are well constrained on using all the lines or only the rest-frame optical lines. Using UV lines in addition to opitcal lines generally improves the constraints but the UV lines alone find it difficult to constrain metallicity and pressure. Using the [\ion{O}{ii}] doublet in addition to the UV lines helps constrain the ISM pressure.
}
\label{fig: pizi_prior}
\end{figure*}

\begin{figure*}
	\centering
	\begin{subfigure}[t]{0.47\textwidth}
		\centering
		\includegraphics[scale=0.42, trim=0.cm 0.0cm 0.0cm 0.0cm,clip=true,angle=0]{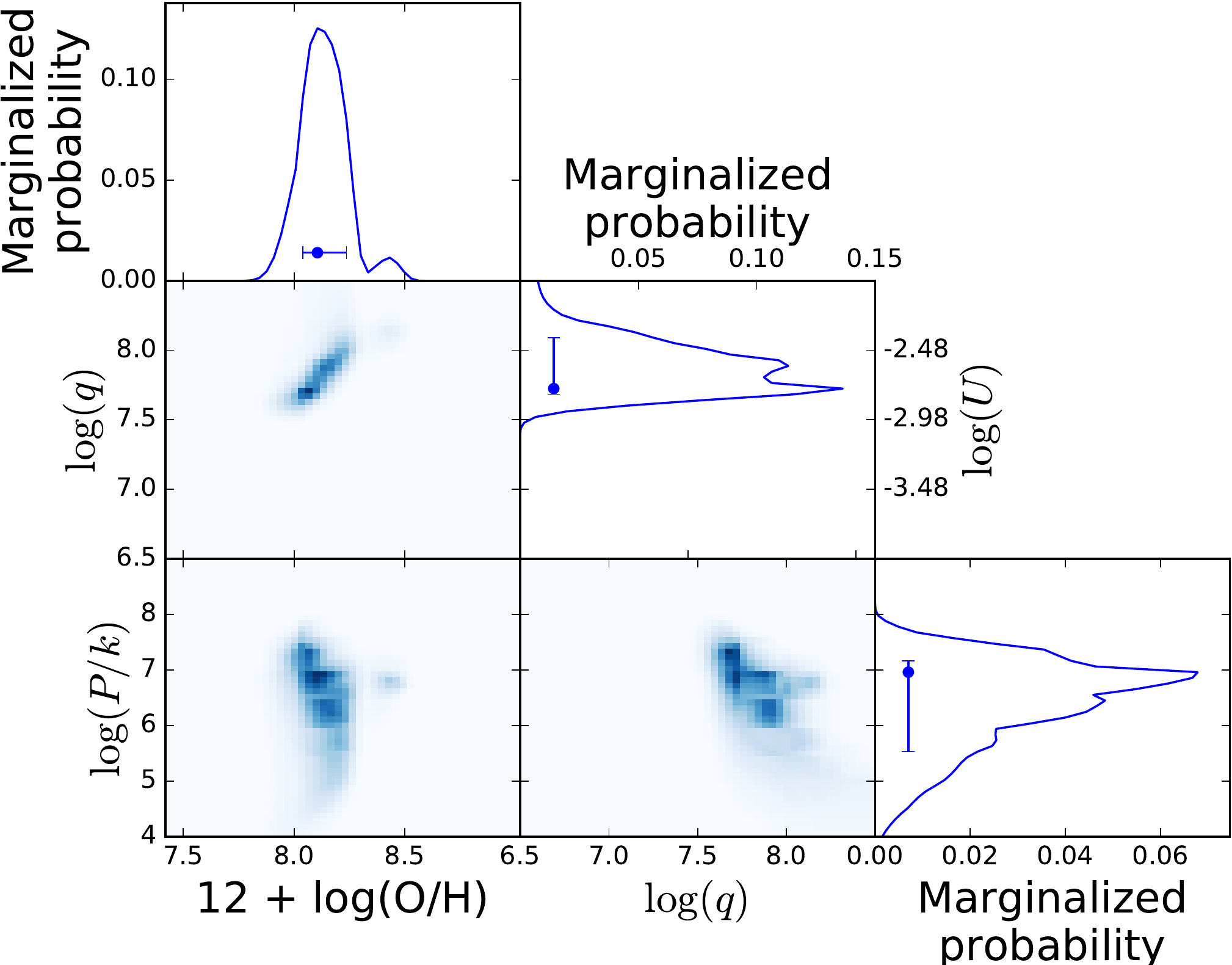}
		\caption{All emission lines used}
		\label{fig: pizi_noprior_all}
	\end{subfigure}
	\hspace{0.5cm}
	\begin{subfigure}[t]{0.47\textwidth}
		\centering
		\includegraphics[scale=0.42, trim=0.cm 0.0cm 0.0cm 0.0cm,clip=true,angle=0]{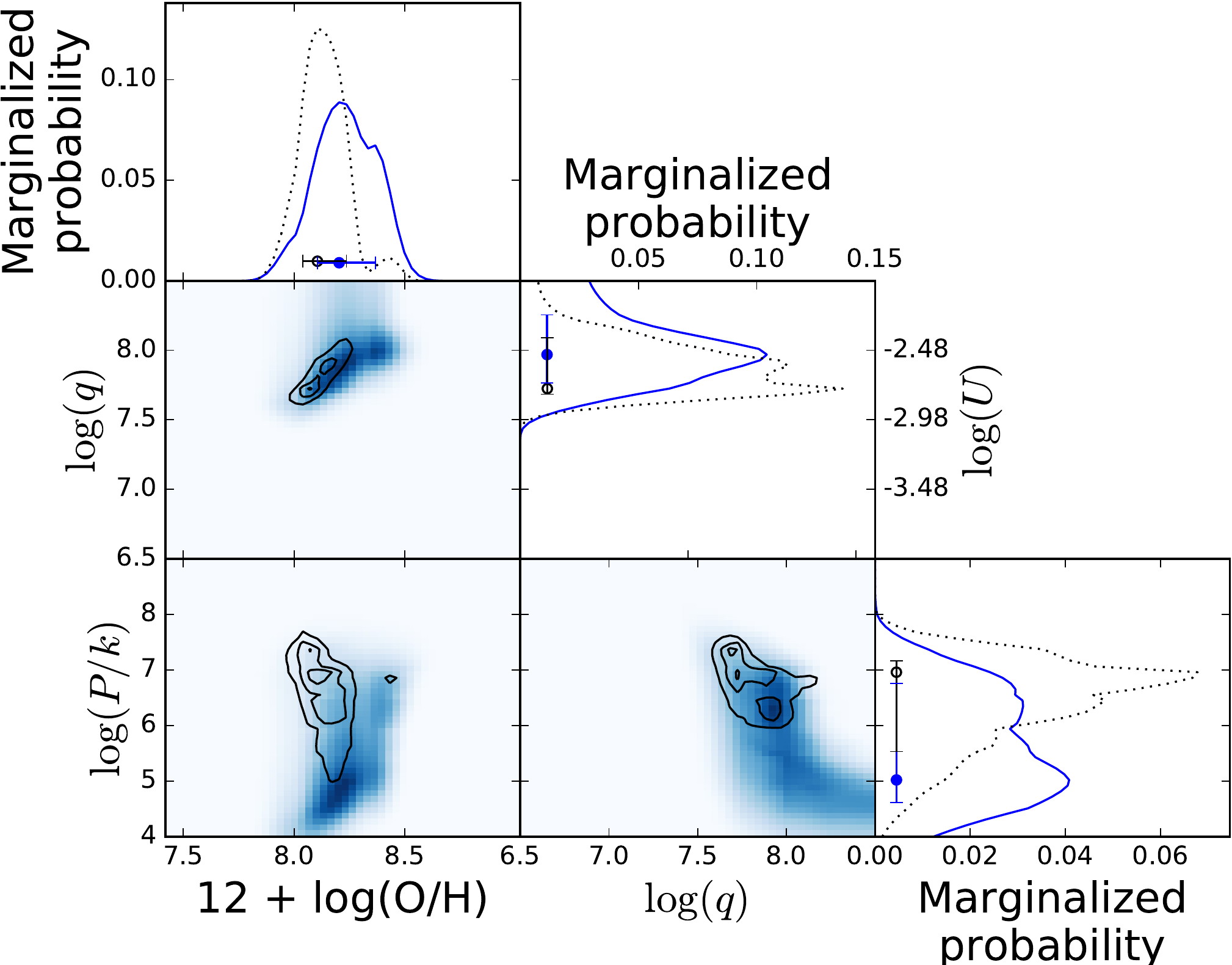}
		\caption{Only optical emission lines used}
		\label{fig: pizi_noprior_onlyopt}
	\end{subfigure}
	
	\caption{
		Same as in Figure~\ref{fig: pizi_prior} but here we use a flat prior on $Z$ even if we include the rest-frame optical emission lines. Figure~\ref{fig: pizi_noprior_all} is shown on Figure~\ref{fig: pizi_noprior_onlyopt} with black contours and dotted histograms for visual aid. On using only the optical lines, we infer a lower ($\sim$2 dex), poorly constrained {\lpok} (right) with a flat prior on $Z$ as compared to using a top-hat prior on $Z$ (Figure~\ref{fig: pizi_prior_onlyopt}).
	}
	\label{fig: pizi_noprior}
\end{figure*}

\subsection{Results from Bayesian methods}\label{sec:izip_res}
We describe our results from the Bayesian method, including the cases where different sets of emission lines were provided to IZIP. Figures~\ref{fig: pizi_prior} and \ref{fig: pizi_noprior} show the marginalised PDFs for the different physical parameters and the last section of Table~\ref{tab:phys_list} quotes the corresponding peak values. Figure~\ref{fig: pizi_noprior_all} represents the case where all available emission lines were provided to IZIP. For panels Figures~\ref{fig: pizi_noprior_onlyopt} and \ref{fig: pizi_noprior_onlyUV} only rest-frame optical and only rest-frame UV (Table~\ref{tab:pizi_lines}) nebular emission lines were used, respectively. The shaded plots denote the 2D PDF (marginalised over the third parameter) and the solid curves denote the 1D PDF for the corresponding physical parameter. The filled circles at the base of each 1D PDF show the peak value of the PDF, with errorbars being the 16$^{th}$ and 84$^{th}$ perecentiles.

\subsubsection{Ionization parameter}
The ionisation parameter, {\logq} = 8.21$^{+0.12}_{-0.12}$ derived by IZIP using all the emission lines (Figure~\ref{fig: pizi_noprior_all}), is $\sim$0.4 dex higher than the mean {\logq} = 7.77$^{+0.01}_{-0.01}$ derived using the emission line diagnostics. The mean {\logq} from the rest-frame optical strong line diagnostics {\logq} =  7.76$^{+0.01}_{-0.01}$ agrees at 1$\sigma$ level of the peak {\logq} = 8.01$^{+0.12}_{-0.12}$ inferred by IZIP using only the rest-frame optical spectra. The {\logq} = 8.01$^{+0.29}_{-0.29}$ inferred by IZIP while using only the UV spectra agrees with the {\logq} = 7.99$^{+0.07}_{-0.07}$ derived using the rest-frame UV diagnostic of K18, within 1$\sigma$ uncertainties. This agreement demonstrates that, at least for the test-case of {\knotE}, the Bayesian approach can be used to reliably determine the ionisation parameter when only the UV lines are available.

\subsubsection{ISM pressure}
IZIP yields a low {\lpok} $\sim$7 (Figure~\ref{fig: pizi_prior_onlyopt}) when only the optical emission lines are provided. However, using only the UV emission lines results in the {\lpok} PDF saturating at the edge of the model parameter space (Figure~\ref{fig: pizi_noprior_onlyUV}). Hence, UV lines alone fail to constrain {\lpok} in the case of {\knotE}. However, the inclusion of the [\ion{O}{ii}]$\lambda\lambda$3727,9 doublet with the UV lines (Figure~\ref{fig: pizi_noprior_onlyUV+O2}), constrains the ISM pressure fairly well ($\sigma <$ 0.4) at a value {\lpok} = $\sim$7 in spite of the inclusion of the high ionisation UV species. This is because the [\ion{O}{ii}]$\lambda\lambda$3727,9 doublet has much higher SNR than the UV lines and hence gets higher weight during the Bayesian analysis. In contrast, when only the rest-optical lines are used, models predict that the [\ion{O}{ii}] lines have comparable strength as the [\ion{S}{ii}]$\lambda\lambda$6717,31 doublet in the high-metallicity regime (such as in {\knotE}) and hence it is not obvious that [\ion{O}{ii}] would dominate the Bayesian inference. Moreover, on using the [\ion{O}{ii}] lines along with the UV lines (Figure~\ref{fig: pizi_noprior_onlyUV+O2}), the PDF is shifted down by $\sim$1.6 dex, implying that the [\ion{O}{ii}] lines probe a lower ISM pressure than the higher ionised UV species (see Section~\ref{disc_press}). Therefore, the inclusion of [\ion{O}{ii}] helps to constrain the pressure but also biases {\lpok} towards lower values than probed by the UV lines.

\subsubsection{Oxygen abundance}
IZIP infers {\logOH} = 8.56$^{+0.07}_{-0.03}$ when the full suite of UV and optical spectra is used (top panel, Figure~\ref{fig: pizi_prior_all}), which agrees within 1$\sigma$ of the mean abundance of all the individual diagnostics. The oxygen abundance is poorly constrained by IZIP when only the rest-frame UV spectra is used, suggesting that it is difficult to constrain the abundance using only UV emission lines. The uncertainties in the Extreme Ultraviolet (EUV) radiation field that is used to produce the {\htr} models is a potential cause for this difficulty. Providing IZIP with only the UV lines and the Balmer lines does not help improve constraints on $\log{(\mathrm{O/H})}$ (see Appendix~\ref{app:izip}). Our work implies that rest-frame optical emission lines are necessary to determine the metallicity. It is difficult to obtain reliable metallicity estimates if only rest-frame UV spectra are present. However, we will further investigate the UV sensitivity to metallicity in our future papers (Acharyya {\etal}, Kewley {\etal}, Byler {\etal} in prep).

\section{Discussion}\label{disc}
In this section we compare our results based on the availability of different parts of the spectra -- UV, optical, UV-optical -- in regards to the Bayesian approach. We also discuss our results in the context of previous studies, the implications of our work on upcoming telescopes, and potential caveats.

\subsection{Comparison Between UV, Optical or UV-optical Bayesian Results}
\paragraph*{Ionization parameter:}
Ionization parameter values are well constrained when only optical emission lines are provided to IZIP, along with an user defined prior on $Z$ based on [\ion{N}{ii}]$\lambda$6584/H$_{\alpha}$ (Figure~\ref{fig: pizi_prior_onlyopt}). However, inferring {\logq} from only the optical lines or only the UV lines, in absence of a prior, show a broad tail (small likelihood) of high {\logq} values and hence put a poorer constraint on {\logq}. The bias towards high {\logq} is more pronounced when the [\ion{O}{ii}]$\lambda$3727 doublet is included in addition to the UV lines, leading to an unconstrained PDF. The $\sim 0.2$ dex higher {\logq} inferred by using all the emission lines, compared to when only the UV or optical lines are used, is not concerning given that the values still agree within 1$\sigma$. The blue and black dotted histograms for {\logq} in Figures~\ref{fig: pizi_prior_onlyopt}, \ref{fig: pizi_noprior_onlyUV} and \ref{fig: pizi_noprior_onlyopt} have considerable overlap to demonstrate the agreement within the uncertainties. We conclude that adding the rest-frame UV information to the existing optical spectra, puts a better ($\sigma \sim$0.2) constraint on {\logq} overall, than when only the optical spectra are used. The UV lines by themselves barely constrain {\logq}, albeit with a larger uncertainty than when all lines are used.

\paragraph*{ISM Pressure:}
In the event of only the UV lines being available, the Bayesian approach would give an ISM pressure value of {\lpok} = 8.90$^{+0.10}_{-1.22}$ whereas, inclusion of the [\ion{O}{ii}]$\lambda\lambda$3727,9 lines would yield a lower {\lpok} = 7.88$^{+0.31}_{-0.20}$). Both these values agree (within 1$\sigma$ uncertainties) with {\lpok} = 8.8$^{+0.2}_{-0.6}$ and {\lpok} = 7.4$^{+0.6}_{-0.2}$ derived from the individual UV and optical emission line diagnostics, respectively. This demonstrates that with a rest-frame UV coverage up to the [\ion{O}{ii}]$\lambda\lambda$3727,9 lines, it is possible to effectively probe the different physical regions in the nebula using the Bayesian approach. Suitably designed future surveys with the {\JWST} should take advantage of this fact.

\paragraph*{Oxygen abundance:}
 When only optical lines are used, IZIP shows a double peaked PDF (Figure~\ref{fig: pizi_noprior_onlyopt}), with the stronger peak being at {\logOH} = 8.20$^{+0.16}_{-0.10}$, reflecting the fact that the optical collisionally excited emission lines are doubled valued with metallicity. For the case where only UV lines are used (Figure~\ref{fig: pizi_noprior_onlyUV}, \ref{fig: pizi_noprior_onlyUV+O2}), the abundance PDF has a broad low metallicity tail, with a peak at {\logOH} = 8.04$^{+0.13}_{-0.26}$. Providing IZIP with the [\ion{O}{ii}]$\lambda$3727 doublet along with the UV lines (Figure~\ref{fig: pizi_noprior_onlyUV+O2}) helps constrain the oxygen abundance to a slightly ($\sim$0.07 dex) higher value {\logOH} = 8.11$^{+0.13}_{-0.26}$ but does not lead to better (narrower) constraints. The broader {\logOH} PDFs when only UV lines are used, imply that it is difficult to infer the oxygen abundance with only the rest-frame UV lines using a Bayesian approach.

\subsection{Comparison With Other Work}

In this section we compare the physical properties of {\knotE} with those of other galaxies from the literature, over a wide range of redshifts. However, one should bear in mind that {\knotE} is a $\sim$100 pc region of vigorous star-formation whereas most of the literature data correspond to spatially integrated spectra of entire galaxies that include both star-forming regions as well as passively evolving stellar populations. As such, some ISM parameters e.g. {\logq} are expected to be higher in {\knotE} than the spatially averaged properties of other galaxies.

\paragraph*{Electron temperature:}
\cite{vanZee:2006aa} studied a sample of nearby dwarf galaxies and reported the ISM {\Te} $\sim$ 1.3$\times$10$^4$ K. \cite{Jones:2015ab} obtained {\Te} $\le$2$\times$10$^4$ K for a sample of 32 $z\sim$0.8 galaxies. On the other hand, \cite{Yuan:2009aa} reported a {\Te} = 2.3$\times$10$^4$ K for a lensed galaxy at a redshift $z \sim $1.7. \citet{Christensen:2012aa} estimated $1.3 \times$10$^4$ K $\gtrsim$ {\Te}$ \gtrsim 2.7\times$10$^4$ K for three lensed galaxies in the redshift range $2 \gtrsim z \gtrsim 3.5$. Further adding to the sample of lensed galaxies, \citet{Stark:2013ab} and \citet{James:2014aa} estimated {\Te} $\sim 1.5$ and $1.7 \times$10$^4$ K for two galaxies at $z \sim 1.4$. \citet{Steidel:2014aa} reported a mean {\Te} of $\sim$ 1.3$\times$10$^4$ K for 3 KBSS-MOSFIRE galaxies at redshift $\sim 2$. \citet{Sanders:2016ab} estimated an [\ion{O}{iii}] T$_e$ = 1.4$^{+0.20}_{-0.14}$ $\times$ 10$^4$ K, and \citet{Bayliss:2014aa} derived an upper limit on {\Te} $\leq$1.4$\times$ 10$^4$ K for two galaxies at $z\sim$3 and 3.6, respectively. Thus, {\knotE} has a [\ion{O}{iii}] T$_e$ (1.2$^{+0.2}_{-0.1}$ $\times 10^4$ K) similar to that of local galaxies, and marginally lower than that of $z\sim 2-3$ galaxies. 
 
 \paragraph*{Ionization parameter:}
 We derive a weighted mean {\logq} = 7.77 $\pm$ 0.01 for {\knotE}, averaging over all the diagnostics. The mean {\logq} agrees with the study of {\z} $\sim$2-3 galaxies by \cite{Steidel:2014aa} where they reported {\logq} values between 7.6 and 8.7 using CLOUDY photoionisation models. Moreover, \citet{Kaasinen:2018aa} analyse spectra from $\sim$200,000 SDSS galaxies and derive a mean {\logq} = 7.4 (using both \citetalias{Kobulnicky:2004aa} and IZI), further supporting the scenario that ionisation parameter at high ($z\sim2$) redshift is higher than that in the local universe \citep{Steidel:2014aa, Strom:2017aa, Kewley:2015ab}.
 
 \paragraph*{Electron density:}
 \citetalias{Sanders:2016aa} measured a mean $\log{(n_e/\mathrm{cm}^{-3})}$ $\approx$ 2.4 for 225 star-forming galaxies at $z \sim 2.3$ with median stellar mass of M$_*$ = 10$^{10}$\,{\Msun} and median SFR = 21.6\,{\Msunpyr}. In agreement with \citetalias{Sanders:2016aa}, \citet{Strom:2017aa} reported a mean $\log{(n_e/\mathrm{cm}^{-3})}$ $\approx$ 2.44 for another high-redshift ($z \simeq 2 - 3$) sample of $\sim$ 380 galaxies, whereas \citet{Kennicutt:1984aa} measured electron densities between 2 $\le \log{(n_e/\mathrm{cm}^{-3})} \le$ 3 for {\htrs} in nearby galaxies. The \citet{Strom:2017aa} sample consists of star-forming galaxies with M$_*$ = 10$^9$ -- 10$^{11.5}$\,{\Msun} and SFR = 3 -- 1000\,{\Msunpyr}, which encompasses the properties of {\galaxy} (M$_*$ = 10$^{10}$\,{\Msun} and SFR = 60{\Msunpyr}). The local galaxies of the \citet{Kennicutt:1984aa} sample, however, are quiescent galaxies spanning M$_*$ = 10$^8$ -- 10$^{11}$\,{\Msun} and SFR = $\leq$ 1\,{\Msunpyr}. All of these studies used rest-frame optical spectra to determine {\Ne}. {\knotE} has a slightly higher electron density ($\log{(n_e)}\simeq$ 3.04) than both the local and $z \simeq 2 - 3$ galaxies. \citet{Kaasinen:2017aa} study a sample of z$\sim$1.5 galaxies with median M$_*$ = 10$^{7.5}$\,{\Msun} and SFR = 15\,{\Msunpyr} and report $\log{(n_e)}\simeq$ 2.05. Their M$_*$- and SFR-matched local analogs yield $\log{(n_e)}\simeq$ 1.99. {\galaxy} being more massive and more rapidly star-forming than the z$\sim$1.5 sample, has a higher ($\sim$1 dex) electron density.
 
 \begin{figure}
 	\centering
 	\includegraphics[scale=0.45, trim=0.0cm 0.0cm 0.0cm 0.0cm,clip=true,angle=0]{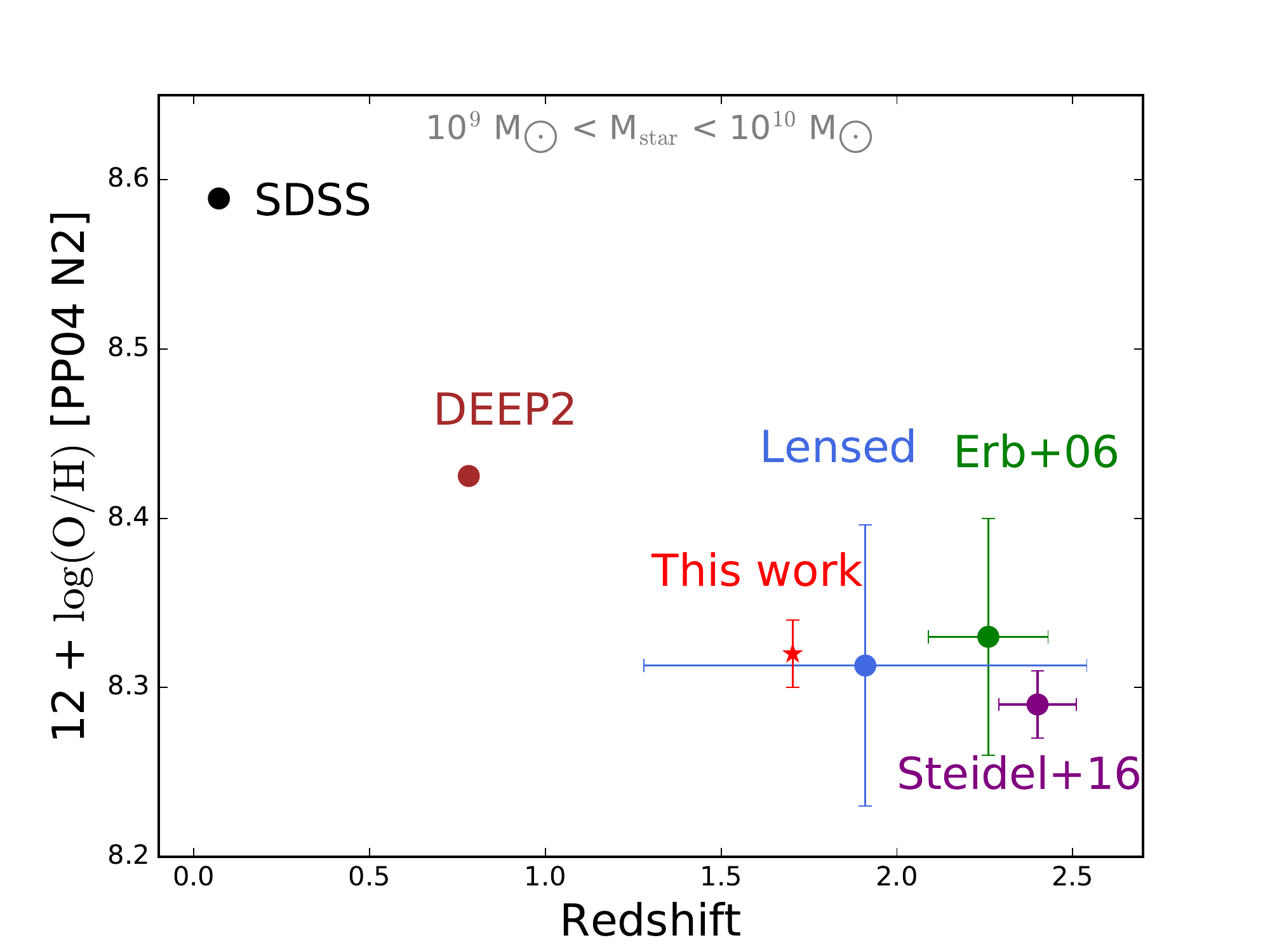}
 	\caption{
 		Redshift evolution of oxygen abundance in the stellar mass bin 10$^{9}$ {\Msun} < M$_*$ < 10$^{10}$ {\Msun}, based on \citet{Yuan:2013ab}. \citet{Yuan:2013ab} obtained the SDSS (black) and DEEP2 (brown) data from \citet{Zahid:2011aa} and the green point from the UV-selected \citet{Erb:2006aa} galaxies. The `Lensed' (blue) data denotes the mean abundance for the lensed sample in \citet{Yuan:2013ab}, which also includes measurements from \citet{Yuan:2011aa}, \citet{Wuyts:2012ab} and \citet{Richard:2011aa}. The {\logOH} we derived using the [\ion{N}{ii}]$\lambda$6584/H$\alpha$ diagnostic of \citetalias{Pettini:2004qe} (PPN2) is shown as a star, while all the other mean abundance values have been taken from Table 2 of \citet{Yuan:2013ab} and are denoted by circles. We show only the PPN2 diagnostic in this plot in order to be consistent with \citet{Yuan:2013ab}. The uncertainties quoted for the SDSS and DEEP2 samples by \citet{Yuan:2013ab} is the 1$\sigma$ standard deviation of the mean from bootstrapping. Uncertainties thus derived are too small to be visible on this scale, because the surveys comprise of a large number of galaxies. In addition to Table 2 of \citet{Yuan:2013ab}, we also include {\logOH} measurement of the composite spectrum from \citet{Steidel:2016aa} as a purple circle.
 	}
 	\label{fig:zcomp}
 \end{figure}
 
 \paragraph*{Oxygen abundance:}Figure~\ref{fig:zcomp} shows the evolution of oxygen abundance as a function of redshift and where, in that evolutionary track, {\galaxy} lies with its stellar mass content of $<$10$^{10}$ {\Msun} \citep{Wuyts:2014aa}. We adopt the mean abundance values in Table 2 of \cite{Yuan:2013ab} to compare with our work. The {\logOH} = 8.32 $\pm$ 0.02 for {\knotE} derived from the [\ion{N}{ii}]$\lambda$6584/H$\alpha$ diagnostic of \citetalias{Pettini:2004qe} (PPN2) is shown Figure~\ref{fig:zcomp} (red star) in order to be consistent with \cite{Yuan:2013ab}. The abundance of {\knotE} (z$\sim$1.7), based on PPN2, is also comparable to those of other high-redshift studies like \citet[][z$\sim$2.0]{Jones:2010qy}, \citet[][z$\sim$2.0]{Shapley:2004aa} and \citet[][z$\sim$2.3]{Steidel:2014aa} that report {\logOH} $\simeq$8.4-8.5 using PPN2. The metallicity of {\knotE} is $\sim$0.2 dex lower than that of the local galaxies of KINGFISH survey, which are reported to have {\logOH} $\sim$ 8.5 (converted to PPN2 frame using \citetalias{Kewley:2008aa}) by \cite{Kennicutt:2011aa}. Using Bayesian methods, \citet{Kaasinen:2018aa} find {\logOH} = 8.7 for local galaxies and {\logOH} = 8.0 for redshift z$\sim$1.5 galaxies, in the mass bin 9 < $\log{(M_*/M_{\bigodot})}$ < 10. Bayesian (using IZIP) abundance estimates for {\knotE} are $\sim$0.2 dex lower and $\sim$0.5 dex higher than the low and high redshift measurements of Kaasinen {\etal}, respectively. Thus, {\knotE} is consistent with the trend of decreasing oxygen abundance with redshift.
 
\subsection{Implications for \JWST}
In this work, we have used the \ion{O}{iii}]$\lambda\lambda$1660,6 lines in the UV (along with [\ion{O}{iii}]$\lambda$5007) to directly determine the oxygen abundance from {{\Te}}. We have also inferred the abundance using a Bayesian approach by providing the UV line and [\ion{O}{iii}]$\lambda$5007 to IZIP (see Appendix~\ref{app:izip}). From both these tests we conclude that it is difficult to reliably estimate metallicity with only the rest-frame UV lines -- the [\ion{O}{iii}]$\lambda$5007 line is required to break the degeneracy. \JWST/NIRSpec will be able to simultaneously capture the [\ion{O}{iii}]$\lambda$5007 line and the [\ion{O}{iii}]$\lambda\lambda$1660,6 doublet within redshift $3.5 \lesssim z \lesssim9$, thereby making it possible to obtain reliable estimates of oxygen abundance at such high redshifts. Currently, rest-frame UV nebular emission line diagnostics for oxygen abundance are scarce. However, \citetalias{Kewley:2019ab} and Byler {\etal} (in prep) investigate the diagnostic power of the UV lines, in detail. 

We demonstrate that {\logq} can be determined using only rest-frame UV spectra, provided at least one of [\ion{C}{iii}]$\lambda\lambda$1906,8/[\ion{C}{ii}]$\lambda$1335, [\ion{C}{iii}]$\lambda\lambda$1906,8/[\ion{C}{ii}]$\lambda$2325, or [\ion{O}{iii}]$\lambda\lambda$1660,6/[\ion{O}{ii}]$\lambda$2470 ratios is available (Section~\ref{qdiag_uv}). These ratios will be within the wavelength coverage of \JWST/NIRSpec for redshifts above $z \sim$ 3.5, 1.6, and 2.6, respectively. However, the [\ion{C}{ii}]$\lambda$1335 group is very weak ($\sim$10$^{-3}$ - 10$^{-4} \times$ H$\beta$ flux for {\logq}=8.0 and $Z$=0.4\,$Z_{\bigodot}$). The unresolved [\ion{C}{ii}]$\lambda$2325 group of lines is very closely spaced in wavelength, thereby making it difficult to estimate the fluxes for moderate resolution spectra. Large uncertainties stemming from line blending could potentially be translated into uncertainties in {\logq}.

\subsection{IZIP with and without priors}\label{sec:disc_pizi}
A point to note about the Bayesian inference method is that collisionally excited emission lines are double valued with metallicity. IZIP, by default, is unable to make an informed choice between the two branches based on specific emission line ratios, and yields bimodal PDFs in many such cases. For instance, one could not use the [\ion{N}{ii}]$\lambda$6584/H$\alpha$ ratio to break the degeneracy while using the R$_{23}$ metallicity indicator. Instead, IZIP would use the R$_{23}$ line flux information and weigh them same as all the other lines, thus leading to a double peaked metallicity PDF. This is not wrong from a Bayesian perspective, because the models indeed predict two probable values of $\log{(\mathrm{O/H})}$ given the emission line ratios. However, the observed {\htr} can only have one or the other abundance; this is where an observer would use other emission line information to decide between the two branches. To facilitate this, IZIP takes into account user defined priors on the physical parameters while computing the posterior distribution. Providing user-defined priors to IZIP based on certain line ratios can help break the degeneracy.

We impose a top-hat prior on the [\ion{N}{ii}]$\lambda$6584/H$\alpha$ ratio, where available (discussed in Section~\ref{izidiag}), to help select the relevant metallicity branch. In the absence of a prior, we infer {\logOH} = 8.53$^{+0.1}_{-0.07}$ when the full suite of UV and optical spectra is used (Figure~\ref{fig: pizi_noprior_all}); and {\logOH} = 8.20$^{+0.16}_{-0.1}$ when only the rest-frame optical lines are used (Figure~\ref{fig: pizi_noprior_onlyopt}). This discrepancy in $\log{(\mathrm{O/H})}$ based on whether UV lines are included could be potentially due to inconsistencies in photoionzation models, as discussed in Section~\ref{sec:caveat}. On using the prior on [\ion{N}{ii}]/H$\alpha$, the oxygen abundance is well constrained to a peak value of {\logOH} = 8.56$^{+0.06}_{-0.03}$ (Figure~\ref{fig: pizi_prior_all}) and {\logOH} = 8.53$^{+0.03}_{-0.03}$ (Figure~\ref{fig: pizi_prior_onlyopt}) by using UV-optical and only optical lines, respectively. In Figure~\ref{fig: pizi_prior_onlyopt}, the top-hat prior used to derive the posterior probability is shown by a black dashed line in the 1D PDF of the metallicity (top) panel. Using the [\ion{N}{ii}]/H$\alpha$ prior leads to narrow constraints, as expected, and closer agreement between the abundances measured with different set of emission lines. Comparing Figure~\ref{fig: pizi_prior_all} with \ref{fig: pizi_noprior_all} and Figure~\ref{fig: pizi_prior_onlyopt} with \ref{fig: pizi_noprior_onlyopt} reveals that usage of a prior on abundance, also has considerable impact on {\logq} and {\lpok}. The high {\logq} and low {\lpok} solution is eliminated on using the prior, leading to much tighter constraints on both these physical properties. We therefore recommend employing priors where possible (e.g. [\ion{N}{ii}]/H$\alpha$ available), to break the degeneracy while using IZIP.

\subsection{ISM pressure -- UV vs optical}\label{disc_press}
\begin{figure}
	\centering
	\includegraphics[scale=0.6, trim=0.3cm 0.0cm 0.5cm 0.0cm,clip=true,angle=0]{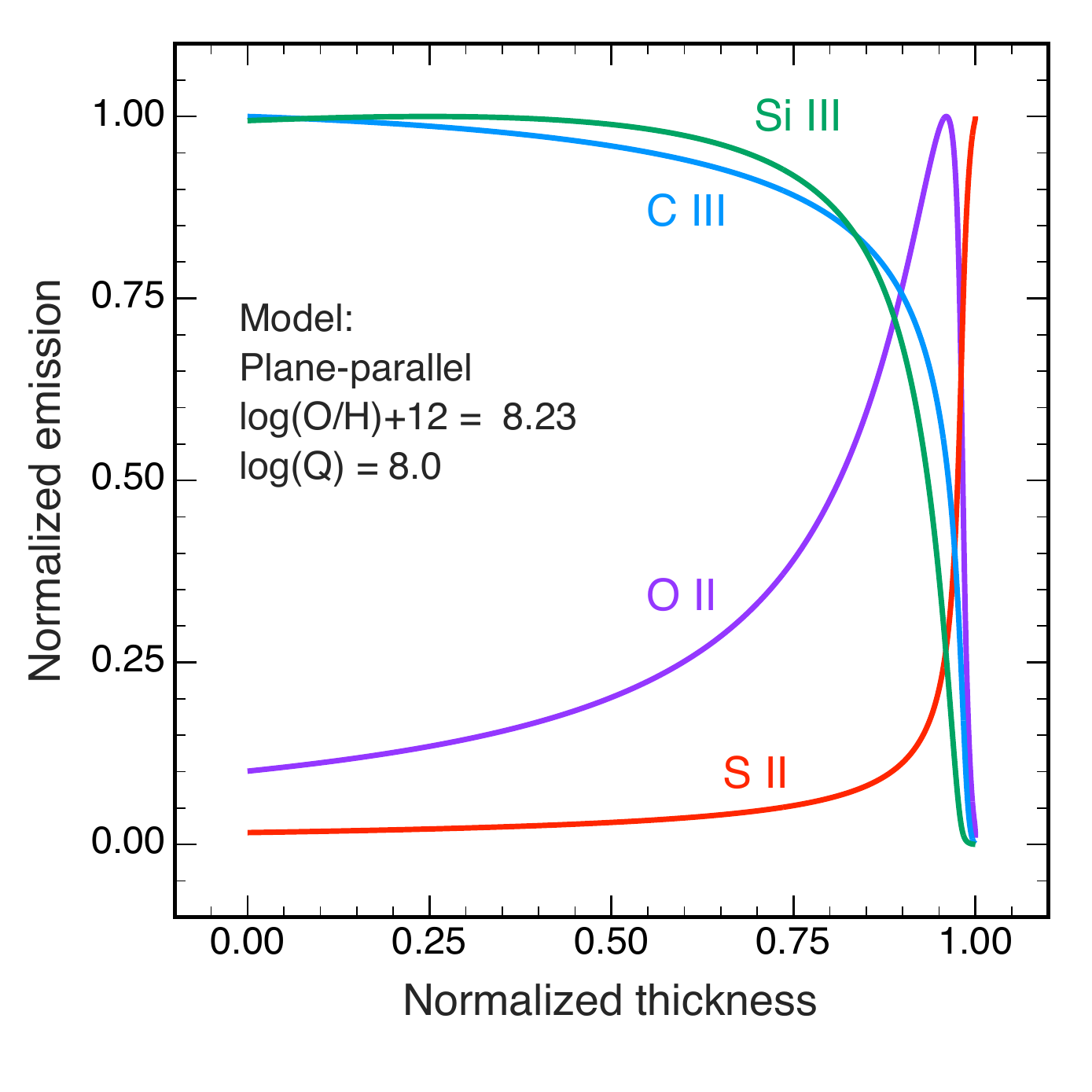}
	\caption{
		Normalised strength of different emission line species as a function of normalised shell thickness. The high ionisation species \ion{C}{iii} and \ion{Si}{iii} originate from inner, denser zones of the nebula whereas the lower ionisation \ion{O}{ii} and \ion{S}{ii} ions trace the less dense peripheral part of the ionised nebular shell. The curves correspond to a MAPPINGS photoionisation {\htr} model, assuming plane parallel geometry, for {\logOH} = 8.23 and {\logq} = 8.0.
	}
	\label{fig:ion_zone}
\end{figure}

ISM pressures measured using the new UV diagnostics are $\sim 1.5$ dex higher than those determined using optical line ratios. This discrepancy can be attributed to the fact that the different emission lines originate from different regions of the ionised ISM. As \citetalias{Kewley:2019aa} point out, the regions closer to the ionising source receive a greater proportion of the UV photon flux resulting in a greater fraction of higher ionisation species. Consequently, the higher ionisation emission lines in the rest-frame UV probe the inner, high pressure region of the nebula, whereas the lower ionisation lines in the optical probe the outer, lower pressure zone. This is due to the different sensitivity of various emission species to different density regimes depending on critical density. It is therefore unsurprising that we would derive a higher pressure and higher density with the UV diagnostics. 

Figure~\ref{fig:ion_zone} shows the emissivity of various line species as a function of the depth into the nebula. Higher ionisation species (rest-frame UV diagnostics) used in this work -- [\ion{C}{iii}] and [\ion{Si}{iii}] are produced throughout the \ion{H}{ii} region, and concentrated in the inner regions ($\lesssim$70\% of nebular shell radius). In contrast, the lower ionisation species (rest-frame optical diagnostics) used in this work -- [\ion{O}{ii}] and [\ion{S}{ii}] are primarily produced in the outer parts of the nebula, specifically, from $\sim$80\% and $\sim$100\% of the nebular shell radius. We might expect that the [\ion{C}{iii}] and [\ion{Si}{iii}] species represent the average physical conditions within $\lesssim$70\% of nebular shell radius, whereas the [\ion{O}{ii}] and [\ion{S}{ii}] species probe the \ion{H}{ii} region at 80\% radius and extreme outskirts, respectively.

Our work demonstrates that, given only the rest-frame UV spectra, it is possible to derive {\lpok} (or {\Ne}) using the SEL diagnostics. Whether the Bayesian methods can reliably constrain {\lpok} using only the UV lines remains unclear. However, one has to bear in mind that the {\lpok} (or {\Ne}) derived from the SEL diagnostics would be representative of the inner regions of the nebula and consequently would be biased towards higher values. There is no clear choice for a ``better" diagnostic between the different UV and optical diagnostics. It is simply a case of different emission line species probing different physical regions in the nebulae. 

Rest-frame optical spectra, when available, probe the outskirts of the nebula. Additionally including rest-frame UV spectra probes the inner physical regions of the nebula as well. \citetalias{Kewley:2019aa} suggest using the \ion{Si}{iii}$\lambda$33$\mu m$ and \ion{Cl}{iii}$\lambda$5518\,\AA\ lines as diagnostics that are more representative of the entire nebula because these emissions originate fairly uniformly throughout the nebula. However, we do not detect either of these emission lines in {\knotE}.

The measured electron densities are not representative of the entire nebula either. Different emission lines have different critical densities, making them sensitive to different density regimes. \citetalias{Kewley:2019aa} point out that the density structure of an {\htr} can be quite complex. Moreover, clumpy star formation knots have been observed in {\galaxy}. As such, it is more sensible to measure the ISM pressure in this case than measuring a constant density which may not be representative of the entire ionised ISM.

The above discussion naturally raises a general (not specific to ISM pressure) concern regarding the combined usage of the emission lines originating in inner nebular regions with those emitted from the outer regions. While this is a valid concern, the extent to which such combined usage would make a difference depends on the particular emission line ratio and the ISM property involved. In case of {\lpok} and {\Ne}, all the lines involved in a particular diagnostic are of the same ionisation species which originate at similar physical regions, but different diagnostics have lines originating from different regions, indicating that each diagnostic clearly probes a different nebular region. As such, it is not sensible to combine the two groups -- UV and optical diagnostics -- for pressure and density measurements and should be considered as probes of distinctly different nebular regions. For abundance and ionisation parameter however, some diagnostics involve emission lines of different ionisation species e.g. the R23, O$_{32}$ and Ne3O2 ratios (see Table~\ref{tab:masterlist}). The [\ion{O}{iii}] lines originate throughout the nebula, implying that the R$_{23}$ and O$_{32}$ diagnostics are representative of the entire gas cloud. Thus, it is sensible to compare diagnostics involving [\ion{O}{iii}] with other diagnostics for {\logOH} or {\logq}. It is difficult to compare Ne3O2 index with other diagnostics as the former yields systematically different values of {\logOH} or {\logq}, due to reasons discussed in Section~\ref{analysis}. Moreover, these diagnostics are based on models for a single {\htr}. whereas in reality, a star-forming knot would comprise of several such {\htrs}. Thus, we probe the emission-weighted average properties of an ensemble of {\htrs}. We conclude that, it is possible to combine rest-frame UV and optical diagnostics; modulo the fact that they probe different physical regions in the case of pressure and density. We refer the reader to \citetalias{Kewley:2019ab}for a review of ionisation parameter and metallicity diagnostics.

\subsection{Caveats}\label{sec:caveat}
One potential reason for the discrepancy between the ISM pressure ($\sim$1.5 dex) and oxygen abundance ($\sim$0.5 dex) derived from UV and optical emission lines, is inconsistencies in the inputs to the \ion{H}{ii} model grid. The currently available stellar atmosphere libraries and stellar evolutionary tracks are based on different abundance standards and do not agree. The O-star models are only sparsely represented in the available model sets. Moreover, the stellar population synthesis model Starburst99 \citep{Leitherer:1999aa} linearly interpolates between these sparsely sampled libraries. 

A second, but less concerning, source of uncertainty comes from the scatter in the available atomic data, which leads to a variation in the ionising energies of crucial nebular emission lines (up to a factor of 2). Thus, there is an intrinsic uncertainty in the input stellar spectra. For instance, the stellar spectra obtained from Starburst99 in different spectral resolution modes do not agree in the rest-frame UV regime. Nevertheless, in the absence of more extensive stellar model sets, the Starburst99 spectra are used by the photoniozation code MAPPINGS v5.1 \citep{Dopita:2013aa} as the driving source of the radiation field leading to all the high ionisation emission lines. The inherent discrepancy in Starburst99 translates to uncertainties in output UV and optical emission line fluxes in the {\htr} model grids of MAPPINGS. The MAPPINGS model grids, in turn, have been used to calibrate the \citetalias{Kewley:2019aa} diagnostics. Thus, there are inherent discrepancies in the rest-frame UV diagnostics. Moreover, our Bayesian analysis is also based on the MAPPINGS model grids. This could potentially lead to discrepancies in the metallicity determined using only UV or only optical emission lines. 

Different stellar population synthesis (SPS) models incorporating different physics could potentially have an impact on the ionising stellar spectra used as an input to the photoionisation models, but such a comparative study is beyond the scope of this paper. However, D'Agostino {\etal} (2019, MNRAS, in press) have recently compared different SPS models and concluded that for most cases the emission line \textit{ratios} do not change considerably. The ionising spectra  were found to be somewhat senstive to the cluster age and SPS codes, extremely sensitive ($\sim 8$ orders of magnitude) to stellar evolutionary tracks, with very little ($\lesssim$ 2\%) dependence on the stellar atmospheres and the inclusion of binaries. D'Agostino {\etal} however, have not investigated the impact of different functional forms for the initial mass function (IMF)  on the spectra.

Work is currently in progress to combine stellar atmosphere models and evolutionary tracks with improved, stochastic stellar population synthesis codes. Once complete, we will have state-of-the-art diagnostics with self-consistent inputs and will be able to determine how much of the observed discrepancy in physical parameters stems from the observed target itself. Given all uncertainties in the models, we can constrain the physical properties remarkably well within the observed level of agreement. Moreover, the caveats discussed in this section are likely to have a smaller effect on the {\lpok} measurements  than the fact that UV and optical lines probe different physical nebular regions.

\section{Summary and Conclusions}\label{sum}
We measure equivalent widths and fluxes of the emission lines in the rest-frame UV and optical spectra of {\knotE}. By applying the full suite of new and existing UV and optical strong emission line (SEL) diagnostics on the dereddened fluxes, we determine the ISM properties of {\knotE}. We show that it is possible to infer some of the ISM properties -- ionisation parameter, electron density and ISM pressure in the inner nebular region -- with only the rest-frame UV emission lines. The rest-frame optical spectra better constrain the abundance and probe the pressure and density at the outer nebular regions.

We develop a new extension of IZI, called IZIP, which uses Bayesian inference method to simultaneously infer {\logOH}, {\logq}, and {\lpok} values. Given a theoretical model grid of emission line fluxes and a set of observed emission lines, IZIP constrains the three physical parameters simultaneously. We run IZIP with four different sets of emission lines -- all available emission lines, only the rest-frame UV lines, only the optical lines and UV+[\ion{O}{ii}]$\lambda\lambda$3727,9 lines -- to mimic observations with different rest-frame wavelength coverage.

By comparing the individual emission line diagnostics and the results from the four different configurations of IZIP, we draw the following conclusions:
\begin{enumerate}
\item The rest-frame UV emission lines infer $\sim$1.5 dex higher ISM pressures than the optical emission lines, because they probe different physical regions. The latter probe the outskirts of the nebula whereas the higher ionisation UV species probe the inner, denser regions. Because it is directly related to ISM pressure, the electron density also exhibits the same behavior.
\item The rest-frame UV emission lines used in this work (see Table~\ref{tab:pizi_lines}) are insufficient to accurately constrain the oxygen abundance for {\knotE}. The [\ion{O}{iii}]$\lambda$5007 emission line, used along with \ion{O}{iii}]$\lambda\lambda$ 1660,6, [\ion{O}{ii}]$\lambda\lambda$3727,9 and H$\beta$, helps constrain the oxygen abundance through the direct ({\Te}) method, and is within range of \JWST/NIRSpec wavelength coverage for redshifts $z\,\lesssim$9.
\item If only rest-frame UV spectra are available, it is possible to derive the ionisation parameter {\logq} as long as at least one of the ([\ion{C}{iii}]$\lambda$1906 + $\lambda$1908)/[\ion{C}{ii}]$\lambda$1335, ([\ion{C}{iii}]$\lambda$1906 + $\lambda$1908)/[\ion{C}{ii}]$\lambda$2325 or ([\ion{O}{iii}]$\lambda$1660 + $\lambda$1666)/[\ion{O}{ii}]$\lambda$2470 ratios are available. \JWST/NIRSpec will be able to capture these ratios in the redshift ranges 3.5$\lesssim\, z\, \lesssim$27, 2.1$\lesssim\, z\, \lesssim$22 and 2.6$\lesssim\, z\, \lesssim$20, respectively.
\item Joint Bayesian analysis is useful to determine {\logq} and {\lpok} when the rest-frame optical lines are available and yield results comparable to SEL diagnostics. Bayesian techniques have the capability to explore non-trivial topology in the PDFs of the inferred parameters e.g. multiple peaks and assymetry. However, when only the UV lines listed in Table~\ref{tab:pizi_lines} are available, it is difficult to constrain the oxygen abundance using Bayesian methods and currently available photoioniation grids. Inclusion of [\ion{O}{ii}]$\lambda\lambda$3727,9 with the UV lines does not make a noticeable difference either. This is a potential problem for {\JWST} at very high redshifts if the [\ion{O}{iii}]$\lambda$5007 line is not available to break the degeneracy.
\item Given rest-frame UV coverage and the optical [\ion{O}{ii}]$\lambda\lambda$3727,9 doublet, it is possible to effectively probe the ISM pressure in different physical regions in the nebula using the Bayesian approach. Future surveys with the {\JWST} will be designed to take advantage of this fact.
\end{enumerate}
In summary, we have demonstrated the power of rest-frame UV emission line diagnostics used in conjunction with rest-frame optical diagnostics, for inferring the ionised gas properties of moderate to high redshift galaxies. The ionisation parameter can be determined with only UV lines, whereas the electron density and ISM pressure additionally require the [\ion{O}{ii}]$\lambda\lambda$3727,9 doublet. We find the UV diagnostics used in this work alone cannot reliably constrain the electron temperature and oxygen abundance. This work paves the way for upcoming large telescopes (e.g. {\JWST}, GMT, TMT, ELT) which will carry out rest-frame UV spectroscopic studies of galaxies out to redshifts exceeding 10.

\section*{Acknowledgements}
We thank the anonymous referee for their thorough review and thoughtful comments that greatly improved this paper. This work includes data gathered with the 6.5 meter Magellan Telescopes located at Las Campanas Observatory, Chile. Some of the data presented herein were obtained at the W.M. Keck Observatory, which is operated as a scientific partnership among the California Institute of Technology, the University of California and the National Aeronautics and Space Administration. The Observatory was made possible by the generous financial support of the W.M. Keck Foundation. We acknowledge the very significant cultural role
and reverence that the summit of Mauna Kea has always had within the indigenous Hawaiian community. We are most fortunate to have the opportunity to conduct observations from
this mountain. Parts of this research were conducted by the Australian Research Council Centre of Excellence for All Sky Astrophysics in 3 Dimensions (ASTRO 3D), through project number CE170100013. L.J.K. gratefully acknowledges the support of an ARC Laureate Fellowship (FL150100113). C.~F.~acknowledges funding provided by the Australian Research Council (Discovery Projects DP150104329 and DP170100603, and Future Fellowship FT180100495), and the Australia-Germany Joint Research Cooperation Scheme (UA-DAAD). G.~B. is supported by CONICYT/ FONDECYT, Programa de Iniciaci\'{o}n, Folio 11150220. This research has made use of the following: NASA's Astrophysics Data System Bibliographic Services, the NASA/IPAC Extragalactic Database (NED), Astropy -- a community-developed core Python package for Astronomy \citep{astropy}, scipy \citep{scipy}, and pandas \citep{pandas} packages.



\bibliographystyle{mnras}
\bibliography{paper} 



\appendix

\section{Additional IZIP analyses}\label{app:izip}

In this section we present the additional tests we conducted with IZIP. No user defined priors are used for these analyses in order to isolate the effects of the individual lines on the Bayesian estimates. Moreover, for computational efficiency, only 10 realisations (refer to Section~\ref{izidiag}) were performed for each of the cases discussed here, unlike the 100 realisations for those discussed before. Each test case broadly converged by 10 realisations, and so our qualitative results and conclusions are robust.

We provide IZIP with the following different sets of emission lines to investigate the impact of the absence of a line or the presence of an additional line in determining the ISM properties. Please refer to Table~\ref{tab:pizi_lines} for the appropriate list of emission lines used.
\begin{enumerate}
	\item All the UV and optical emission lines \textit{except} the [\ion{Si}{iii}]$\lambda\lambda$1882,92 doublet.
	\item Only the UV lines except the [\ion{Si}{iii}]$\lambda\lambda$1882,92 doublet.
	\item All the UV and optical emission lines \textit{except} the [\ion{C}{iii}]$\lambda\lambda$1906,8 doublet.
	\item Only the UV lines except the [\ion{C}{iii}]$\lambda\lambda$1906,8 doublet.
	\item Only the UV lines and the H$\alpha$ line.
	\item Only the UV lines and the H$\beta$ line.
	\item Only the UV lines along with the [\ion{O}{iii}]$\lambda$5007 line.
\end{enumerate}

We also conduct tests with IZIP by providing only the emission lines sensitive to a single ISM parameter, at a time, as follows. We select the lines sensitive to a given ISM parameter as those used in the SEL diagnostics for that parameter (refer to Table~\ref{tab:masterlist}).
\begin{enumerate}
	\item All emission lines (from Table~\ref{tab:masterlist}) that are used for {\Te} diagnostic
	\item All emission lines (from Table~\ref{tab:masterlist}) that are used for {\logOH} diagnostic	
	\item All emission lines (from Table~\ref{tab:masterlist}) that are used for {\lpok} diagnostic
	\item All emission lines (from Table~\ref{tab:masterlist}) that are used for {\logq} diagnostic
\end{enumerate}

\begin{figure}
	\centering
	\begin{subfigure}[t]{0.47\textwidth}
		\centering
		\includegraphics[scale=0.42,trim=0.0cm 0.0cm 0.0cm 0.0cm,clip=true,angle=0]{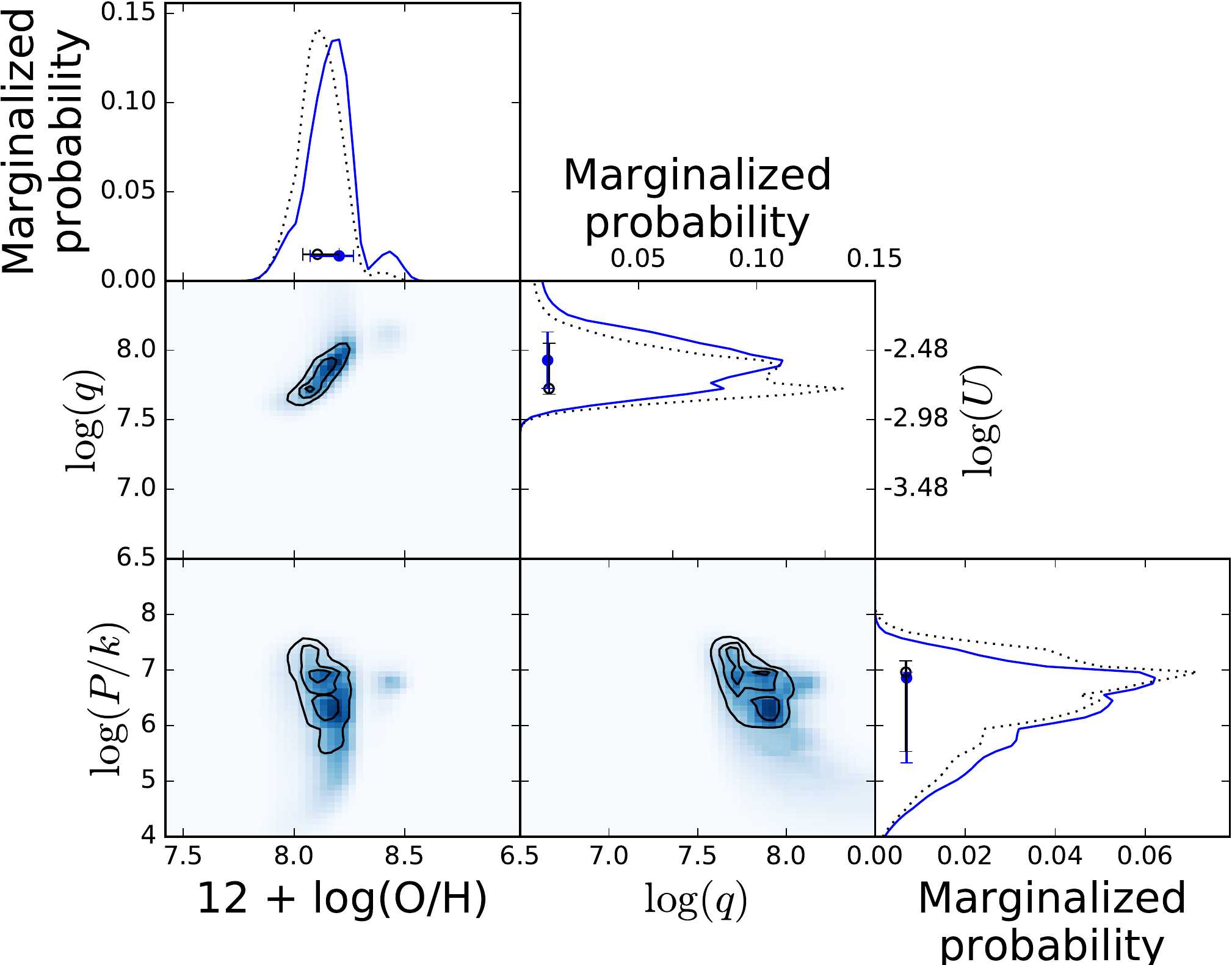}
	\caption{All except [\ion{Si}{iii}]$\lambda\lambda$1882,92 lines used}
	\end{subfigure}
	
	\caption{Same as in Figure~\ref{fig: pizi_noprior} but here we investigate the effect of excluding the [\ion{Si}{iii}]$\lambda\lambda$1882,92 pair. The black contours and dotted histograms denote the fiducial case i.e. when all the lines are used. We assume flat prior in both these cases because the objective of this test was to isolate the effect of the [\ion{Si}{iii}] doublet. We find that {\logq} is constained better on not using the [\ion{Si}{iii}] lines.
	}
	\label{fig:pizi_noprior_all_without_Siiii}
\end{figure}

\begin{figure}
	\centering
	\begin{subfigure}[t]{0.47\textwidth}
		\centering
		\includegraphics[scale=0.42,trim=0.0cm 0.0cm 0.0cm 0.0cm,clip=true,angle=0]{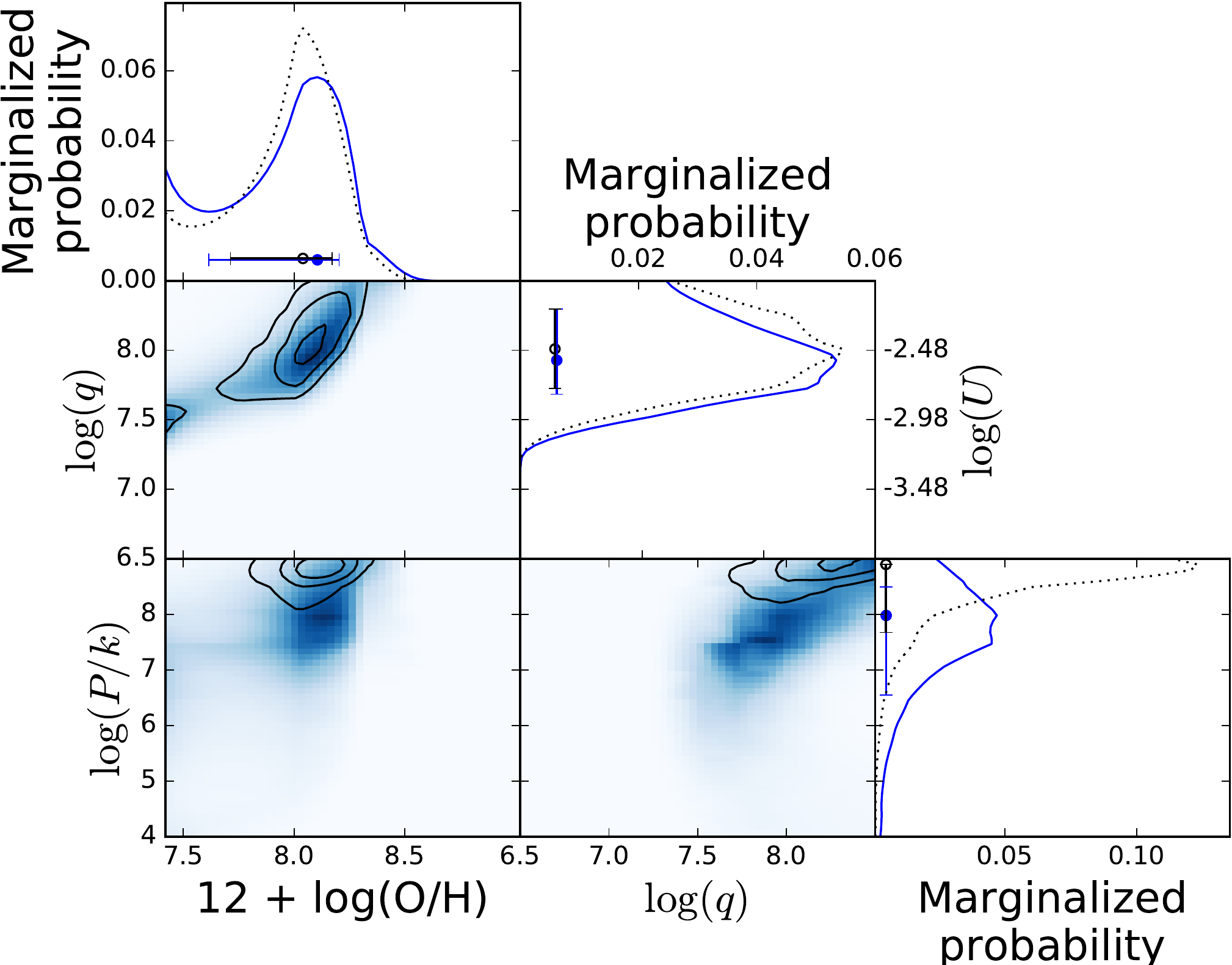}
		\caption{All UV lines except [\ion{Si}{iii}]$\lambda\lambda$1882,92 used}
	\end{subfigure}	
	\caption{Same as in Figure~\ref{fig: pizi_noprior} but this time using only the UV lines except the [\ion{Si}{iii}]$\lambda\lambda$1882,92 doublet. The black contours and dotted histograms denote the case all the UV lines are used. Similar to \autoref{fig:pizi_noprior_all_without_Siiii}, we find that {\logq} is constained better on excluding the [\ion{Si}{iii}] lines.
	}
	\label{fig:pizi_noprior_UV_without_Siiii}
\end{figure}

\begin{figure}
	\centering
	\begin{subfigure}[t]{0.45\textwidth}
	\centering
	\includegraphics[scale=0.42, trim=0.0cm 0.0cm 0.0cm 0.0cm,clip=true,angle=0]{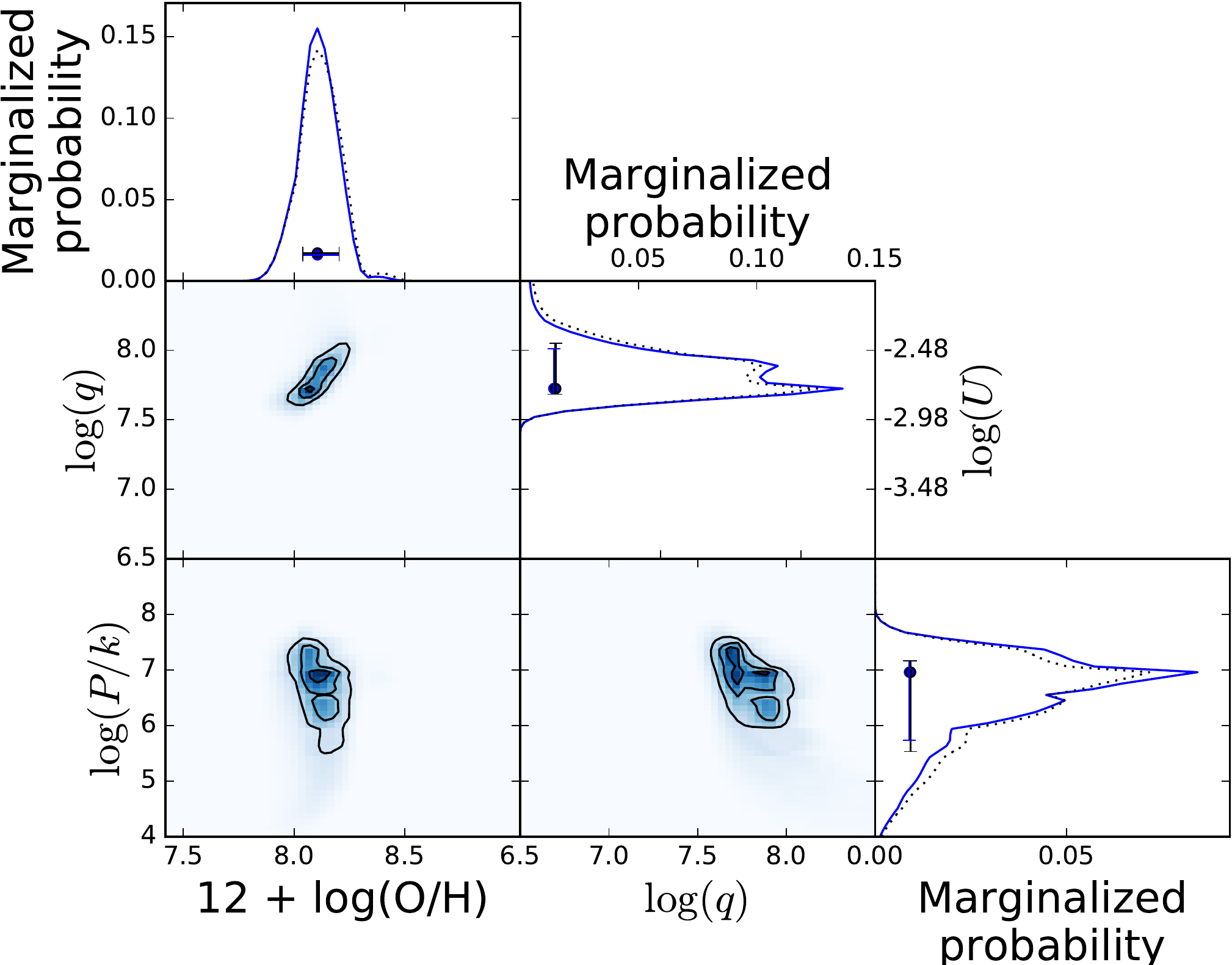}
	\caption{All but [\ion{C}{iii}]$\lambda\lambda$1906,8 lines used}
	\end{subfigure}
	\caption{
		Same as in Figure~\ref{fig: pizi_noprior} but this time without using all emission lines except [\ion{C}{iii}]$\lambda\lambda$1906,8. The black contours and dotted histograms denote the fiducial case i.e. when all the lines are used. Absence of [\ion{C}{iii}] does not have any discernible impact on any of the ISM properties.
	}
	\label{fig: pizi_noprior_all_without_CIII}
\end{figure}

\begin{figure}
	\centering
	\begin{subfigure}[t]{0.45\textwidth}
		\centering
		\includegraphics[scale=0.42, trim=0.0cm 0.0cm 0.0cm 0.0cm,clip=true,angle=0]{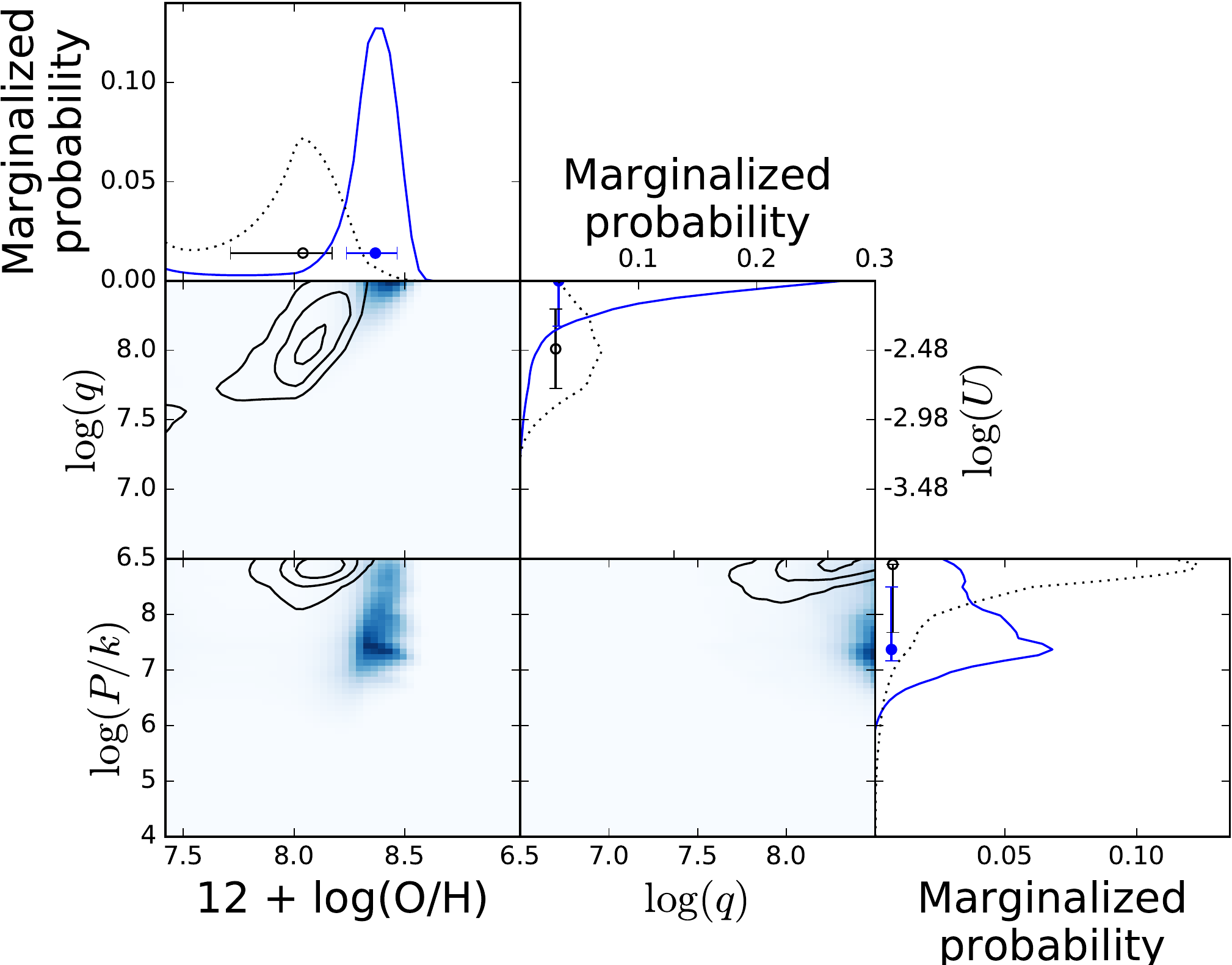}
		\caption{All UV lines but [\ion{C}{iii}]$\lambda\lambda$1906,8 lines used}
	\end{subfigure}
	\caption{
		Same as in Figure~\ref{fig: pizi_noprior} but this time without using only UV emission lines except [\ion{C}{iii}]$\lambda\lambda$1906,8. The black contours and dotted histograms denote the case when all the UV lines are used. Although the {\logOH} estimate is improved by dropping [\ion{C}{iii}], {\logq} is now  unconstrained.
	}
	\label{fig: pizi_noprior_UV_without_CIII}
\end{figure}

\begin{figure}
	\centering
	\begin{subfigure}[t]{0.45\textwidth}
		\centering
		\includegraphics[scale=0.42, trim=0.0cm 0.0cm 0.0cm 0.0cm,clip=true,angle=0]{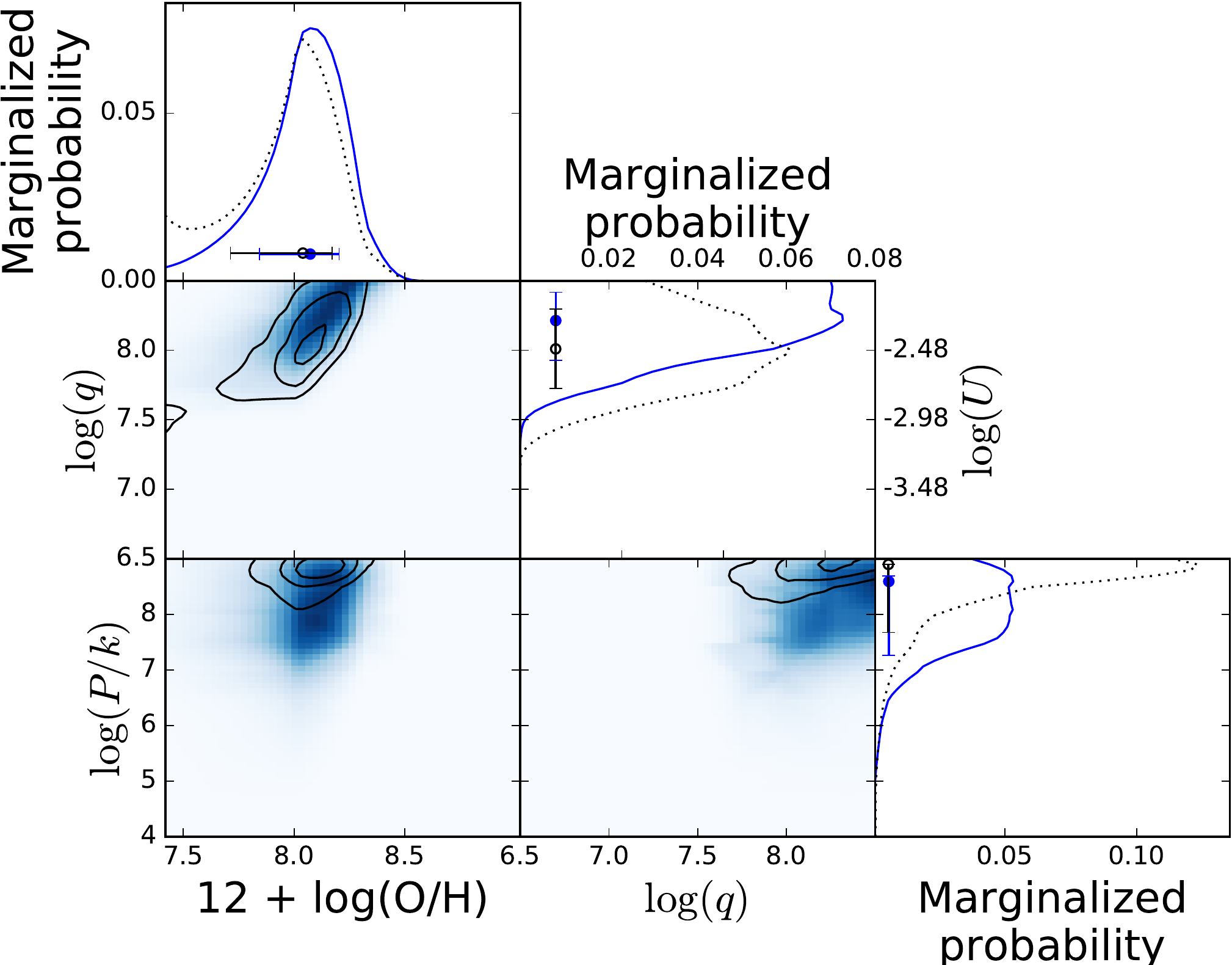}
		\caption{All UV lines used along with [\ion{O}{iii}]$\lambda$5007}
	\end{subfigure}
	\caption{
		Same as in Figure~\ref{fig: pizi_noprior} but this time using the [\ion{O}{iii}]$\lambda$5007 line along with rest-frame UV lines. The black contours and dotted histograms indiicate the case when only the UV lines are used. Inclusion of [\ion{O}{iii}] helps to break the degeneracy of the metallicity branch, but fails to constrain {\logq} or {\lpok}.
		}
	\label{fig: pizi_noprior_UV_with_OIII}
\end{figure}

\begin{figure*}
	\centering
	\begin{subfigure}[t]{0.47\textwidth}
		\centering
		\includegraphics[scale=0.42, trim=0.cm 0.0cm 0.0cm 0.0cm,clip=true,angle=0]{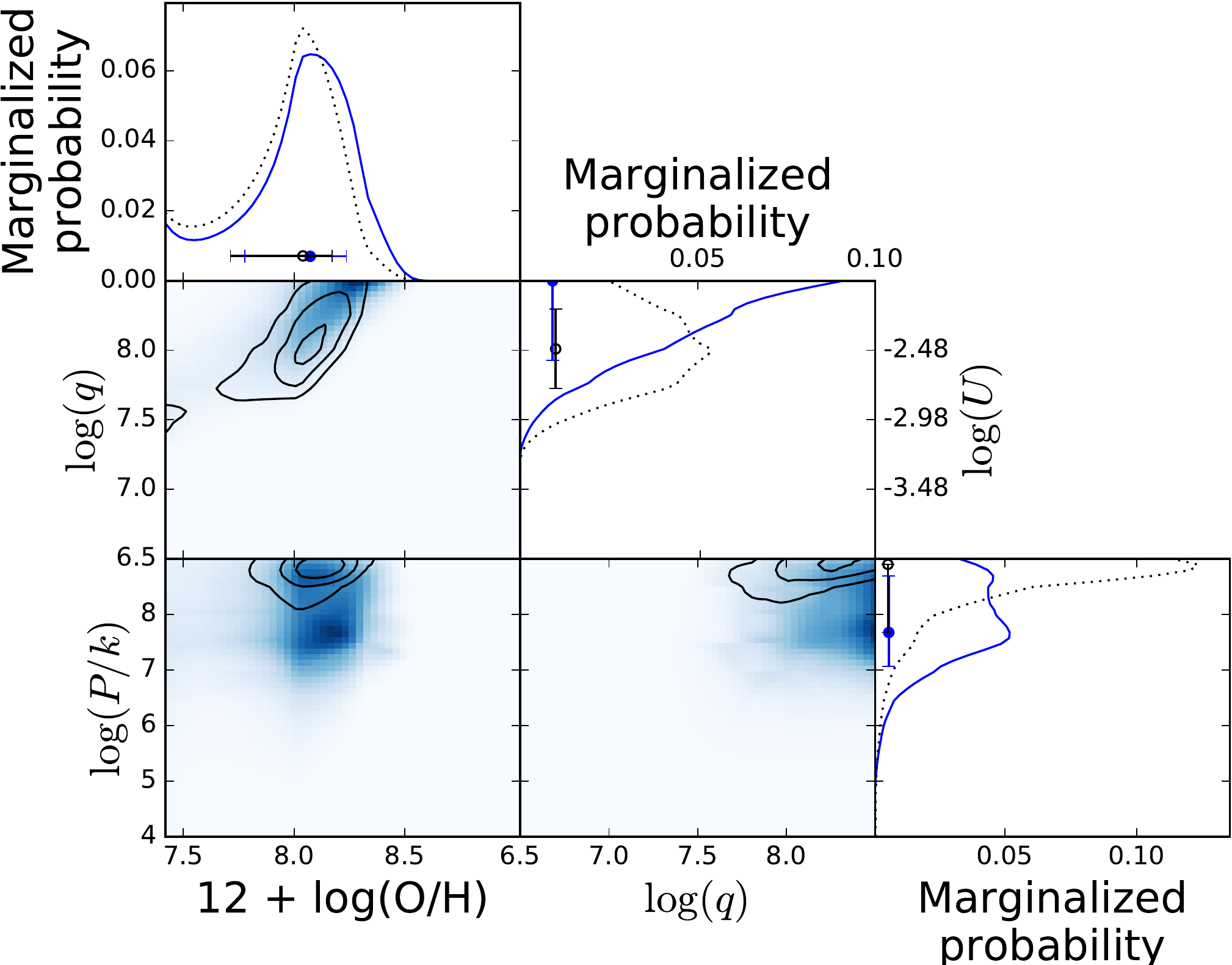}
		\caption{All UV lines lines used along with H$\alpha$}
	\end{subfigure}
	\hspace{0.5cm}
	\begin{subfigure}[t]{0.47\textwidth}
		\centering
		\includegraphics[scale=0.42, trim=0.cm 0.0cm 0.0cm 0.0cm,clip=true,angle=0]{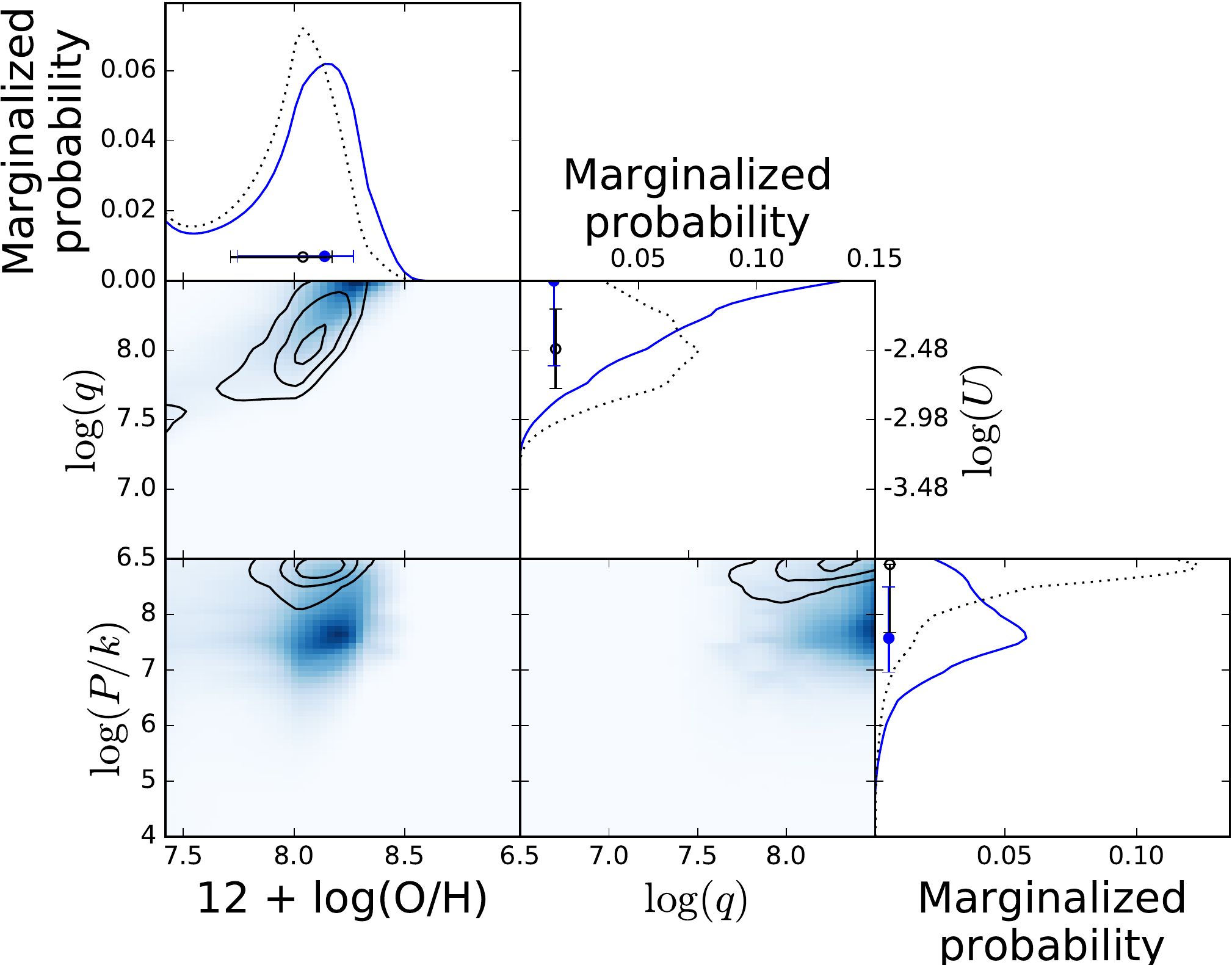}
		\caption{All UV lines lines used along with H$\beta$}
	\end{subfigure}
	\caption{
		Same as in Figure~\ref{fig: pizi_noprior} but this time without using only UV emission lines along with a Balmer line. Inclusion of Balmer lines do not have an appreciable impact on the metallicity constraints. Moreover, the {\logq} and {\lpok} are unconstrained now, compared to when only UV lines were used (black contours and dotted histograms), thus suggesting that lack of Balmer lines in the UV spectra is not the main source of uncertainty.
	}
	\label{fig: pizi_noprior_UV_with_H}
\end{figure*}

\begin{figure*}
	\centering
	\begin{subfigure}[t]{0.47\textwidth}
		\centering
		\includegraphics[scale=0.42, trim=0.cm 0.0cm 0.0cm 0.0cm,clip=true,angle=0]{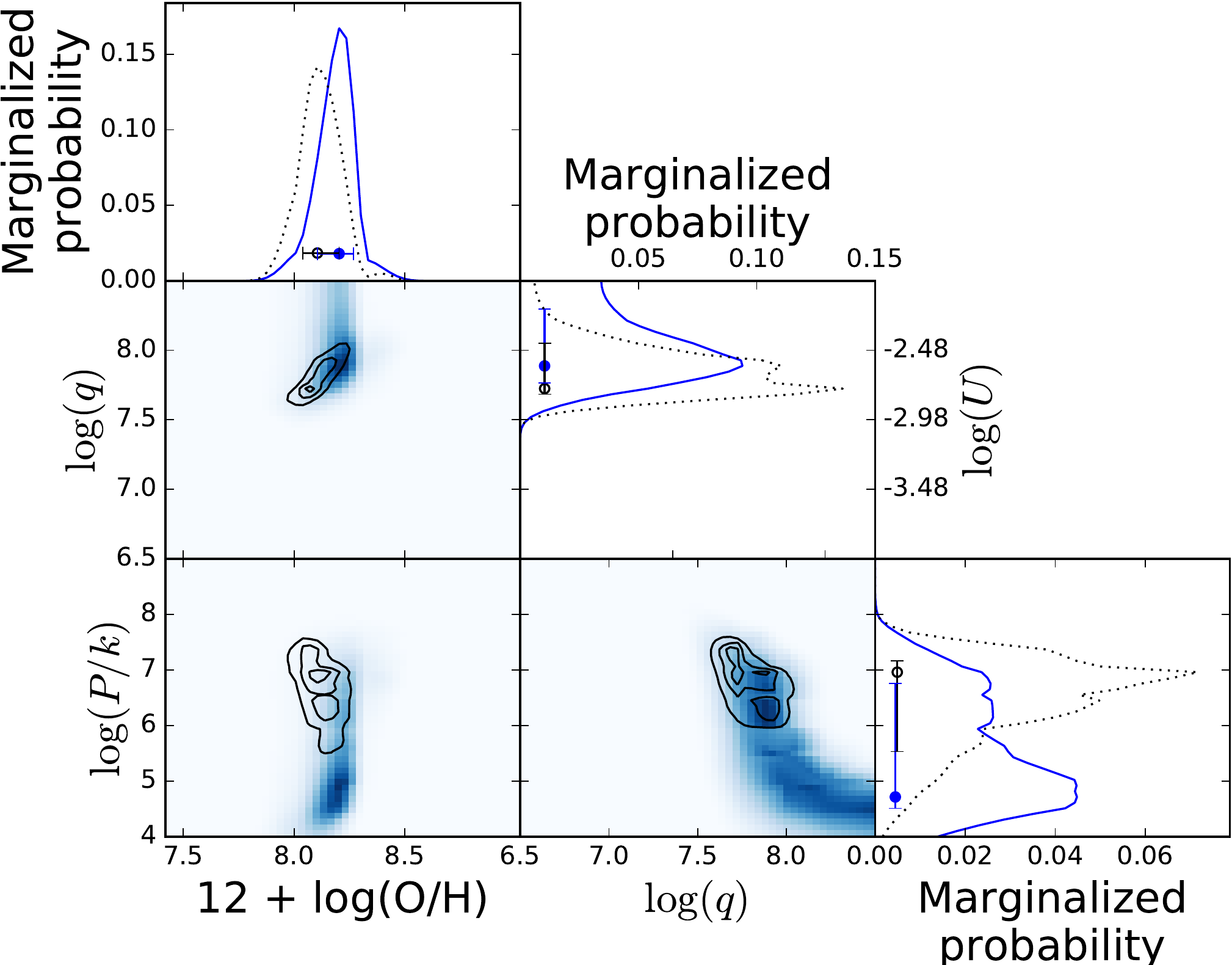}
		\caption{Using all lines involved in SEL diagnostics for {\logOH}}
	\end{subfigure}
	\hspace{0.5cm}
	\begin{subfigure}[t]{0.47\textwidth}
		\centering
		\includegraphics[scale=0.42, trim=0.cm 0.0cm 0.0cm 0.0cm,clip=true,angle=0]{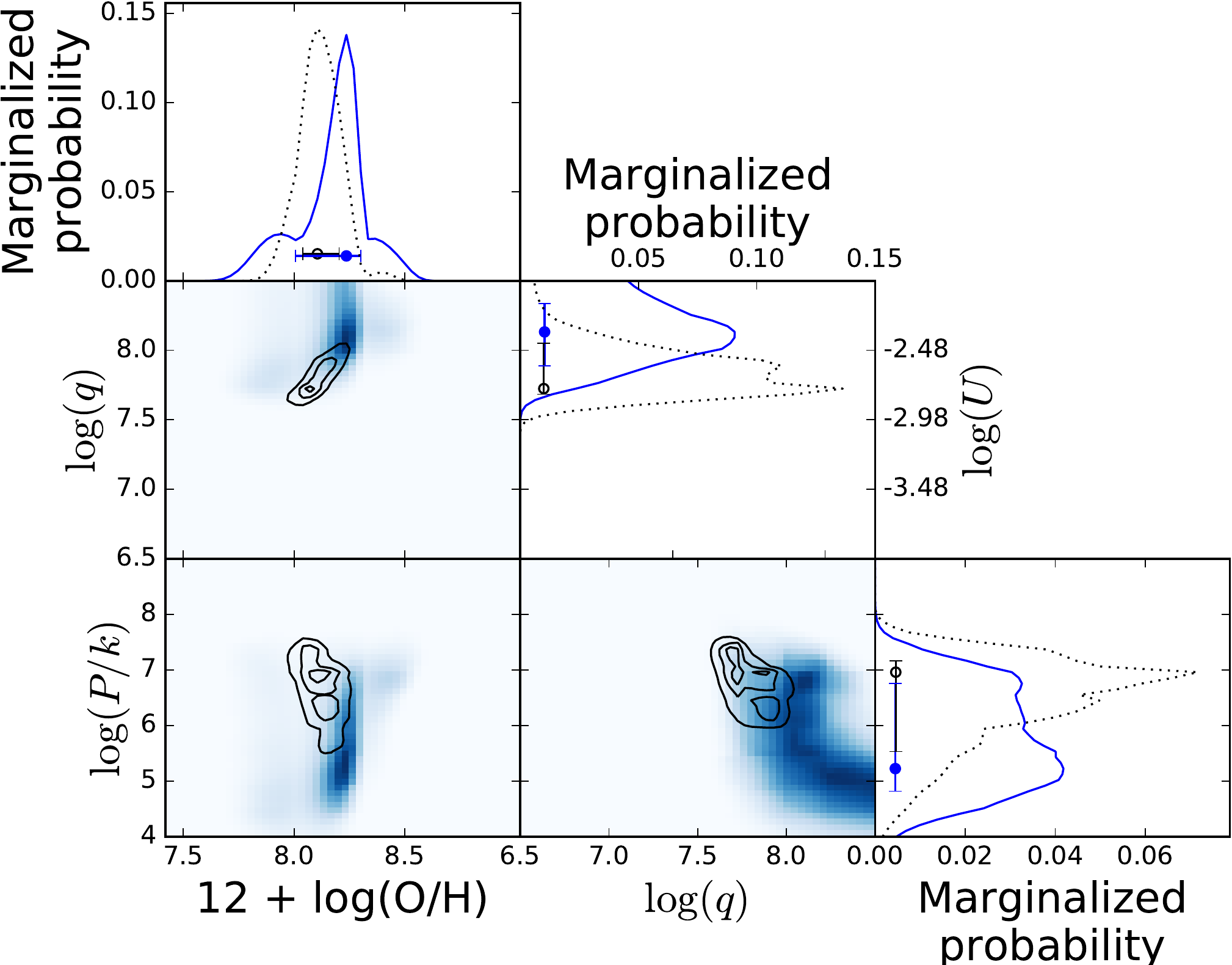}
		\caption{Using all lines involved in SEL diagnostics for {\logq}}
	\end{subfigure}
	
	\vspace{1cm}
	
	\begin{subfigure}[t]{0.45\textwidth}
		\centering
		\includegraphics[scale=0.42, trim=0.cm 0.0cm 0.0cm 0.0cm,clip=true,angle=0]{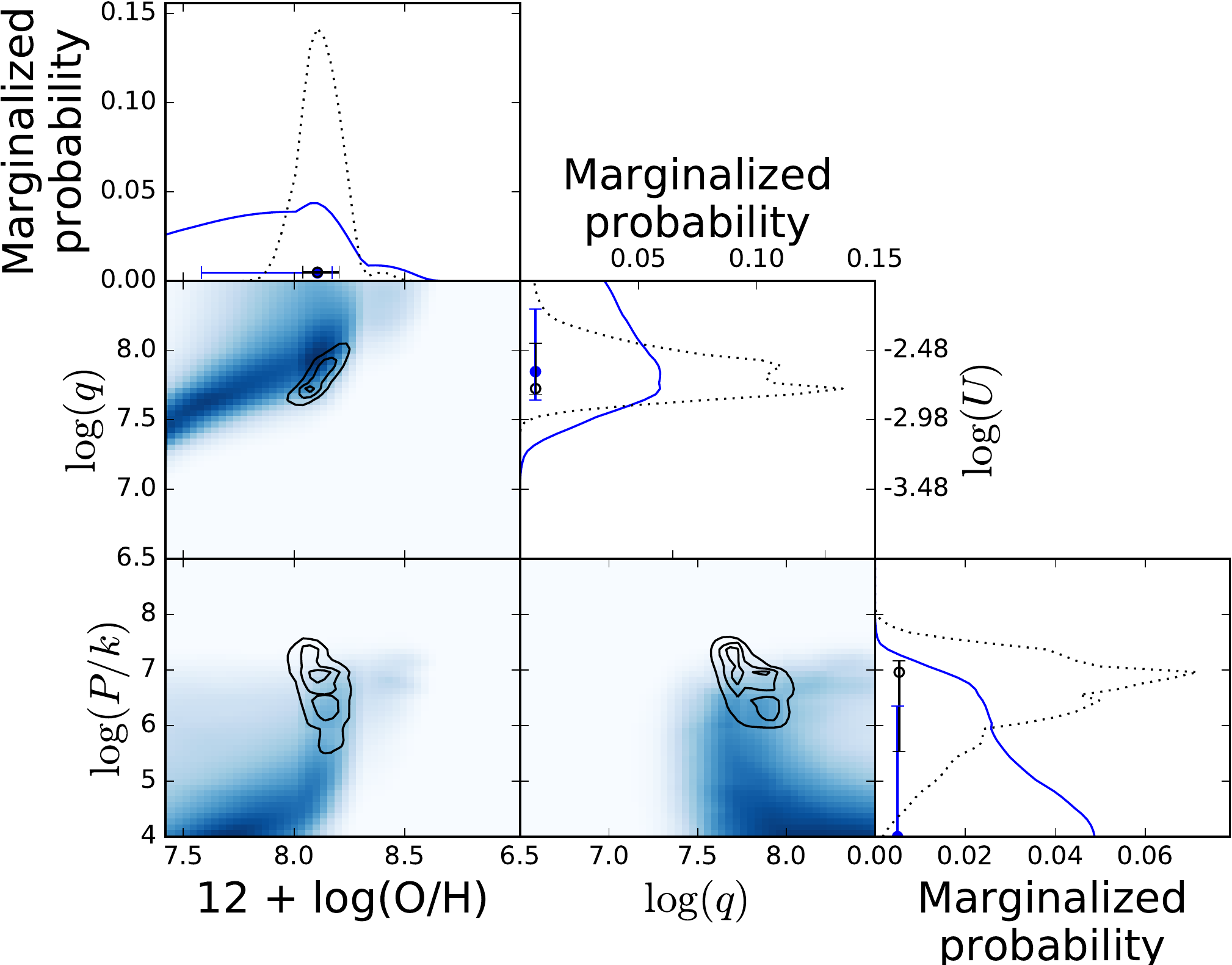}
		\caption{Using all lines involved in SEL diagnostics for {\lpok}}
	\end{subfigure}
	\hspace{0.5cm}
	\begin{subfigure}[t]{0.45\textwidth}
	\centering
	\includegraphics[scale=0.42, trim=0.cm 0.0cm 0.0cm 0.0cm,clip=true,angle=0]{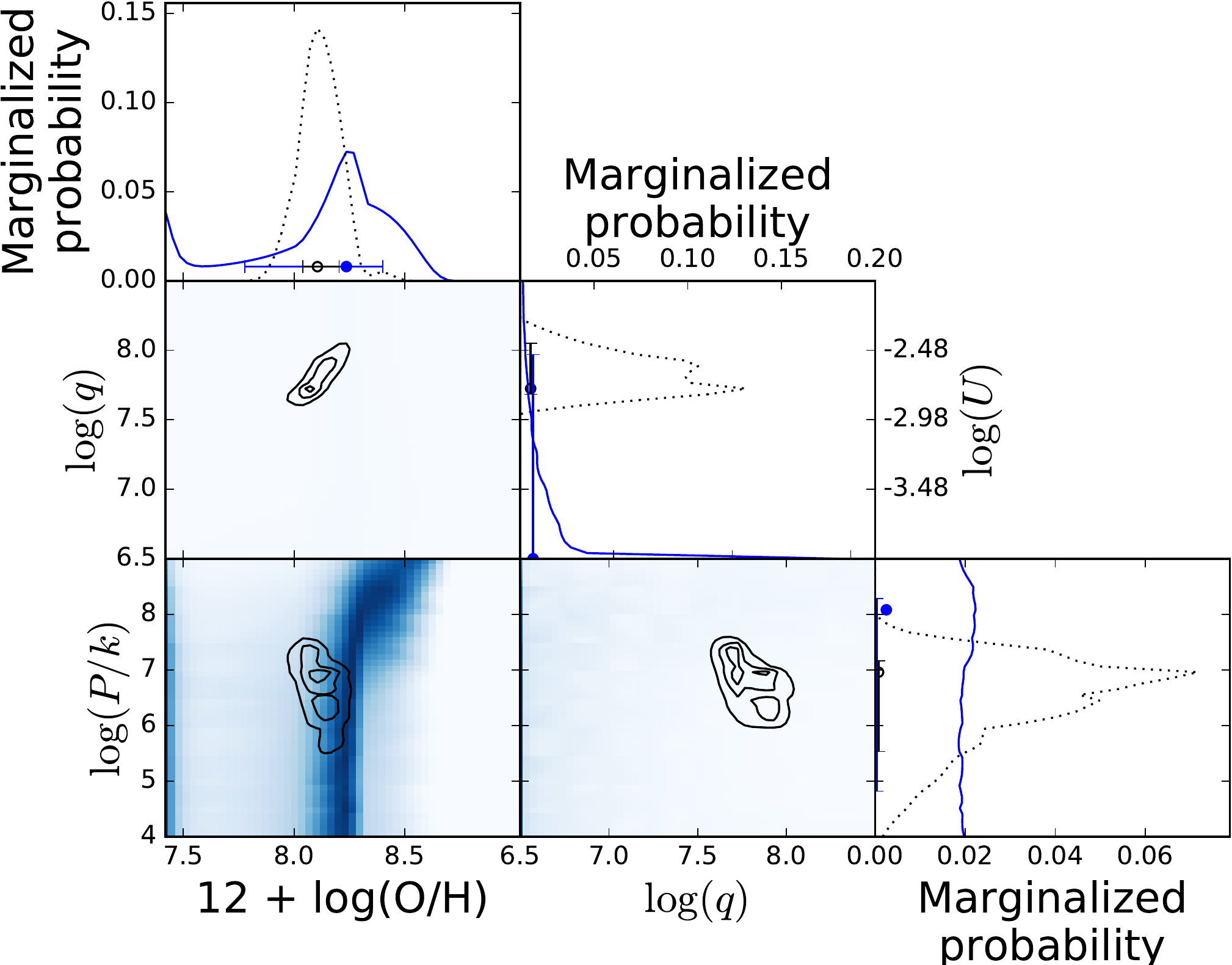}
	\caption{Using all lines involved in SEL diagnostics for {\Te}}
\end{subfigure}
	\caption{
		Same as in Figure~\ref{fig: pizi_noprior} but this time using only those emission lines that are sensitive to a specific ISM property. The black contours and dotted histograms represent the fiducial case i.e. when all the emission lines are used, without using a [\ion{N}{ii}]/H$\alpha$ based prior. In none of these cases, are all three parameters well constrained. In the cases where only {\logq} sensitive and only {\lpok} sensitive lines were provided, the respective parameters themselves are not constrained. We conclude that the parameters are inter-dependent and hence absence of the lines sensitive to one parameter leads to failure of the Bayesian estimate of another parameter even if the lines sensitive to the latter parameter is present. The only exception to this is when all metallicity sensitive lines are used, the metallicity and ionisation parameter is well constrained. This is because the metallicity sensitive lines (in Table~\ref{tab:masterlist}) already include the most of the (optical) {\logq} sensitive lines.
	}
	\label{fig: pizi_noprior_diag_lines}
\end{figure*}

Figures~\ref{fig:pizi_noprior_all_without_Siiii} to \ref{fig: pizi_noprior_diag_lines} show the results of the above tests and Table~\ref{tab:app_izip} quotes the corresponding values. 

Exclusion of the [\ion{Si}{iii}]$\lambda\lambda$1882,92 lines from the UV-optical suite of emission lines constrains {\logq} (Figure~\ref{fig:pizi_noprior_all_without_Siiii}) which otherwise hits the model grid boundaries. Depletion of Si from the gas phase onto dust grains or erosion of dust grains by shocks can have a considerable impact on the abundacne of Si in the ISM, which in turn may influence the [\ion{Si}{iii}] flux \citep{Jones:2000aa}. This is a potential source of discrepancy for diagnostics that use the [\ion{Si}{iii}] lines if the effects of dust have not been appropriately accounted for in the photoionisation models (Byler {\etal} \textit{in prep}). However, excluding the [\ion{Si}{iii}] doublet from the set of UV lines (Figure~\ref{fig:pizi_noprior_UV_without_Siiii}) makes the PDFs worse (compared to using all UV lines) i.e. {\logq} now hits the model grid boundaries. This contradictory behavior for UV-optical and only UV lines could be because the [\ion{Si}{iii}] doublet is one of the few strong emission lines in the UV regime and removing it forces the Bayesian method to work with considerably less amount of information. Excluding the [\ion{Si}{iii}]$\lambda\lambda$1882,92 doublet makes very little difference to {\lpok} in either case (Figures~\ref{fig:pizi_noprior_all_without_Siiii} and \ref{fig:pizi_noprior_UV_without_Siiii}) and yields a slightly lower ($\sim 0.3$ dex) value of the inferred abundance but similar widths of the $Z$ PDFs. 

We find that absence of [\ion{C}{iii}]$\lambda\lambda$1906,8 does not impact the {\logq} measurement when all lines are used (Figure~\ref{fig: pizi_noprior_all_without_CIII}) but pushes it against the model grid ceiling when only rest-frame UV lines are used (Figure~\ref{fig: pizi_noprior_UV_without_CIII}). Such a dissimilarity exists because a considerably larger set of emission lines have been used in the former case than the latter, implying that the former configuration had more information available whereas the latter case did not have enough information to constrain the PDF. However, the abundance is much better constrained ($\sigma \sim$0.2 dex) by only the UV lines on excluding [\ion{C}{iii}]. A similar effect is observed on including the [\ion{O}{iii}]$\lambda$5007 line with the UV lines (including [\ion{C}{iii}]), in that the oxygen abundance is better constrained but the {\logq} and {\lpok} estimates fail. The {\logq} and {\lpok} PDFs hit the model grid boundaries and we have tested that simply extending the models towards higher values leads to unphysically high solutions for {\logq} and {\lpok}. The improved abundance estimate suggests that  [\ion{O}{iii} is necessary to break the degeneracy in the metallicity branch, when using only UV lines.

Including Balmer lines with the UV spectra does not help improve the constraints. Similarly, providing IZIP with emission lines sensitive to one ISM parameter at a time, does not yield reliable estimates of the other parameters, or at times even the same parameters which the input lines are sensitive to. This is because each parameter depends on the other and, in absence of spectral lines sensitive to the other two parameters, fails to reliably infer the concerned parameter as well. For instance, in Figure~\ref{fig: pizi_noprior_diag_lines}c, although only the lines sensitive to ISM pressure has been provided to IZIP, the absence of $Z$ or {\logq} sensitive lines lead to unconstrained $Z$ and {\logq} and consequently, fails to constrain {\lpok} which is dependent on the other two parameters.

\begin{table*}
	\centering
	\caption{Inferred physical parameters by providing different sets of emission lines to IZIP. For computational efficiency, we performed only 10 realisations (refer to Section~\ref{izidiag}) for each of these cases (instead of 100, as in Table~\ref{tab:phys_list}). However, each case was satisfactorily converged by then, and adding any more realisation would not change the results qualitatively.}
	\label{tab:app_izip}
	\begin{tabular}{lccc}
		\hline
		Lines provided to IZIP & Oxygen abundance & Ionization parameter & ISM pressure \\
		\hline
		&  {\logOH} & $\log{(q (cm/s))}$ & $\log{(P/k (K/cm^3))}$ \\
		\hline
		All lines except [\ion{Si}{iii}]$\lambda\lambda$1882,92 & 8.20 $^{+0.07}_{-0.13}$ & 7.93 $^{+0.20}_{-0.20}$ & 6.86 $^{+0.10}_{-1.53}$ \\
		UV lines except [\ion{Si}{iii}]$\lambda\lambda$1882,92 & 8.11 $^{+0.10}_{-0.49}$ & 7.93 $^{+0.37}_{-0.24}$ & 7.98 $^{+0.51}_{-1.43}$ \\
		All lines except [\ion{C}{iii}]$\lambda\lambda$1906,8 &  8.11 $^{+0.1}_{-0.06}$ & 7.72 $^{+0.29}_{-0.04}$ & 6.96 $^{+0.20}_{-1.22}$ \\
		UV lines except [\ion{C}{iii}]$\lambda\lambda$1906,8 &  8.37 $^{+0.1}_{-0.13}$ & unconstrained & 7.37 $^{+1.22}_{-0.20}$ \\
		UV lines + H$\alpha$ &  8.07 $^{+0.16}_{-0.29}$ & unconstrained & 7.67 $^{+1.02}_{-0.61}$ \\
		UV lines + H$\beta$ &  8.14 $^{+0.13}_{-0.39}$ & unconstrained & 7.57 $^{+0.92}_{-0.61}$ \\
		UV lines + [\ion{O}{iii}]$\lambda$5007 &  8.07 $^{+0.13}_{-0.23}$ & unconstrained & unconstrained \\
		All lines used for {\Te} diagnostic$^a$ &  8.24 $^{+0.16}_{-0.46}$ & unconstrained & unconstrained \\
		All lines used for {\logOH} diagnostic$^a$ &  8.20 $^{+0.07}_{-0.1}$ & 7.89 $^{+0.41}_{-0.12}$ & 4.71 $^{+2.04}_{-0.20}$ \\
		All lines used for {\lpok} diagnostic$^a$ &  unconstrained& 7.85 $^{+0.45}_{-0.20}$ & unconstrained \\
		All lines used for {\logq} diagnostic$^a$ &  8.24 $^{+0.07}_{-0.23}$ & 8.13 $^{+0.20}_{-0.24}$ & 5.22 $^{+1.53}_{-0.41}$ \\
		\hline
		\multicolumn{4}{l}{$^a$ As per Table~\ref{tab:masterlist}}\\
	\end{tabular}	
\end{table*}

\bsp	
\label{lastpage}
\end{document}

%% file: Tables/lineflux_restopt_detected.tex
\caption{Keck/NIRSPEC line flux measurements. These measurements include updated (relative to \citet{Wuyts:2014ab}) tweak factors to take into account the cross-filter flux calibration. flux$_{\mathrm{obs}}$ and $\delta$ flux$_{\mathrm{obs}}$ denote the observed flux and uncertainty respectively. flux$_{\mathrm{dereddened}}$ is the dereddened flux using E(B-V) = 0.4 $\pm$ 0.07. All fluxes and uncertainties are in units of 10$^{-17}$ ergs/s/cm$^2$}
\begin{tabular}{llrrr}
\toprule
         Line ID & $\lambda_{\mathrm{rest}}$ (\AA) & flux$_{\mathrm{obs}}$ & $\delta$ flux$_{\mathrm{obs}}$ & flux$_{\mathrm{dereddened}}$ \\
\midrule
\hline
\multicolumn{5}{c}{Filter N1} \\ 
\hline 
 {[}O II] 3727,9 &                        3728.483 &                 165.3 &                            7.3 &                       958.36 \\
   {[}O II] 3727 &                        3727.092 &                  87.8 &                              3 &                       509.23 \\
   {[}O II] 3729 &                        3729.875 &                  76.6 &                              4 &                       443.94 \\
 {[}Ne III] 3869 &                        3869.860 &                  25.4 &                            2.6 &                       141.33 \\
\hline
\multicolumn{5}{c}{Filter N2} \\ 
\hline 
        H$\zeta$ &                        3890.166 &                  17.8 &                              2 &                       98.408 \\
       H$\delta$ &                        3971.198 &                    16 &                            4.1 &                        86.09 \\
     H$\epsilon$ &                        4102.892 &                     7 &                              3 &                       35.882 \\
\hline
\multicolumn{5}{c}{Filter N3} \\ 
\hline 
       H$\gamma$ &                        4341.692 &                  49.8 &                              2 &                       231.64 \\
      O III 4363 &                        4364.435 &                   5.5 &                              2 &                        25.34 \\
  {[}Ar IV] 4741 &                        4741.449 &                     0 &                              0 &                            0 \\
        H$\beta$ &                        4862.691 &                   119 &                            1.6 &                       449.53 \\
  {[}O III] 4959 &                        4959.895 &                   190 &                            1.4 &                       693.78 \\
  {[}O III] 5007 &                        5008.239 &                   613 &                              3 &                       2202.5 \\
       H$\alpha$ &                        6564.632 &                 465.2 &                           23.2 &                       1183.4 \\
   {[}N II] 6584 &                        6585.273 &                    55 &                            3.9 &                        139.4 \\
   {[}S II] 6717 &                        6718.294 &                  26.3 &                            2.3 &                       65.114 \\
       S II 6731 &                        6732.674 &                  19.4 &                            6.2 &                        47.91 \\
 {[}Ar III] 7136 &                        7137.770 &                  10.8 &                            3.1 &                       24.824 \\
\bottomrule
\end{tabular}

%% file: Tables/lineflux_restUV_detected.tex
\caption{MagE/Magellan line flux measurements. flux$_{\mathrm{obs}}$ and $\delta$ flux$_{\mathrm{obs}}$ denote     the observed flux and uncertainty respectively. flux$_{\mathrm{dereddened}}$ is the dereddened flux using     E(B-V) = 0.4 $\pm$ 0.07. W$_\mathrm{r,fit}$ and $\delta$ W$_{\mathrm{r,fit}}$ denote the rest-frame     equivalent width measured and the corresponding uncertainty in \AA\, respectively. For cases of     non-detection (i.e. < 3 $\sigma$ detection), the 3 $\sigma$ upper limit on equivalent widths and fluxes are quoted.     Uncertainty estimates for these entries are not quoted because they do not provide any meaningful information.}
\begin{tabular}{lrllllll}
\toprule
         Line ID &  $\lambda_{\mathrm{rest}}$ & W$_{\mathrm{r,fit}}$ & $\delta$ W$_{\mathrm{r,fit}}$ & W$_{\mathrm{r,signi}}$ & flux$_{\mathrm{obs}}$ & $\delta$ flux$_{\mathrm{obs}}$ & flux$_{\mathrm{dereddened}}$ \\
& (\AA) & (\AA) & (\AA) & (\AA) & (10$^{-17}$ ergs/s/cm$^2$) &                 (10$^{-17}$ ergs/s/cm$^2$) & (10$^{-17}$ ergs/s/cm$^2$) \\
\midrule
      Ly$\alpha$ &                  1215.6700 &             >-0.5706 &                            .. &                     .. &               $<$0.43 &                             .. &                     $<$23.32 \\
        O I 1304 &                  1304.8576 &             >-0.2609 &                            .. &                     .. &               $<$0.25 &                             .. &                      $<$5.26 \\
        O I 1306 &                  1306.0286 &             >-0.2589 &                            .. &                     .. &               $<$0.24 &                             .. &                      $<$5.22 \\
      Si II 1309 &                  1309.2757 &              -0.6397 &                          0.15 &                   7.37 &                  1.62 &                           0.37 &                        34.82 \\
      C II 1335a &                  1334.5770 &             >-0.1547 &                            .. &                     .. &               $<$0.15 &                             .. &                      $<$3.16 \\
      C II 1335b &                  1335.6630 &             >-0.1575 &                            .. &                     .. &               $<$0.15 &                             .. &                      $<$3.22 \\
      C II 1335c &                  1335.7080 &             >-0.1575 &                            .. &                     .. &               $<$0.15 &                             .. &                      $<$3.22 \\
      N II] 1430 &                  1430.4100 &             >-0.1131 &                            .. &                     .. &               $<$0.10 &                             .. &                      $<$2.21 \\
      N II] 1431 &                  1430.9730 &              >-0.116 &                            .. &                     .. &               $<$0.11 &                             .. &                      $<$2.26 \\
      N IV] 1486 &                  1486.5000 &              -0.2242 &                          0.09 &                   5.92 &                  0.55 &                           0.23 &                        11.17 \\
      Si II 1533 &                  1533.4312 &              -0.3645 &                          0.05 &                  10.83 &                  0.86 &                           0.12 &                        16.84 \\
      He II 1640 &                  1640.4170 &              -0.4451 &                          0.07 &                  15.06 &                  1.03 &                           0.16 &                        18.56 \\
     O III] 1660 &                  1660.8090 &              -0.1209 &                          0.07 &                   4.24 &                  0.28 &                           0.16 &                         4.94 \\
     O III] 1666 &                  1666.1500 &              -0.4817 &                          0.07 &                  15.88 &                  1.10 &                           0.16 &                        19.52 \\
     N III] 1750 &                  1749.7000 &              -0.3396 &                          0.05 &                  13.45 &                  0.73 &                           0.11 &                        12.82 \\
  {[}Si II] 1808 &                  1808.0130 &             >-0.0896 &                            .. &                     .. &               $<$0.07 &                             .. &                      $<$1.22 \\
  {[}Si II] 1816 &                  1816.9280 &              -0.3036 &                          0.06 &                   9.83 &                  0.62 &                           0.13 &                        11.14 \\
 {[}Si III] 1882 &                  1882.7070 &              -0.3104 &                          0.05 &                  11.59 &                  0.61 &                           0.10 &                        11.75 \\
    Si III] 1892 &                  1892.0290 &              -0.4239 &                          0.06 &                  15.22 &                  0.82 &                           0.12 &                        16.10 \\
  {[}C III] 1906 &                  1906.6800 &              -1.5011 &                          0.32 &                  52.55 &                  2.89 &                           0.61 &                        57.99 \\
     C III] 1908 &                  1908.7300 &              -1.1148 &                          0.11 &                  38.19 &                  2.14 &                           0.21 &                        43.16 \\
      N II] 2140 &                  2139.6800 &             >-0.1038 &                            .. &                     .. &               $<$0.06 &                             .. &                      $<$2.24 \\
  {[}O III] 2320 &                  2321.6640 &             >-0.1132 &                            .. &                     .. &               $<$0.06 &                             .. &                      $<$1.39 \\
      C II] 2323 &                  2324.2140 &             >-0.1023 &                            .. &                     .. &               $<$0.05 &                             .. &                      $<$1.24 \\
     C II] 2325c &                  2326.1130 &              -0.7885 &                          0.20 &                  13.72 &                  1.08 &                           0.27 &                        25.58 \\
     C II] 2325d &                  2327.6450 &              -0.3731 &                          0.10 &                  10.39 &                  0.51 &                           0.13 &                        12.01 \\
      C II] 2328 &                  2328.8380 &              -0.1435 &                          0.08 &                   4.10 &                  0.20 &                           0.11 &                         4.59 \\
    Si II] 2335a &                  2335.1230 &             >-0.1064 &                            .. &                     .. &               $<$0.05 &                             .. &                      $<$1.22 \\
    Si II] 2335b &                  2335.3210 &             >-0.1064 &                            .. &                     .. &               $<$0.05 &                             .. &                      $<$1.22 \\
      Fe II 2365 &                  2365.5520 &              -0.5465 &                          0.07 &                  14.59 &                  0.71 &                           0.10 &                        14.68 \\
     Fe II 2396a &                  2396.1497 &             >-0.1234 &                            .. &                     .. &               $<$0.06 &                             .. &                      $<$1.07 \\
     Fe II 2396b &                  2396.3559 &             >-0.1279 &                            .. &                     .. &               $<$0.06 &                             .. &                      $<$1.11 \\
   {[}O II] 2470 &                  2471.0270 &              -0.9819 &                          0.07 &                  25.09 &                  1.20 &                           0.08 &                        18.01 \\
      Fe II 2599 &                  2599.1465 &             >-0.1064 &                            .. &                     .. &               $<$0.04 &                             .. &                      $<$0.50 \\
      Fe II 2607 &                  2607.8664 &             >-0.1694 &                            .. &                     .. &               $<$0.07 &                             .. &                      $<$0.77 \\
      Fe II 2612 &                  2612.6542 &               -0.732 &                          0.20 &                  16.06 &                  0.77 &                           0.21 &                         8.89 \\
      Fe II 2614 &                  2614.6051 &             >-0.1453 &                            .. &                     .. &               $<$0.06 &                             .. &                      $<$0.65 \\
      Fe II 2618 &                  2618.3991 &             >-0.1384 &                            .. &                     .. &               $<$0.05 &                             .. &                      $<$0.61 \\
      Fe II 2621 &                  2621.1912 &             >-0.1351 &                            .. &                     .. &               $<$0.05 &                             .. &                      $<$0.59 \\
      Fe II 2622 &                  2622.4518 &             >-0.1569 &                            .. &                     .. &               $<$0.06 &                             .. &                      $<$0.69 \\
      Fe II 2626 &                  2626.4511 &              -0.9937 &                          0.10 &                  21.09 &                  1.04 &                           0.11 &                        11.67 \\
      Fe II 2629 &                  2629.0777 &             >-0.1376 &                            .. &                     .. &               $<$0.05 &                             .. &                      $<$0.59 \\
      Fe II 2631 &                  2631.8321 &             >-0.1259 &                            .. &                     .. &               $<$0.05 &                             .. &                      $<$0.54 \\
      Fe II 2632 &                  2632.1081 &             >-0.1245 &                            .. &                     .. &               $<$0.05 &                             .. &                      $<$0.53 \\
     Mg II 2797b &                  2798.7550 &               -1.089 &                          0.07 &                  23.33 &                  1.08 &                           0.07 &                        10.00 \\
     Mg II 2797d &                  2803.5310 &              -0.4567 &                          0.06 &                  10.97 &                  0.45 &                           0.06 &                         4.19 \\
       He I 2945 &                  2945.1030 &             >-0.1515 &                            .. &                     .. &               $<$0.07 &                             .. &                      $<$0.55 \\
\bottomrule
\end{tabular}

%% file: Tables/lineflux_restUV_detected_interv.tex
\caption{Intervening absorption lines in \knotE Magellan/MagE spectrum. W$_{\mathrm{r,fit}}$ denotes the rest-frame         equivalent width measured, in \AA. $z$ an $\Delta z$ are the redshift and corresponding uncertainty respectively, as         measured from our line fitting code.}
\begin{tabular}{lrrrrr}
\toprule
    Line ID &  $\lambda_{\mathrm{rest}}$ &  W$_{\mathrm{r,fit}}$ &  $\delta$ W$_{\mathrm{r,fit}}$ &  zz$_{\mathrm{}}$ &  zz$_{\mathrm{u}}$ \\
& (\AA) & (\AA) & (\AA) & & \\
\midrule
 Fe II 2344 &                  2344.2140 &                0.4791 &                         0.0612 &           0.98285 &            0.00006 \\
 Fe II 2383 &                  2382.7650 &                0.5997 &                         0.0578 &           0.98298 &            0.00004 \\
 Al II 1670 &                  1670.7874 &                0.0572 &                         0.0609 &           1.87880 &            0.00150 \\
 Fe II 2586 &                  2586.6500 &                0.3040 &                         0.0456 &           0.98290 &            0.00004 \\
 Fe II 2600 &                  2600.1729 &                0.8633 &                         0.3090 &           0.98295 &            0.00003 \\
 Mg II 2796 &                  2796.3520 &                1.0939 &                         0.0671 &           0.98293 &            0.00003 \\
 Mg II 2803 &                  2803.5310 &                1.2589 &                         0.0637 &           0.98295 &            0.00003 \\
  Mg I 2853 &                  2852.9640 &                0.2608 &                         0.0519 &           0.98299 &            0.00006 \\
\bottomrule
\end{tabular}